\documentclass{article}

\usepackage[preprint]{neurips_2025}

\usepackage[utf8]{inputenc} 
\usepackage[T1]{fontenc}    
\usepackage{hyperref}       
\usepackage{url}            
\usepackage{booktabs}       
\usepackage{amsfonts}       
\usepackage{nicefrac}       
\usepackage{microtype}      
\usepackage{xcolor}         

\newcommand{\myparatight}[1]{\smallskip\noindent{\bf {#1}:}~}
\usepackage{shortcut}
\usepackage{amsthm}
\usepackage{multirow}
\usepackage{makecell}
\usepackage{mathtools}
\usepackage{float}
\usepackage{graphics}
\usepackage{graphicx}
\usepackage[skip=0pt]{caption}
\usepackage{algorithm}
\usepackage{algorithmicx}
\usepackage{algpseudocode}
\usepackage{bm}
\usepackage{color}
\usepackage{multirow}
\usepackage{multicol}
\usepackage{makecell}
\usepackage{balance}
\usepackage{pdfx}
\usepackage{pifont}
\usepackage{tcolorbox}
\usepackage{colortbl}
        
\usepackage{amssymb}
\usepackage{bbm}
\usepackage{booktabs}
\usepackage{subcaption}

\usepackage[normalem]{ulem}

\usepackage{pifont}

\usepackage{wasysym}

\usepackage{wrapfig}

\allowdisplaybreaks

\definecolor{PineGreen}{RGB}{0,139,114}
\definecolor{BrickRed}{RGB}{140,55,62}
\definecolor{RedD}{RGB}{255,242,242}
\definecolor{RedD}{RGB}{255,242,242}
\definecolor{greyD}{RGB}{242,242,242}
\definecolor{BlueD}{RGB}{242,242,255}
\definecolor{GreenD}{RGB}{242,255,242}
\newcommand{\cmark}{{\color{PineGreen}\ding{51}}}%
\newcommand{\xmark}{{\color{BrickRed}\ding{55}}}%

\definecolor{myurlcolor}{rgb}{0.1, 0.2, 0.8}

\usepackage{hyperref}
\hypersetup{
	colorlinks=true,
	linkcolor=blue,
	filecolor=magenta,      
        urlcolor=myurlcolor,
        citecolor=blue,
	pdftitle={Overleaf Example},
	pdfpagemode=FullScreen,
}

\algnewcommand\algorithmicforpara{\textbf{for}}
\algnewcommand\algorithmicdoinparallel{\textbf{do in parallel}}
\algdef{S}[FOR]{ForParallel}[1]{\algorithmicforpara\ #1\ \algorithmicdoinparallel}

\title{Benchmarking Poisoning Attacks against Retrieval-Augmented Generation}

\author{
    Baolei Zhang$^{1\dagger}$,
    Haoran Xin$^{1\dagger}$,
    Jiatong Li$^{1\dagger}$,
    Dongzhe Zhang$^{1\dagger}$, \\
    \textbf{Minghong Fang$^{2*}$,
    Zhuqing Liu$^3$,
    Lihai Nie$^{1*}$, 
    Zheli Liu$^1$} \\
    $^1$Nankai University \\
    $^2$University of Louisville \\
    $^3$University of North Texas \\
    \texttt{\{zhangbaolei,haoranxin,lijiatong,zhangdongzhe\}@mail.nankai.edu.cn}\\
    \texttt{minghong.fang@louisville.edu, zhuqing.liu@unt.edu} \\
    \texttt{\{NLH, liuzheli\}@nankai.edu.cn } \\
}

\begin{document}

\footnotetext[2]{~Equal Contribution.}
\footnotetext[1]{~Corresponding Author.}

\maketitle

\begin{abstract}

Retrieval-Augmented Generation (RAG) has proven effective in mitigating hallucinations in large language models by incorporating external knowledge during inference. However, this integration introduces new security vulnerabilities, particularly to poisoning attacks. Although prior work has explored various poisoning strategies, a thorough assessment of their practical threat to RAG systems remains missing. To address this gap, we propose the first comprehensive benchmark framework for evaluating poisoning attacks on RAG. Our benchmark covers 5 standard question answering (QA) datasets and 10 expanded variants, along with 13 poisoning attack methods and 7 defense mechanisms, representing a broad spectrum of existing techniques.
Using this benchmark, we conduct a comprehensive evaluation of all included attacks and defenses across the full dataset spectrum. Our findings show that while existing attacks perform well on standard QA datasets, their effectiveness drops significantly on the expanded versions.
Moreover, our results demonstrate that various advanced RAG architectures, such as sequential, branching, conditional, and loop RAG, as well as multi-turn conversational RAG, multimodal RAG systems, and RAG-based LLM agent systems, remain susceptible to poisoning attacks. Notably, current defense techniques fail to provide robust protection, underscoring the pressing need for more resilient and generalizable defense strategies.

\end{abstract}


\section{Introduction} \label{sec:intro}

Despite the remarkable capabilities of large language models (LLMs)~\cite{achiam2023gpt,anil2023palm,brown2020language}, they often generate factually incorrect or nonsensical outputs, commonly referred to as hallucinations. Retrieval-Augmented Generation (RAG)~\cite{borgeaud2022improving, chen2024benchmarking, gao2023retrieval,jiang2023active,karpukhin2020dense, lewis2020retrieval,salemi2024evaluating,thoppilan2022lamda} has emerged as a promising framework to mitigate hallucinations by incorporating external knowledge at inference time. A typical RAG pipeline comprises three key components: a knowledge database, a retriever, and an LLM. Given a user query, the retriever selects the top-$K$ relevant documents from the database, which are then used to condition the LLM’s response generation.

As RAG techniques continue to evolve, several benchmarks~\cite{chen2024benchmarking,cuconasu2024power,ru2024ragchecker,yang2024crag,zeng2025worse} have been proposed to evaluate their performance in terms of generation accuracy and robustness against naturally occurring noise or counterfactual content in the knowledge database. However, the security risks to RAG systems, particularly poisoning attacks, remain largely underexplored. Since RAG knowledge databases are typically built by aggregating content from publicly available sources such as Wikipedia~\cite{thakur2021beir}, they present an opportunity for the attacker to inject malicious content. This poisoned content may later be retrieved and influence the system’s output. Although SafeRAG~\cite{liang2025saferag} introduces four specific attack tasks aimed at testing security bypasses, it does not provide a thorough evaluation of existing poisoning techniques. Therefore, there is a clear need for more comprehensive research and systematic assessment of poisoning attacks on RAG systems.

To bridge the gap in understanding the security vulnerabilities of RAG systems, we introduce RAG Security Bench (RSB), a unified benchmark that systematically evaluates a broad range of poisoning attacks and defense mechanisms across diverse RAG architectures. 
The RSB benchmark includes 13 representative poisoning attacks, classified into three categories based on their adversarial objectives: targeted poisoning, denial-of-service (DoS), and trigger-based DoS. Targeted poisoning attacks~\cite{chang2025one,liu2024formalizing,roychowdhury2024confusedpilot,su2024corpus,tan2024glue,zhang2025practical,zhang2024hijackrag,zou2024poisonedrag} aim to manipulate the system’s output for specific queries through the injection of maliciously crafted content. DoS attacks~\cite{shafran2024machine,suo2025hoist} attempt to suppress the model’s response for certain inputs, while trigger-based DoS attacks~\cite{chaudhari2024phantom,chen2024agentpoison,xue2024badrag} extend this threat by causing the model to refuse any query containing predefined triggers.
On the defense side, RSB incorporates 7 representative methods, organized into three categories. Process-optimized defenses~\cite{jain2023baseline,wang2024astute,wei2024instructrag,xiang2024certifiably,zou2024poisonedrag} aim to improve system resilience by optimizing prompts and retrieval procedures. Detection-based defenses~\cite{jelinek1980interpolated,shafran2024machine,xue2024badrag,zhang2025practical,zhang2025traceback,zhong2023poisoning} focus on identifying and removing poisoned entries from the knowledge base. Hybrid defenses~\cite{zhou2025trustrag} combine detection techniques with prompt engineering or system-level adaptations to enhance robustness.
RSB further extends its evaluation to 6 advanced RAG frameworks spanning four categories: sequential RAG with fixed retrieval-generation pipelines, branching RAG that explores multiple retrieval paths, conditional RAG that adapts retrieval based on the query content, and loop RAG that performs iterative refinement through repeated retrieval and generation cycles. 
In addition to these frameworks, RSB also encompasses evaluations of advanced architectures, including multi-turn conversational RAG, multimodal RAG systems, and RAG-based LLM agent systems, to provide a more complete picture of security risks across emerging RAG paradigms.

\myparatight{Empirical findings}%
Leveraging the RSB framework, we conduct comprehensive evaluations of all poisoning attacks across 15 datasets, which include five widely-used QA datasets along with their challenging expanded versions. These expanded datasets introduce a higher density of correct-answer texts that are semantically close to the target queries, increasing both retrieval redundancy and contextual coverage. Based on these evaluations, we summarize the following key findings.

\emph{Effectiveness.} 1) Most poisoning attacks retain strong effectiveness on the original datasets, highlighting the inherent susceptibility of RAG systems to adversarial content injection. 2) However, their effectiveness significantly declines on the challenging expansions, suggesting that enriching the knowledge database with more diverse and redundant correct-answer texts can passively reduce the impact of poisoning, offering a simple yet effective layer of defense. 3) Notably, attacks that are explicitly optimized for individual poisoned texts, such as CRAG-AS, maintain high success rates even in the more robust settings, demonstrating the strength of fine-grained optimization in overcoming increased retrieval resilience.

\emph{Defenses.} 1) Process-optimized defenses are effective in mitigating DoS attacks but offer limited protection against targeted poisoning attacks. 
2) Detection-based defenses tend to have negligible impact across most attack types, suggesting that current detection methods are inadequate for identifying well-crafted poisoned content. 
3) Hybrid defenses consistently achieve better performance than other methods, yet they can only partially mitigate the effects of poisoning attacks.
These results reveal critical limitations in existing defense mechanisms and underscore the pressing need for more robust and comprehensive solutions.

\emph{Ablation studies.}  1) State-of-the-art LLMs continue to show significant vulnerability to poisoned contextual inputs, revealing a critical limitation of current alignment techniques that prioritize prompt-level control while overlooking harmful content embedded in retrieved context.
2) All evaluated retrievers exhibit substantial susceptibility to poisoning attacks, underscoring the need for retriever training approaches that explicitly improve adversarial robustness.
3) Retrieval based on dot product similarity is particularly vulnerable, as it provides a larger optimization space for adversaries compared to cosine similarity. This suggests that developing more robust similarity metrics may serve as an effective line of defense.
4) On the original NQ dataset, most attacks remain highly effective regardless of the number of retrieved texts, challenging the assumption that increasing retrieval depth inherently enhances robustness.
5) In contrast, on the challenging expansions, increasing the number of retrieved texts does not correspond to higher attack success rates despite greater recall of poisoned content. This indicates that the presence of abundant, semantically relevant correct-answer texts can neutralize the influence of adversarial inputs.

\emph{Transferability.} 1) Poisoned texts originally designed for naive RAG systems exhibit unexpected transferability, effectively influencing outputs in several advanced RAG architectures.
2) Attack success drops markedly in frameworks with adaptive retrieval strategies, revealing a key weakness of existing poisoning techniques and the inherent challenge of manipulating queries that do not reliably initiate retrieval in such systems.

\emph{Multi-turn RAG.} Existing poisoning attacks show reduced effectiveness when targeting multi-turn conversational RAG systems, revealing the limitations of current attack strategies. This highlights the inherent difficulty of executing successful poisoning attacks in multi-turn conversational settings.

\emph{Multimodal RAG.} 1) Multimodal RAG systems also exhibit vulnerability to poisoning attacks due to their reliance on retrieval and context augmentation processes similar to those in standard RAG, indicating that current multimodal retrievers and vision-language models lack robustness against malicious textual inputs. 
2) Furthermore, the weak semantic correspondence between images and texts in the knowledge database makes it easier for the attacker to carry out successful attacks on multimodal RAG systems.

\emph{RAG-based LLM agents.} 1) RAG-based LLM agent systems remain susceptible to poisoning attacks, where both previously known and slightly modified attack strategies perform similarly well.
2) The added complexity of these agents does not hinder attack execution; instead, their reliance on retrieval mechanisms makes it straightforward to adapt existing poisoning techniques.


\section{Preliminaries and Related Work}

\subsection{Retrieval-Augmented Generation (RAG)}

\begin{wrapfigure}{r}{0.4\textwidth}
  \vspace{-12mm}
\centering
    \centering  
    \includegraphics[width=0.4\textwidth]{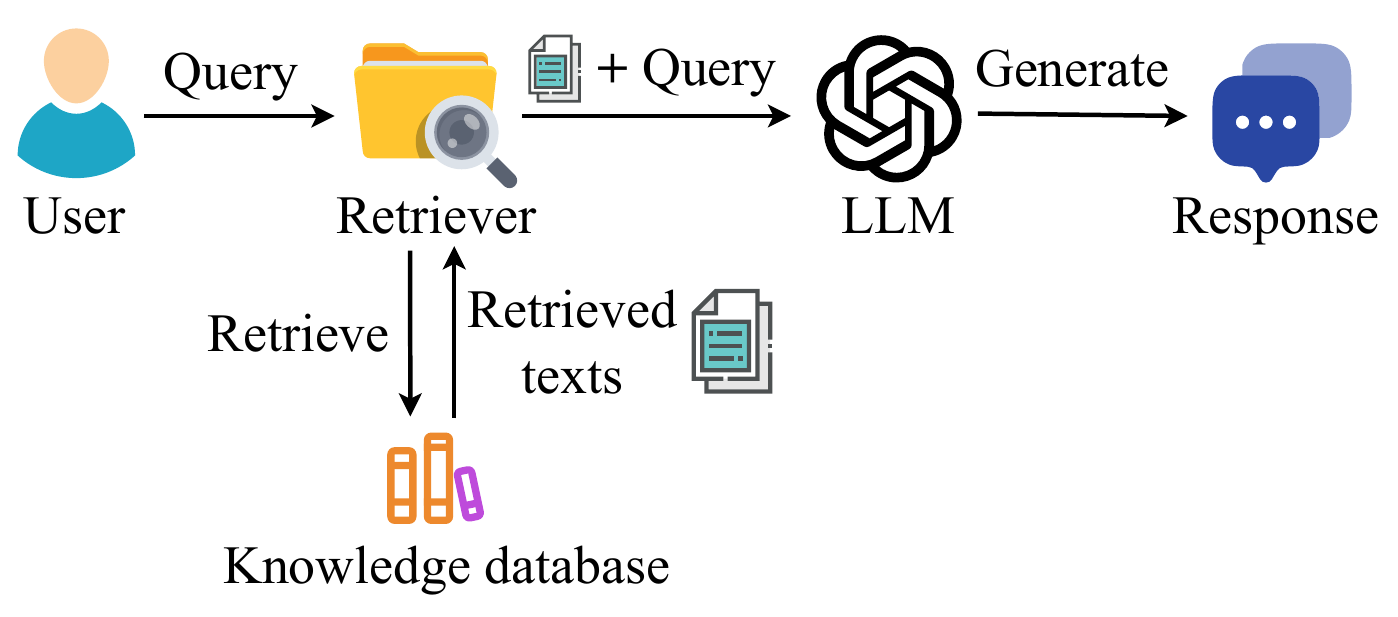} 
    \caption{Illustration of the standard workflow in a RAG system.}  
    \label{fig:rag_overview}  
\end{wrapfigure}

As shown in Fig.~\ref{fig:rag_overview}, a typical RAG framework consists of three core components: a {\em knowledge database}, a {\em retriever}, and an {\em LLM}. The knowledge database $\mathcal{D}$ is a collection of textual content sourced from diverse domains, including encyclopedic entries~\cite{thakur2021beir}, news reports~\cite{soboroff2019trec}, and domain-specific documents such as financial records~\cite{loukas2023making}. Given a user query $q$, the system follows a two-stage process to generate a response:

\begin{list}{\labelitemi}{\leftmargin=1em \itemindent=-0.0em \itemsep=.2em}

\item \textbf{Knowledge retrieval:}
In the first stage, given a user query $q$, the retriever encodes it into an embedding vector and computes its similarity to the embeddings of all entries in the knowledge database $\mathcal{D}$. The top-$K$ most relevant texts are then retrieved based on these similarity scores, denoted as $\mathcal{E}_K(q, \mathcal{D})$.

\item \textbf{Response generation:} 
In this stage, the retrieved texts $\mathcal{E}_K(q, \mathcal{D})$ are concatenated with the query $q$ using a system prompt $p_\text{sys}$ (an example is provided in Appendix~\ref{appendix:system_prompts}), and the resulting input is fed into the LLM to generate the final output $R$, expressed as $R = \operatorname{LLM}\left(p_\text{sys}, q,\mathcal{E}_K(q,\mathcal{D})\right)$.

\end{list}

\subsection{Related Work}

\myparatight{Threat landscape of RAG systems}%
Recent research has identified three primary attack vectors against RAG systems: (1) {\em privacy inference attacks}, which extract confidential information through carefully crafted queries~\cite{jiang2024rag,li2024seeing,naseh2025riddle}; (2) {\em trigger-based retriever attacks}, which manipulate the retriever using embedded triggers to influence downstream generation~\cite{cheng2024trojanrag, long2024backdoor}; and (3) {\em poisoning attacks}, which inject adversarial content into the knowledge database~\cite{chaudhari2024phantom,chen2024agentpoison, shafran2024machine,su2024corpus, xue2024badrag,zhang2025practical,zou2024poisonedrag}.
Privacy inference attacks are conceptually aligned with jailbreaking techniques that have been extensively examined in the broader LLM security literature. Trigger-based retriever 
 attacks typically assume strong attacker capabilities, such as modify the retriever model. This assumption is often unrealistic in practice, especially for systems that rely on closed-source or publicly trusted retrievers.
In contrast, poisoning attacks directly exploit the unique architecture of RAG systems, particularly the reliance on a textual knowledge database composed of content from diverse, and often uncurated, sources. This makes the injection of malicious content both practical and feasible for real-world adversaries. Despite their relevance, poisoning attacks on RAG have not been systematically studied under a unified framework.
To address this gap, our work presents a dedicated benchmark that focuses on poisoning attacks and their defenses in RAG systems. 

\myparatight{Distinctions from prior RAG benchmarking efforts}%
Existing benchmarks~\cite{chen2024benchmarking, cuconasu2024power,ru2024ragchecker,yang2024crag,zeng2025worse,zheng2025retrieval} mainly assess RAG systems in non-adversarial settings, focusing on performance under natural noise or misleading information. For instance, RGB~\cite{chen2024benchmarking}, CRAG~\cite{yang2024crag}, and RAGuard~\cite{zeng2025worse} evaluate robustness to imperfect retrieval rather than malicious interference. 
SafeRAG~\cite{liang2025saferag} and~\cite{zhou2024trustworthiness} introduce several security-oriented tasks but do not offer a thorough evaluation of poisoning-based threats.
Our benchmark fills this gap by systematically analyzing poisoning attacks, standardizing threat models and metrics, and enabling fair comparison of existing defenses.

\section{Threat Model}
\label{Threat_model}

We summarize the objectives, assumptions, and capabilities of the attacker in existing RAG poisoning attacks~\cite{chaudhari2024phantom,chen2024agentpoison, shafran2024machine,su2024corpus, xue2024badrag,zhang2025practical,zou2024poisonedrag}.

\myparatight{Attacker's objectives}%
We categorize the objectives of existing poisoning attacks into three escalating types,
from explicit manipulation to more systematic disruption. 
The first category is {\em targeted poisoning attack}, where the attacker aims to make the RAG return a specific, attacker-chosen response to a particular query. For example, when asked ``Who is the CEO of OpenAI?'', the system might be manipulated to answer ``Tim Cook''. The second category is {\em denial-of-service (DoS) attack}, which aim to render the system unusable by causing it to refuse to answer general user queries, such as responding with ``I don’t know''. The final category is {\em trigger-based DoS attack}, where the attacker poisons the system to refuse responses only when specific trigger phrases, such as ``OpenAI'', appear, thereby selectively blocking functionality on targeted topics.

\begin{wraptable}{r}{0.4\textwidth}
  \vspace{-5mm}
\tiny
\centering
\addtolength{\tabcolsep}{-5.0pt}
\caption{Poisoning attacks against RAG by attacker’s background knowledge.}
\label{tab:summary_poisoning_attacks}
\begin{tabular}{l|l|cccc}
\toprule
Category & Attack & \begin{tabular}[c]{@{}c@{}}Knowledge\\ database\end{tabular} & Retriever & LLM & \begin{tabular}[c]{@{}c@{}}Targeted\\ query\end{tabular} \\ \midrule
\rowcolor{BlueD}& BPRAG~\cite{zou2024poisonedrag} & 

\xmark & \xmark & \xmark & \cmark \\
 \rowcolor{BlueD}& WPRAG~\cite{zou2024poisonedrag} & \xmark & \cmark & \xmark & \cmark \\
 \rowcolor{BlueD}& BPI~\cite{liu2024formalizing,zou2024poisonedrag} & \xmark & \xmark & \xmark & \cmark \\
 \rowcolor{BlueD}& WPI~\cite{liu2024formalizing,zou2024poisonedrag} & \xmark & \cmark & \xmark & \cmark \\
\rowcolor{BlueD} & AGGD~\cite{su2024corpus} & \xmark & \cmark & \xmark & \cmark \\
 \rowcolor{BlueD}& CRAG-AS~\cite{zhang2025practical} & \xmark & \xmark & \xmark & \cmark \\
 \rowcolor{BlueD}\multirow{-7}{*}{\begin{tabular}[c]{@{}c@{}}Targeted\\poisoning \end{tabular}}& CRAG-AK~\cite{zhang2025practical} & \xmark & \xmark & \xmark & \cmark \\ \midrule

  \rowcolor{GreenD}& JamInject~\cite{shafran2024machine} & \cellcolor{green!5}\xmark & \xmark & \xmark & \cmark \\
  \rowcolor{GreenD}& JamOracle~\cite{shafran2024machine} & \xmark & \xmark & \xmark & \cmark \\
 \rowcolor{GreenD}\multirow{-3}{*}{\ \ \ \ DoS} & JamOpt~\cite{shafran2024machine} & \xmark & \xmark & \xmark & \cmark \\ \midrule

\rowcolor{RedD}
& AP~\cite{chen2024agentpoison}     & \cmark & \cmark & \xmark & \xmark \\
\rowcolor{RedD}
  & BadRAG~\cite{xue2024badrag}       & \xmark & \cmark & \xmark & \xmark \\
\rowcolor{RedD}
  \multirow{-3}{*}{
\begin{tabular}[c]{@{}c@{}}Trigger-\\ based\\ DoS\end{tabular}}   & Phantom~\cite{chaudhari2024phantom} & \xmark & \cmark & \xmark & \xmark \\
 \bottomrule
\end{tabular}
\end{wraptable}

\myparatight{Attacker's background knowledge}%
We examine the attacker's background knowledge with respect to the three main components of a RAG system. For the knowledge database, we consider both cases where the attacker knows or does not know its contents. For the LLM, we adopt a practical threat model in which the attacker has no access to its internal parameters. For the retriever, we consider both {\em black-box} and {\em white-box} settings: in the black-box case, the attacker cannot access or query the retriever; in the white-box case, the attacker can access the retriever's parameters but cannot modify them. Additionally, we account for scenarios where the attacker may or may not know the targeted queries, that is, the user queries the attacker intends to influence.

\myparatight{Attacker's capabilities}%
We assume the attacker can inject arbitrary text into the knowledge database by modifying the data sources from which the database is constructed. For instance, if the database is populated from Wikipedia, the attacker may edit relevant pages to embed malicious content, following recent poisoning techniques such as those in~\cite{carlini2024poisoning}. However, we do not assume the attacker can alter user queries, as doing so would fall under the category of jailbreak attacks rather than data poisoning.


\section{Formalizing Poisoning Attacks Against RAG}
\label{sec:attacks}

As outlined in Section~\ref{Threat_model}, poisoning attacks against RAG fall into three categories based on the attacker's objectives: targeted poisoning, denial-of-service (DoS), and trigger-based DoS. In this section, we provide formal mathematical definitions for each of these attack types.

\myparatight{Definition 1: Targeted Poisoning Attack}
\textit{In a targeted poisoning attack, the attacker selects a set of targeted queries $\mathcal{Q}$ and injects $M$ poisoned texts into the knowledge database $\mathcal{D}$ for each query $q_i \in \mathcal{Q}$, aiming to make the RAG system return a predefined answer $a_i$ when queried with $q_i$.}

Formally, the attacker aims to maximize the following objective:
\begin{align}
\label{eq:targeted_attack}
\frac{1}{|\mathcal{Q}|} \mathbbm{1} \left( \operatorname{LLM}(p_\text{sys}, q_i, \mathcal{E}_K(q_i, \mathcal{D}_\text{poison})) = a_i \right),
\end{align}
where $\mathbbm{1}$ is the indicator function, $p_\text{sys}$ is the system prompt, and $\mathcal{E}_K(q_i, \mathcal{D}_\text{poison})$ denotes the top-$K$ texts retrieved from the poisoned knowledge database for the query $q_i$. 
The poisoned knowledge database is defined as $\mathcal{D}_\text{poison} = \mathcal{D} \cup \mathcal{P}$, where $\mathcal{P} = \{ \mathcal{P}_i^j \mid i = 1, \dots, |\mathcal{Q}|,\ j = 1, \dots, M \}$ represents the set of poisoned texts injected for all targeted queries. Here, $\mathcal{P}_i^j$ denotes the $j$-th poisoned text associated with the $i$-th targeted query.

State-of-the-art targeted poisoning attacks include Black-box PoisonedRAG (BPRAG)~\cite{zou2024poisonedrag}, White-box PoisonedRAG (WPRAG)~\cite{zou2024poisonedrag}, Black-box prompt injection (BPI)~\cite{liu2024formalizing, zou2024poisonedrag}, White-box prompt injection (WPI)~\cite{liu2024formalizing, zou2024poisonedrag}, AGGD~\cite{su2024corpus}, CorruptRAG-AS (CRAG-AS)~\cite{zhang2025practical}, and CorruptRAG-AK (CRAG-AK)~\cite{zhang2025practical}. See Appendix~\ref{appendix_sec:targeted_poisoning_attacks} for further details on these attacks.

\myparatight{Definition 2: Denial-of-Service (DoS) Attack}%
\textit{In a DoS attack, the attacker selects a set of targeted queries $\mathcal{Q}$ and injects $M$ poisoned texts into the knowledge database $\mathcal{D}$ for each query $q_i \in \mathcal{Q}$, with the goal of causing the RAG system to produce a refusal response $a$ (e.g., ``I don't know'') when queried with $q_i$.}

Formally, the attacker aims to maximize:
\begin{align}
\label{eq:targeted_attack}
\frac{1}{\left|\mathcal{Q}\right|}\mathbbm{1}\left(\operatorname{LLM}\left(p_\text{sys}, q_i,\mathcal{E}_K(q_i,\mathcal{D}_\text{poison})\right)=a\right).
\end{align}

Prominent DoS attacks include Jamming-based instruction injection (JamInject)~\cite{shafran2024machine}, Jamming-based oracle generation (JamOracle)~\cite{shafran2024machine}, and Jamming-based black-box optimization (JamOpt)~\cite{shafran2024machine}. Additional details on these attacks are provided in Appendix~\ref{appendix_sec:dos_attacks}.

\myparatight{Definition 3: Trigger-based DoS Attack}%
\textit{Given a query distribution $\pi_{q^t}$, where each user query $q^t$ contains a specific trigger string $t$, the attacker performs a trigger-based DoS attack by injecting $M$ poisoned texts into the knowledge database $\mathcal{D}$, aiming to cause the RAG system to output a refusal response $a$ for any query $q^t \sim \pi_{q^t}$.}

Formally, the attacker's objective is to maximize:
\begin{align}
\label{eq:trigger_dos_attack}
\mathbb{E}_{q^t \sim \pi_{q^t}}\left[\mathbbm{1}\left(\operatorname{LLM}\left(p_\text{sys}, q^t,\mathcal{E}_K(q^t,\mathcal{D}_\text{poison})\right)=a\right)\right].
\end{align}

Trigger-based DoS attacks include AgentPoison (AP)~\cite{chen2024agentpoison}, BadRAG~\cite{xue2024badrag}, and Phantom~\cite{chaudhari2024phantom}. Further details on these attacks are provided in Appendix~\ref{appendix_sec:trigger_dos_attacks}.

Table~\ref{tab:summary_poisoning_attacks} provides a comprehensive summary of the aforementioned poisoning attacks against RAG, categorized by the attacker’s level of knowledge.
Appendix~\ref{appendix_other_attacks} provides a list of additional attacks that were not included in our experiments, along with explanations for their exclusion from our benchmarks.

\section{Evaluation}  
 \label{sec:exp}

\subsection{Experimental Setup (Datasets, Evaluation Metrics, Targeted queries, RAG Settings)} 
\label{sec:experimental_setup}

Our benchmark uses fifteen QA datasets, including five established ones: Natural Questions (NQ)~\cite{kwiatkowski2019natural}, HotpotQA~\cite{yang2018hotpotqa}, MS-MARCO~\cite{nguyen2016ms}, SQuAD~\cite{rajpurkar2016squad}, and BoolQ~\cite{clark2019boolq}. Each dataset contains queries paired with ground-truth relevant texts, embedded within large knowledge databases. To evaluate attacks under different information densities, we introduce two expansions for each dataset by adding extra ground-truth texts at medium (EX-M) and large (EX-L) scales. Full dataset statistics and construction details are in Appendix~\ref{appendix:detail_datasets} and Table~\ref{tab:datasets} (Appendix).
We use three metrics: accuracy (ACC), attack success rate (ASR), and F1-score. ACC and ASR measure the rates of correct and targeted answers, evaluated using GPT-4o-mini. 
F1-score measures the accuracy of retrieval.
See Appendix~\ref{appendix:details_of_metrics} for metric details.
Existing studies use different sets of targeted queries, making comparisons difficult. To enable fair evaluation, we construct a unified set of 100 targeted queries and corresponding targeted answers for each attack category (see Section~\ref{sec:attacks}). Their construction is detailed in Appendix~\ref{appendix:details_of_targeted_answers}. 
These queries are selected such that the RAG system does not yield the attacker-chosen targeted answer without attack.
Answer accuracy of the RAG system under non-attack conditions are shown in Table~\ref{tab:performanc_100_ques} (Appendix).
Our benchmark uses the FlashRAG~\cite{jin2024flashrag} framework to build the RAG system, with Contriever~\cite{izacard2021unsupervised} as the retriever and GPT-4o-mini as the LLM by default. 
We retrieve the top-5 texts ($K=5$) by cosine similarity and prepend them to the query using a system prompt (Appendix~\ref{appendix:system_prompts}).
We conducted all experiments on NVIDIA A800 GPUs, running each test five times and reporting the average results. The variance of results was small, so we omit it.

\begin{table*}[!t]
\tiny
\centering
\addtolength{\tabcolsep}{-4.0pt}
\caption{The results of all poisoning attacks on various datasets.}
\label{tab:comparision}
\begin{tabular}{l|c|ccccccc|ccc|ccc}
\toprule
\multirow{2}{*}{Dataset} & \multirow{2}{*}{Metric} & \multicolumn{7}{c|}{Targeted poisoning attack} & \multicolumn{3}{c|}{DoS attack} & \multicolumn{3}{c}{Trigger-based DoS attack} \\ \cmidrule{3-15} 
 &  & BPRAG & WPRAG & BPI & WPI & AGGD & CRAG-AS & CRAG-AK & JamInject & JamOracle & JamOpt & AP & BadRAG & Phantom \\ \midrule
\multirow{3}{*}{NQ} & ACC & 0.27 & 0.25 & 0.02 & 0.01 & 0.33 & 0.06 & 0.04 & 0.15 & 0.13 & 0.29 & 0.01 & 0.65 & 0.99 \\
 & ASR & 0.62 & 0.64 & 0.94 & 0.97 & 0.51 & 0.89 & 0.88 & 0.85 & 0.87 & 0.59 & 0.99 & 0.35 & 0.00 \\
 & F1-score & 0.96 & 0.96 & 0.91 & 0.93 & 0.78 & 0.86 & 0.95 & 0.75 & 0.83 & 0.76 & 1.00 & 0.37 & 0.00 \\ \midrule
\rowcolor{greyD} 
\cellcolor{greyD}   & ACC & 0.13 & 0.15 & 0.00 & 0.00 & 0.15 & 0.00 & 0.00 & 0.00 & 0.05 & 0.19 & 0.00 & 0.69 & 0.97 \\
 \rowcolor{greyD} 
\cellcolor{greyD} & ASR & 0.81 & 0.79 & 0.99 & 1.00 & 0.82 & 1.00 & 0.99 & 1.00 & 0.95 & 0.69 & 1.00 & 0.30 & 0.00 \\
 \rowcolor{greyD} 
   \cellcolor{greyD}\multirow{-3}{*}{HotpotQA}& F1-score & 1.00 & 1.00 & 1.00 & 1.00 & 0.98 & 1.00 & 1.00 & 0.96 & 1.00 & 0.80 & 1.00 & 0.40 & 0.00 \\ \midrule
 \multirow{3}{*}{MS-MARCO} & ACC & 0.06 & 0.10 & 0.08 & 0.05 & 0.25 & 0.19 & 0.02 & 0.32 & 0.33 & 0.57 & 0.00 & 0.94 & 0.97 \\
 & ASR & 0.81 & 0.78 & 0.90 & 0.93 & 0.63 & 0.76 & 0.96 & 0.68 & 0.62 & 0.35 & 1.00 & 0.06 & 0.00 \\
 & F1-score & 0.93 & 0.84 & 0.84 & 0.83 & 0.66 & 0.68 & 0.95 & 0.62 & 0.56 & 0.48 & 1.00 & 0.05 & 0.00 \\ \midrule

 \rowcolor{greyD} 
\cellcolor{greyD} & ACC & 0.20 & 0.22 & 0.06 & 0.03 & 0.21 & 0.13 & 0.07 & 0.20 & 0.26 & 0.37 & 1.00 & 0.97 & 0.99 \\
 \rowcolor{greyD} 
\cellcolor{greyD}  & ASR & 0.65 & 0.68 & 0.94 & 0.96 & 0.68 & 0.84 & 0.91 & 0.80 & 0.71 & 0.51 & 0.00 & 0.00 & 0.00 \\
   \rowcolor{greyD} 
\cellcolor{greyD} \multirow{-3}{*}{BoolQ} & F1-score & 0.96 & 0.94 & 0.89 & 0.87 & 0.89 & 0.78 & 0.97 & 0.73 & 0.85 & 0.70 & 1.00 & 0.00 & 0.00 \\ \midrule
 \multirow{3}{*}{SQuAD} & ACC & 0.08 & 0.07 & 0.00 & 0.00 & 0.08 & 0.00 & 0.02 & 0.00 & 0.01 & 0.06 & 0.00 & 0.97 & 0.99 \\
 & ASR & 0.91 & 0.89 & 1.00 & 0.99 & 0.91 & 0.99 & 0.96 & 1.00 & 0.99 & 0.80 & 1.00 & 0.00 & 0.00 \\
 & F1-score & 0.95 & 0.95 & 0.97 & 0.96 & 0.91 & 0.97 & 0.96 & 0.92 & 0.90 & 0.80 & 1.00 & 0.00 & 0.01 \\ \midrule

\rowcolor{greyD} 
\cellcolor{greyD}  & ACC & 0.65 & 0.59 & 0.77 & 0.80 & 0.87 & 0.81 & 0.36 & 0.97 & 0.94 & 0.98 & 0.94 & 0.99 & 1.00 \\
 \rowcolor{greyD} 
\cellcolor{greyD}  & ASR & 0.11 & 0.15 & 0.16 & 0.14 & 0.04 & 0.14 & 0.50 & 0.03 & 0.05 & 0.01 & 0.05 & 0.01 & 0.00 \\
  \rowcolor{greyD} 
\cellcolor{greyD}\multirow{-3}{*}{NQ-EX-M} & F1-score & 0.48 & 0.53 & 0.20 & 0.26 & 0.20 & 0.07 & 0.40 & 0.06 & 0.25 & 0.08 & 0.09 & 0.00 & 0.00 \\ \midrule
\multirow{3}{*}{HotpotQA-EX-M} & ACC & 0.64 & 0.72 & 0.88 & 0.92 & 0.88 & 0.87 & 0.33 & 1.00 & 0.90 & 1.00 & 0.99 & 1.00 & 1.00 \\
 & ASR & 0.20 & 0.20 & 0.09 & 0.05 & 0.03 & 0.10 & 0.47 & 0.00 & 0.10 & 0.00 & 0.01 & 0.00 & 0.00 \\
 & F1-score & 0.50 & 0.46 & 0.08 & 0.09 & 0.15 & 0.06 & 0.36 & 0.02 & 0.40 & 0.00 & 0.00 & 0.00 & 0.00 \\ \midrule

  \rowcolor{greyD} 
\cellcolor{greyD}& ACC & 0.34 & 0.49 & 0.63 & 0.63 & 0.72 & 0.88 & 0.15 & 0.99 & 0.97 & 0.99 & 0.97 & 0.99 & 1.00 \\
  \rowcolor{greyD} 
\cellcolor{greyD} & ASR & 0.30 & 0.29 & 0.33 & 0.32 & 0.13 & 0.07 & 0.66 & 0.00 & 0.01 & 0.02 & 0.03 & 0.01 & 0.00 \\
  \rowcolor{greyD} 
\cellcolor{greyD} \multirow{-3}{*}{MS-MARCO-EX-M} & F1-score & 0.57 & 0.51 & 0.27 & 0.32 & 0.26 & 0.05 & 0.69 & 0.01 & 0.09 & 0.04 & 0.06 & 0.00 & 0.00 \\ \midrule
 \multirow{3}{*}{BoolQ-EX-M} & ACC & 0.49 & 0.54 & 0.77 & 0.66 & 0.74 & 0.94 & 0.33 & 0.99 & 1.00 & 0.97 & 1.00 & 0.99 & 0.99 \\
 & ASR & 0.35 & 0.34 & 0.24 & 0.28 & 0.14 & 0.05 & 0.48 & 0.00 & 0.00 & 0.03 & 0.00 & 0.00 & 0.00 \\
 & F1-score & 0.53 & 0.54 & 0.17 & 0.18 & 0.28 & 0.04 & 0.55 & 0.02 & 0.10 & 0.06 & 0.02 & 0.00 & 0.00 \\ \midrule

  \rowcolor{greyD} 
\cellcolor{greyD} & ACC & 0.52 & 0.58 & 0.68 & 0.76 & 0.77 & 0.79 & 0.28 & 1.00 & 0.87 & 1.00 & 1.00 & 0.99 & 0.98 \\
  \rowcolor{greyD} 
\cellcolor{greyD}  & ASR & 0.21 & 0.24 & 0.24 & 0.18 & 0.10 & 0.18 & 0.65 & 0.00 & 0.13 & 0.00 & 0.00 & 0.00 & 0.00 \\
  \rowcolor{greyD} 
\cellcolor{greyD} \multirow{-3}{*}{SQuAD-EX-M} & F1-score & 0.44 & 0.41 & 0.25 & 0.24 & 0.25 & 0.09 & 0.51 & 0.04 & 0.46 & 0.00 & 0.00 & 0.00 & 0.00 \\ \midrule

\multirow{3}{*}{NQ-EX-L} & ACC & 0.89 & 0.87 & 0.90 & 0.86 & 0.97 & 0.97 & 0.74 & 0.98 & 0.94 & 0.98 & 0.95 & 0.99 & 1.00 \\
 & ASR & 0.03 & 0.05 & 0.08 & 0.10 & 0.00 & 0.00 & 0.20 & 0.02 & 0.04 & 0.01 & 0.03 & 0.01 & 0.00 \\
 & F1-score & 0.19 & 0.24 & 0.08 & 0.11 & 0.07 & 0.00 & 0.17 & 0.02 & 0.14 & 0.02 & 0.00 & 0.00 & 0.00 \\ \midrule

\rowcolor{greyD} 
\cellcolor{greyD} & ACC & 0.88 & 0.87 & 0.97 & 0.98 & 0.98 & 0.99 & 0.77 & 1.00 & 0.97 & 1.00 & 0.98 & 1.00 & 1.00 \\
  \rowcolor{greyD} 
\cellcolor{greyD}  & ASR & 0.05 & 0.05 & 0.03 & 0.02 & 0.02 & 0.01 & 0.12 & 0.00 & 0.03 & 0.00 & 0.01 & 0.00 & 0.00 \\
\rowcolor{greyD} 
\cellcolor{greyD}\multirow{-3}{*}{HotpotQA-EX-L} & F1-score & 0.18 & 0.17 & 0.02 & 0.02 & 0.05 & 0.00 & 0.13 & 0.00 & 0.17 & 0.00 & 0.00 & 0.00 & 0.00 \\ \midrule
\multirow{3}{*}{MS-MARCO-EX-L} & ACC & 0.68 & 0.74 & 0.87 & 0.91 & 0.83 & 0.96 & 0.36 & 1.00 & 0.98 & 0.99 & 0.97 & 0.99 & 1.00 \\
 & ASR & 0.08 & 0.13 & 0.10 & 0.07 & 0.06 & 0.02 & 0.44 & 0.00 & 0.00 & 0.00 & 0.02 & 0.01 & 0.00 \\
 & F1-score & 0.27 & 0.28 & 0.08 & 0.12 & 0.12 & 0.00 & 0.48 & 0.00 & 0.02 & 0.00 & 0.02 & 0.00 & 0.00 \\ \midrule

\rowcolor{greyD} 
\cellcolor{greyD}& ACC & 0.84 & 0.75 & 0.91 & 0.92 & 0.87 & 0.97 & 0.62 & 1.00 & 0.98 & 0.96 & 1.00 & 0.98 & 0.99 \\
 
\rowcolor{greyD} 
\cellcolor{greyD}& ASR & 0.11 & 0.21 & 0.10 & 0.10 & 0.05 & 0.03 & 0.18 & 0.00 & 0.01 & 0.02 & 0.00 & 0.00 & 0.00 \\
 
\rowcolor{greyD} 
\cellcolor{greyD}\multirow{-3}{*}{BoolQ-EX-L} & F1-score & 0.23 & 0.28 & 0.07 & 0.06 & 0.12 & 0.02 & 0.25 & 0.00 & 0.04 & 0.03 & 0.01 & 0.00 & 0.00 \\ \midrule
\multirow{3}{*}{SQuAD-EX-L} & ACC & 0.89 & 0.89 & 0.96 & 0.95 & 0.97 & 0.96 & 0.60 & 1.00 & 0.95 & 1.00 & 0.99 & 1.00 & 1.00 \\
 & ASR & 0.04 & 0.04 & 0.03 & 0.03 & 0.01 & 0.01 & 0.33 & 0.00 & 0.05 & 0.00 & 0.00 & 0.00 & 0.00 \\
 & F1-score & 0.12 & 0.11 & 0.06 & 0.05 & 0.05 & 0.01 & 0.27 & 0.00 & 0.23 & 0.00 & 0.00 & 0.00 & 0.00 \\ 
 \bottomrule
\end{tabular}
 \vspace{-0.5cm}
\end{table*}

\subsection{Benchmark Poisoning Attacks}

\subsubsection{Effectiveness}
We evaluate poisoning attacks across fifteen datasets, including five existing ones (NQ, HotpotQA, MS-MARCO, SQuAD, and BoolQ) and ten expansions. Table~\ref{tab:comparision} presents the results, highlighting key vulnerabilities in RAG systems. Our main findings are:

\myparatight{Most poisoning attacks demonstrate considerable effectiveness on existing datasets}%
Targeted poisoning and DoS attacks generally achieve high ASRs, with simple methods like BPI often rivaling more complex ones. This underscores the threat posed by low-complexity attacks.
BPRAG and WPRAG yield lower ASRs in our benchmark compared to prior reports~\cite{zou2024poisonedrag}, likely due to differences in LLM settings. Since the originally used PaLM 2 model used in~\cite{zou2024poisonedrag} is now deprecated and its API unavailable, we re-run the experiments under matching settings. Results are shown in Table~\ref{tab:results_of_poisonedrag} and further analyzed in Appendix~\ref{appendix_sec:details_of_poisonedrag_results}.
Trigger-based DoS attacks show mixed results: AP attack performs well, likely due to its effective trigger design, while BadRAG and Phantom fall short of their original reports, despite faithful reproduction.
We conduct additional experiments, with results in Tables~\ref{tab:results_of_badrag} and~\ref{tab:results_of_phantom}, and detailed analysis in Appendix~\ref{appendix_sec:details_of_effectiveness}.

\myparatight{All poisoning attacks show significantly reduced effectiveness on challenging expansions}%
As shown in Table~\ref{tab:comparision}, ASRs consistently drop as the knowledge database expands from the original to EX-M and EX-L. To investigate this trend, we measured how many correct-answer texts appear in the top-5 retrieved results for each target query under non-attack conditions (see Fig.~\ref{fig:number_of_correct_texts_nq} in Appendix and detailed in Appendix~\ref{appendix_sec:details_of_correct_in_top5}). In the original NQ, most queries retrieve only one correct text, leaving room for poisoned texts. In contrast, EX-M and EX-L versions retrieve more correct texts with higher similarity, offering stronger signals to the LLM and reducing attack success. This suggests that enriching the knowledge database with relevant content can passively improve RAG robustness.

\myparatight{CRAG-AK demonstrates superior effectiveness on challenging expansions compared to other attacks}%
CRAG-AK shows higher effectiveness on challenging expansions than other attacks, due to its optimization strategy that maximizes the impact of each poisoned text under a fixed budget. As a result, even with similar F1-scores, it achieves significantly higher ASRs. This suggests that optimizing per-text effectiveness can sustain attack success in information-rich settings.

\begin{table*}[!t]
\tiny
\centering
\caption{The results of all poisoning attacks against various defenses on NQ dataset.}
\label{tab:defense_nq}
\begin{tabular}{l|c|cccccccc}
\toprule
Attack & Metric & No defense & Paraphrasing & InstructRAG & RobustRAG & AstuteRAG & PPL & Norm & TrustRAG \\ \midrule
 & ACC & 0.27 & 0.25 & 0.39 & 0.43 & 0.39 & 0.27 & 0.27 & 0.68 \\
\multirow{-2}{*}{BPRAG} & ASR & 0.62 & 0.62 & 0.57 & 0.31 & 0.58 & 0.60 & 0.61 & 0.05 \\ 
\midrule
\rowcolor{greyD} 
\cellcolor{greyD} & ACC & 0.25 & 0.27 & 0.39 & 0.43 & 0.43 & 0.24 & 0.24 & 0.72 \\
\rowcolor{greyD} 
\cellcolor{greyD}\multirow{-2}{*}{WPRAG} & ASR & 0.64 & 0.63 & 0.56 & 0.28 & 0.57 & 0.63 & 0.64 & 0.06 \\ 
\midrule
 & ACC & 0.02 & 0.01 & 0.54 & 0.39 & 0.42 & 0.03 & 0.03 & 0.72 \\
 \multirow{-2}{*}{BPI}& ASR & 0.94 & 0.93 & 0.41 & 0.28 & 0.42 & 0.94 & 0.94 & 0.03 \\ 
\midrule
\rowcolor{greyD} 
\cellcolor{greyD} & ACC & 0.01 & 0.00 & 0.57 & 0.44 & 0.47 & 0.00 & 0.00 & 0.72 \\
 \rowcolor{greyD} 
\cellcolor{greyD}\multirow{-2}{*}{WPI}& ASR & 0.97 & 0.94 & 0.36 & 0.28 & 0.33 & 0.98 & 0.97 & 0.05 \\ 
\midrule
 & ACC & 0.33 & 0.27 & 0.44 & 0.46 & 0.48 & 0.34 & 0.34 & 0.71 \\
\multirow{-2}{*}{AGGD} & ASR & 0.51 & 0.60 & 0.42 & 0.23 & 0.46 & 0.50 & 0.50 & 0.04 \\ 
\midrule
 \rowcolor{greyD} 
\cellcolor{greyD}& ACC & 0.06 & 0.00 & 0.46 & 0.43 & 0.35 & 0.06 & 0.07 & 0.70 \\
\rowcolor{greyD} 
\cellcolor{greyD} \multirow{-2}{*}{CRAG-AS}& ASR & 0.89 & 0.96 & 0.53 & 0.25 & 0.59 & 0.90 & 0.90 & 0.03 \\ 
\midrule
 & ACC & 0.04 & 0.04 & 0.37 & 0.46 & 0.23 & 0.04 & 0.04 & 0.70 \\
 \multirow{-2}{*}{CRAG-AK}& ASR & 0.88 & 0.83 & 0.62 & 0.26 & 0.69 & 0.89 & 0.89 & 0.04 \\ 
\midrule
\rowcolor{greyD} 
\cellcolor{greyD} & ACC & 0.15 & 0.11 & 0.79 & 0.42 & 0.78 & 0.15 & 0.15 & 0.75 \\
\rowcolor{greyD} 
\cellcolor{greyD}\multirow{-2}{*}{JamInject} & ASR & 0.85 & 0.88 & 0.07 & 0.53 & 0.08 & 0.85 & 0.85 & 0.01 \\ 
\midrule
 & ACC & 0.13 & 0.02 & 0.78 & 0.78 & 0.73 & 0.12 & 0.12 & 0.74 \\
\multirow{-2}{*}{JamOracle} & ASR & 0.87 & 0.97 & 0.00 & 0.11 & 0.01 & 0.88 & 0.87 & 0.01 \\ 
\midrule
\rowcolor{greyD} 
\cellcolor{greyD} & ACC & 0.29 & 0.37 & 0.81 & 0.76 & 0.82 & 0.26 & 0.27 & 0.76 \\
 \rowcolor{greyD} 
\cellcolor{greyD}\multirow{-2}{*}{JamOpt}& ASR & 0.59 & 0.55 & 0.00 & 0.11 & 0.00 & 0.63 & 0.63 & 0.00 \\ 
\midrule
 & ACC & 0.01 & 0.44 & 0.71 & 0.40 & 0.67 & 0.00 & 0.00 & 0.67 \\
\multirow{-2}{*}{AP} & ASR & 0.99 & 0.48 & 0.04 & 0.49 & 0.08 & 1.00 & 1.00 & 0.02 \\ 
\midrule
\rowcolor{greyD} 
\cellcolor{greyD}& ACC & 0.65 & 0.66 & 0.84 & 0.67 & 0.90 & 0.65 & 0.64 & 0.80 \\
\rowcolor{greyD} 
\cellcolor{greyD} \multirow{-2}{*}{BadRAG} & ASR & 0.35 & 0.34 & 0.15 & 0.30 & 0.01 & 0.35 & 0.36 & 0.01 \\ 
\midrule
 & ACC & 0.99 & 0.69 & 0.98 & 0.84 & 0.95 & 0.96 & 0.96 & 0.82 \\
 \multirow{-2}{*}{Phantom}& ASR & 0.00 & 0.30 & 0.00 & 0.14 & 0.00 & 0.04 & 0.03 & 0.00 \\
\bottomrule
\end{tabular}
 \vspace{-0.4cm}
\end{table*}

\subsubsection{Defenses}

Recent studies have introduced three main categories of defenses. Process-optimized defenses~\cite{jain2023baseline,wang2024astute,wei2024instructrag,xiang2024certifiably,yao2025ecosaferag,zou2024poisonedrag} improve system robustness through prompt engineering and architectural adjustments. Detection-based defenses~\cite{jelinek1980interpolated,shafran2024machine,xue2024badrag,zhang2025practical,zhong2023poisoning,zou2024poisonedrag} aim to filter poisoned texts. 
Hybrid approaches~\cite{zhou2025trustrag} combine filtering with system-level interventions. 
The details of these defenses are provided in Appendix~\ref{appendix_sec:details_of_defenses}.
We systematically evaluate these defenses using our benchmark. Results (ACC and ASR) on NQ dataset are shown in Table~\ref{tab:defense_nq}, with results for the other 14 datasets provided in Tables~\ref{tab:defense_hotpotqa}-\ref{tab:defense_squad_ex_l} in Appendix. 
Detailed analysis are provided in Appendix~\ref{appendix_sec:details_of_defense_results}.
Detection performance, accuracy (DACC), false positive rate (FPR), and false negative rate (FNR), for detection-based defenses is reported in Tables~\ref{tab:detection_nq}-\ref{tab:detection_squad_ex_l} in Appendix, with metric definitions in Appendix~\ref{appendix_sec:details_of_detection_metrics}.

Our evaluation yields several important observations. First, defense performance varies by attack type. Process-optimized methods such as InstructRAG and AstuteRAG are highly effective against DoS attacks, for instance, reducing JamOracle's ASR from 87\% to 1\% on NQ dataset, but are less effective against targeted poisoning. Second, detection-based methods like PPL and Norm generally fail to detect sophisticated poisoned content, showing limited overall effectiveness. 
%
%
Third, hybrid defenses like TrustRAG consistently surpass other methods in performance, but their ability to counter poisoning attacks remains limited (see Appendix~\ref{appendix_sec:details_of_defense_results} for analysis). These results highlight the need for developing more effective defenses to enhance the security of RAG systems.

\subsection{Ablation Studies}

\begin{wrapfigure}[11]{r}{0.4\textwidth}
    \vspace{-5mm}
\centering
    \centering  
    \includegraphics[width=0.4\textwidth]{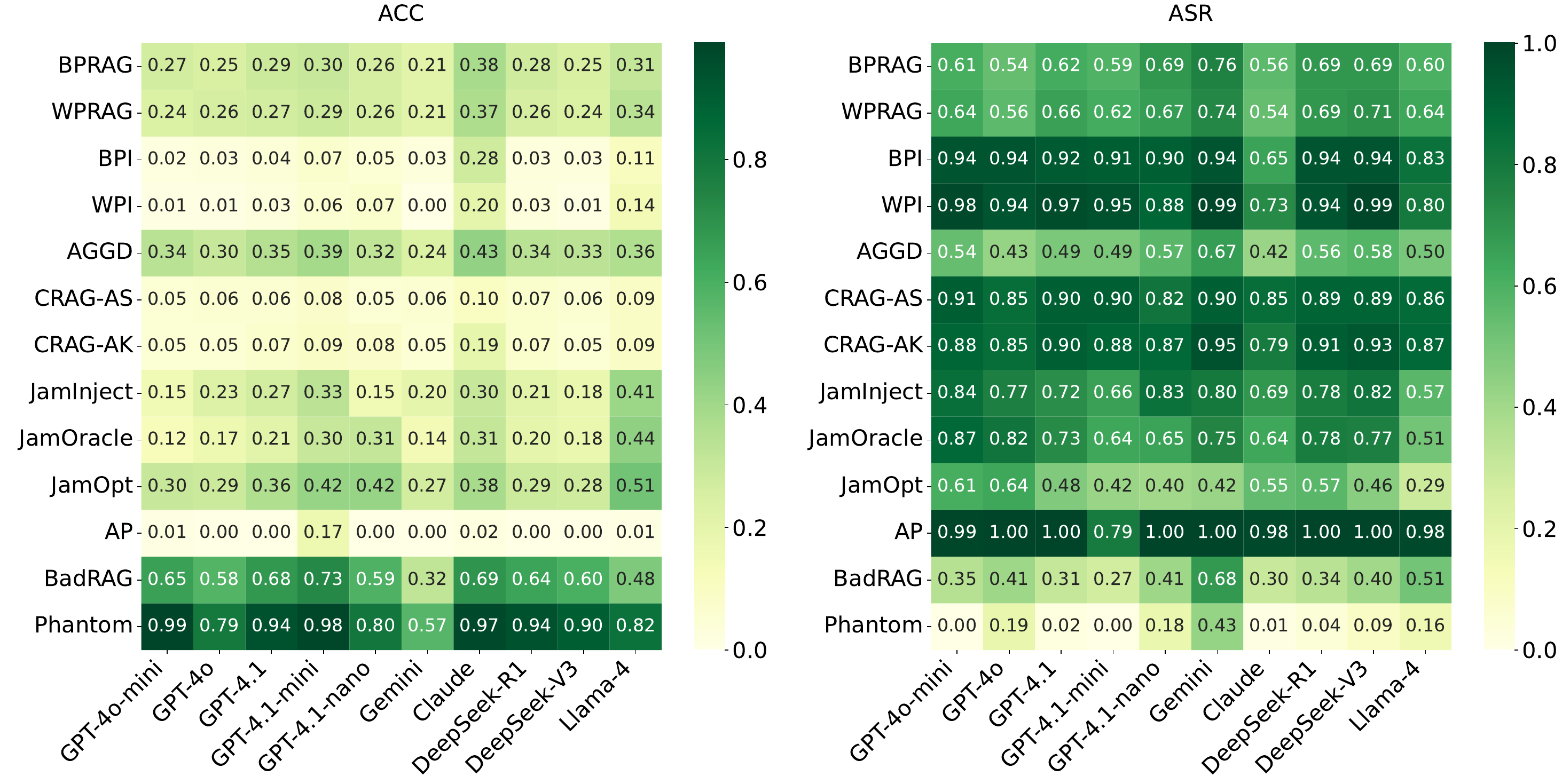} 
    \caption{Results of poisoning attacks under different LLMs of RAG on NQ dataset. LLM versions in Appendix~\ref{appendix_sec:details_of_llms}.}  
    \label{fig:rag_llm_nq}  
\end{wrapfigure}

\myparatight{Impact of LLMs}%
We conduct extensive experiments to investigate how LLMs influence the vulnerability of RAG systems to poisoning attacks across ten state-of-the-art LLM models~\cite{claude,gemini,gpt41,llama4,qwq,guo2025deepseek,hurst2024gpt,liu2024deepseek}. The results on NQ dataset are presented in Fig.~\ref{fig:rag_llm_nq}, and results on the NQ-EX-M and NQ-EX-L datasets are shown in Fig.~\ref{fig:rag_llm_nq_ex} in the Appendix. These experiments reveal two key findings.
First, despite extensive alignment training, all models exhibit substantial vulnerability when processing poisoned context. This exposes a critical limitation in current alignment methods, which primarily target direct prompt inputs rather than harmful content embedded within retrieved context. Second, we observe that Claude demonstrates markedly stronger resistance to poisoning attacks than other models, especially under targeted poisoning scenarios. This suggests that LLMs can be fortified to maintain robustness even when the input context is compromised.
These findings highlight an important direction for defense: enhancing LLMs' ability to identify and disregard malicious contextual content. Such improvements would provide a foundational layer of defense against RAG poisoning, complementing retrieval-based and prompt-based protection strategies.

\myparatight{Impact of retrievers}%
We perform a comprehensive evaluation to examine how different retrievers influence the susceptibility of RAG systems to poisoning attacks, using three state-of-the-art retrievers~\cite{izacard2021unsupervised,xiong2020approximate}. Results on the NQ, NQ-EX-M, and NQ-EX-L datasets (Table~\ref{tab:impact_of_retrievers_nq} in Appendix) reveal a consistent vulnerability across all retrievers. This vulnerability stems from their training objective, which focuses on maximizing similarity to ground-truth texts without accounting for poisoned content. These findings emphasize the need for adversarial training to improve retrievers' ability to detect and resist poisoning attempts.

\myparatight{Impact of similarity measurements}%
We conduct experiments to assess how different similarity measures affect RAG’s vulnerability to poisoning attacks, focusing on two commonly used methods: dot product and cosine similarity.
Results on the NQ, NQ-EX-M, and NQ-EX-L datasets (Table~\ref{tab:impact_of_similarity_nq} in Appendix) show that the dot product is more susceptible than cosine similarity, particularly under white-box attack settings. This increased vulnerability is likely due to the absence of normalization in dot product, which allows a larger optimization space for attackers. These results suggest a promising defense direction: designing more robust similarity functions, such as hybrid retrieval methods that combine multiple measures, to better resist adversarial manipulation.

\begin{wrapfigure}{r}{0.5\textwidth}
      \vspace{-5mm}
\centering
    \centering  
    \includegraphics[width=0.5\textwidth]{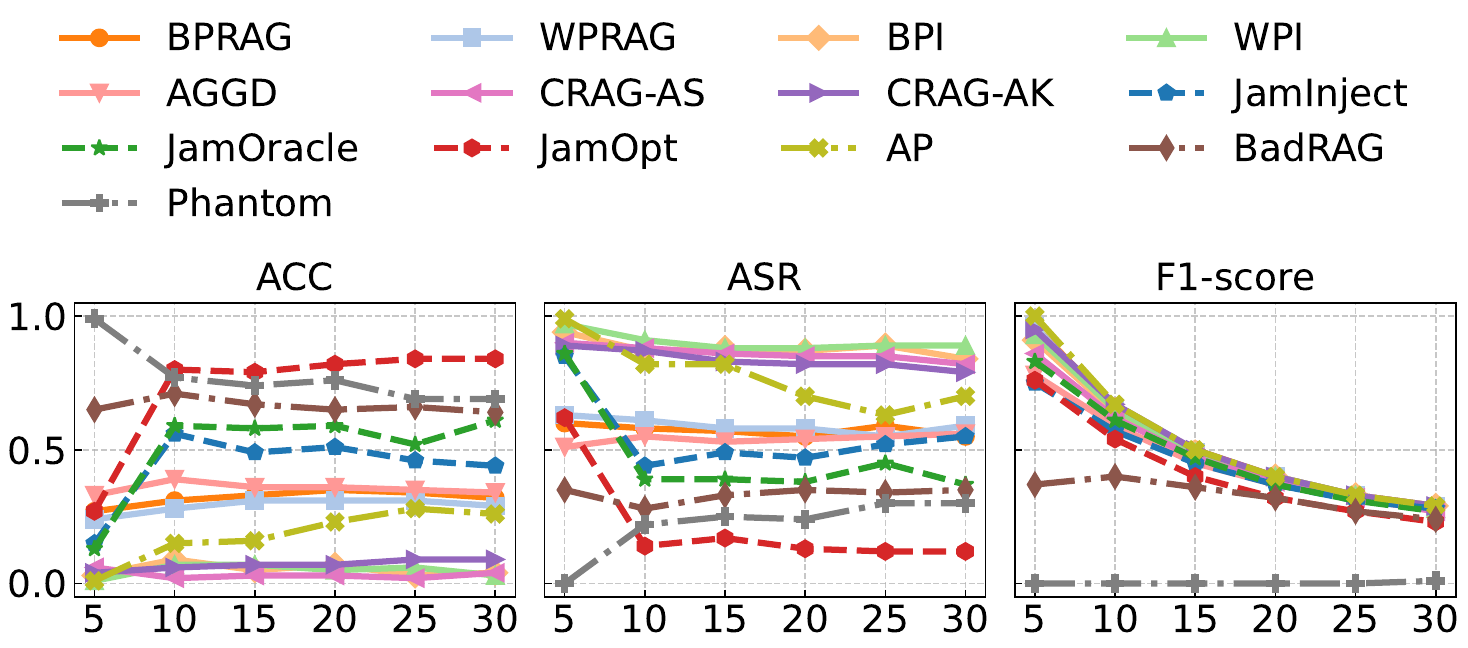} 
    \caption{The results of poisoning attacks under different top-$K$ of RAG on NQ dataset.}  
    \label{fig:k_nq_base}  
\end{wrapfigure}

\myparatight{Impact of $K$}%
We perform experiments to examine how varying the top-$K$ retrieved texts influences RAG's vulnerability to poisoning attacks. Results on NQ are presented in Fig.~\ref{fig:k_nq_base}, and results on NQ-EX-M and NQ-EX-L datasets appear in Fig.~\ref{fig:k_nq_exs} in Appendix, revealing three main findings. First, on the original NQ dataset, most attacks remain highly effective regardless of $K$, as increasing $K$ adds mostly irrelevant content due to the scarcity of correct-answer texts. Second, on the expanded datasets, higher $K$ values increase the recall of poisoned texts, but attacks do not become more effective, since the inclusion of correct answers provides the LLM with enough reliable information to resist manipulation. Third, CRAG-AS and CRAG-AK stand out on NQ-EX-M, showing improved effectiveness with larger $K$. 
Their budget strategy produces strong poisoned texts that remain effective amid many correct ones.

\subsection{Transferability Studies}

Most existing poisoning attacks focus on naive RAG, leaving their impact on advanced frameworks largely unexplored. To address this, our benchmark includes four categories of advanced RAG systems—sequential RAG~\cite{yu2023augmentation}, branching RAG~\cite{kim2024sure}, conditional RAG~\cite{jeong2024adaptive}, and loop RAG~\cite{chan2024rq,jiang2023active,trivedi2022interleaving}—covering six frameworks in total (see Appendix~\ref{appendix_sec:details_of_advanced_rag_frameworks} for details). Evaluation results in Tables~\ref{tab:framework_nq}-\ref{tab:framework_nq_ex_l} in Appendix provide insights into how architectural design affects vulnerability.
We identify two key findings. First, poisoned texts designed for naive RAG transfer effectively to many advanced frameworks, as they still rely on retrieved context for generation. This shows that architectural complexity alone does not eliminate the threat. Second, frameworks with adaptive retrieval, such as FLARE, demonstrate strong robustness by skipping retrieval when unnecessary, thereby reducing exposure to poisoned content. This highlights adaptive retrieval as a promising direction for defense.


\section{Discussion}

\subsection{Poisoning Attacks against Multi-turn RAG}

Existing poisoning attacks are mostly designed for single-turn RAG systems, overlooking the more practical multi-turn conversational RAG scenarios~\cite{aliannejadi2024trec,cheng2024coral,katsis2025mtrag,kuo2024rad,mo2024survey,xu2023recomp}. To bridge this gap, we implement a naive multi-turn RAG system that rewrites each user query using the full dialogue history before retrieving relevant texts. To evaluate poisoning effectiveness, we simulate multi-turn conversations by decomposing a targeted query into natural sub-questions with an LLM, using the final turn’s query to compute ASR and ACC (see Appendix~\ref{appendix_sec:details_of_multi_turn_rag}).
Results in Table~\ref{tab:multi_turn_results} in Appendix show reduced attack effectiveness in the multi-turn setting, underscoring limitations of attacks. We attribute this to the query rewriting, which alters the retrieval and hinders the retrieval of poisoned texts crafted for original queries. These findings highlight that poisoning in multi-turn RAG must overcome both retrieval constraints and the LLM’s context integration over dynamic dialogue history.

\subsection{Poisoning Attacks against Multimodal RAG}

We investigate the vulnerability of multimodal RAG systems~\cite{chen2022murag, lahiri2024alzheimerrag, xia2024mmed, xia2024rule, xue2024enhanced, yu2024visrag} to poisoning attacks targeting both image and text modalities. These systems use multimodal retrievers to select relevant image-text pairs as input to vision-language models (VLMs). Prior work~\cite{liu2025poisoned} introduced the Poisoned-MRAG attack, showing that injected malicious pairs can manipulate outputs.
Our benchmark includes Poisoned-MRAG and extends existing single-modality attacks to the multimodal setting (see Appendix~\ref{appendix_sec:details_of_multi_modal_rag}). As shown in Table~\ref{tab:multi_modal_results} in Appendix, multimodal RAG remains vulnerable due to its reliance on retrieval and augmentation strategies similar to naive RAG. Current retrievers and VLMs lack robustness to poisoned content. 
Additionally, weak image-text alignment allows attackers to fix the image and manipulate the text, effectively reducing the task to a text-based attack.

\subsection{Poisoning Attacks against RAG-based LLM Agent Systems}

We investigate the vulnerability of RAG-based LLM agent systems~\cite{mao2023language,shi2024ehragent, yao2023react, yuan2024rag} to poisoning attacks. These systems retrieve relevant query-solution pairs from a memory database and use them to guide the generation of new solution paths. Recent work~\cite{chen2024agentpoison} introduced the AgentPoison attack, showing that malicious entries can manipulate the agent’s behavior, such as triggering specific tool calls.
Our benchmark implements AgentPoison and adapts RAG poisoning attacks to the LLM agent setting (details in Appendix~\ref{appendix_sec:details_of_llm_agent}). Results in Table~\ref{tab:llm_agent_results} in Appendix confirm that LLM agents are highly vulnerable: both AgentPoison and adapted attacks achieve strong success rates. Notably, the added complexity of LLM agents does not hinder attacks. Because retrieval depends mainly on query similarity, existing poisoning methods like PoisonedRAG can be applied with minimal changes. These findings highlight the urgent need for defenses specifically designed for LLM agent systems.

\section{Conclusion, Limitations, and Future Work}
\label{sec:conclusion}

In this work, we introduced RSB, a benchmark for evaluating poisoning attacks on RAG systems, covering 13 attacks, 7 defenses, and 6 advanced RAG frameworks. We conducted extensive evaluations, including analyses on multi-turn RAG, multimodal RAG, and RAG-based LLM agents, revealing key security insights. Currently, our evaluation is limited to RAG-based agents; future work will explore more complex LLM agents based on the model context protocol~\cite{mcp}.

\bibliographystyle{plain}
\bibliography{refs}

\newpage

\appendix

\section*{Appendix}

The appendix is structured as follows.

\begin{itemize}
    \item Appendix~\ref{appendix:system_prompts}: System Prompt.
    
    \item Appendix~\ref{appendix_sec:details_of_poisoning_attacks}: Details of Poisoning Attacks.

    \item Appendix~\ref{appendix_other_attacks}: Other Attacks Not Considered in Experiments.

    \item Appendix~\ref{appendix_sec:setup_details}: Experimental Setup Details.
    
    \item Appendix~\ref{appendix_sec:details_of_poisonedrag_results}: Details of Results for BPRAG and WPRAG.

    \item Appendix~\ref{appendix_sec:details_of_effectiveness}: Details of Results for BadRAG and Phantom.

    \item Appendix~\ref{appendix_sec:details_of_correct_in_top5}: Details of Measuring the Number of Correct-Answer Texts Appearing in the Top-5 Retrieved Results.

    \item Appendix~\ref{appendix_sec:details_of_defenses}: Details of Defenses.

    \item Appendix~\ref{appendix_sec:details_of_defense_results}: Details of Defense Results.

    \item Appendix~\ref{appendix_sec:details_of_detection_metrics}: Details of Detection Metrics.

    \item Appendix~\ref{appendix_sec:details_of_llms}: Details of LLMs.

    \item Appendix~\ref{appendix_sec:details_of_advanced_rag_frameworks}: Details of Advanced RAG Frameworks.

    \item Appendix~\ref{appendix_sec:details_of_multi_turn_rag}: Details of Poisoning Attacks against Multi-turn RAG.

    \item Appendix~\ref{appendix_sec:details_of_multi_modal_rag}: Details of Poisoning Attacks against Multimodal RAG.

    \item Appendix~\ref{appendix_sec:details_of_llm_agent}: Details of Poisoning Attacks against LLM Agent Systems.

\end{itemize}

\section{System Prompt}
\label{appendix:system_prompts}

\begin{tcolorbox}[colback=gray!10,
                  colframe=black!80,
                  width=\linewidth,
                  arc=1mm, auto outer arc,
                  boxrule=1pt,
                  title =  System prompt of RAG
                 ]
You are a helpful assistant, below is a query from a user and some relevant contexts. Answer the question given the information in those contexts. Provide only the final answer, with no explanations or extra context. Keep your answer as short and concise as possible. \\
\textbf{Context:} [top-$K$ relevant texts] \\
\textbf{Query:} [user query]
\end{tcolorbox}

\section{Details of Poisoning Attacks}
\label{appendix_sec:details_of_poisoning_attacks}
\subsection{Targeted Poisoning Attack}
\label{appendix_sec:targeted_poisoning_attacks}

\myparatight{Black-box PoisonedRAG (BPRAG)~\cite{zou2024poisonedrag}}%
In this attack, the attacker has only \emph{black-box} access to the retriever. Each poisoned text is strategically divided into two functional sub-texts, designed to separately satisfy the requirements of retrieval and generation components. The generation sub-text is crafted by prompting an auxiliary LLM to create content that, when provided as context, will lead the auxiliary LLM to generate the targeted answer. Meanwhile, the retrieval sub-text contains the exact targeted query, ensuring that the poisoned text will be retrieved among the top-$K$ relevant texts when the targeted query is submitted to the system.

\myparatight{White-box PoisonedRAG (WPRAG)~\cite{zou2024poisonedrag}}%
This attack assumes the attacker has \emph{white-box} access to the retriever. Unlike the black-box variant, the attacker can leverage this privileged access to optimize the retrieval sub-text, maximizing the similarity between the poisoned text and the targeted query to ensure higher retrieval probability.

\myparatight{Black-box prompt injection (BPI)~\cite{liu2024formalizing,zou2024poisonedrag}}%
Prompt injection was originally proposed to manipulate LLM outputs by embedding malicious instructions into user inputs. This technique has been adapted for RAG systems in recent research~\cite{zou2024poisonedrag}. The key difference from BPRAG attack is that the generation sub-text contains an explicit malicious instruction that directly prompts the LLM to generate the targeted answer for the targeted query, rather than relying on contextual manipulation.

\myparatight{White-box prompt injection (WPI)~\cite{liu2024formalizing,zou2024poisonedrag}}%
This attack combines the optimization capabilities of WPRAG attack with the BPI attack. The retrieval sub-text is optimized using white-box access to the retriever, while the generation sub-text contains a malicious instruction that explicitly directs the LLM to produce the targeted answer.

\myparatight{AGGD~\cite{su2024corpus}}%
In this attack, the attacker has \emph{white-box} access to the retriever. AGGD differs from WPRAG in its more efficient optimization method for the retrieval sub-text. Specifically, it enhances gradient utilization by identifying and selecting the highest-ranked token across all possible token positions.

\myparatight{CorruptRAG-AS (CRAG-AS)~\cite{zhang2025practical}}%
This attack introduces a budget-aware objective function to enhance the independent effectiveness of each poisoned text. Specifically, the attacker divides the poisoned text into two sub-texts: the first contains the targeted query to ensure retrieval effectiveness of the entire poisoned text. The second sub-text is designed using adversarial principles, creating a template that generates the final poisoned text by filling the correct answer and the targeted answer for the targeted query.

\myparatight{CorruptRAG-AK (CRAG-AK)~\cite{zhang2025practical}}%
This attack is an enhanced variant of CRAG-AS. Specifically, the attacker leverages an auxiliary LLM to transform the second sub-text of the poisoned content into knowledge-like text, thereby improving its generalizability and stealthiness.

\subsection{DoS Attack}
\label{appendix_sec:dos_attacks}

\myparatight{Jamming-based instruction injection (JamInject) \cite{shafran2024machine}}%
This attack requires no specific background knowledge of the RAG system. The structure of the poisoned text resembles that of the BPRAG attack, but with a crucial difference: the generation sub-text contains an explicit instruction designed to prompt the LLM to produce a refusal response rather than helpful content.

\myparatight{Jamming-based on oracle generation (JamOracle) \cite{shafran2024machine}}%
This attack employs an oracle LLM to create the generation sub-text. The sub-text is specifically crafted so that, when provided as context to the oracle LLM, it will lead the model to generate a refusal response, effectively denying service to legitimate users.

\myparatight{Jamming-based on black-box optimization (JamOpt) \cite{shafran2024machine}}%
This approach introduces a black-box optimization method that iteratively refines the generation sub-text. The optimization objective is to maximize the similarity between a predetermined refusal answer and the actual output of the RAG system when the poisoned text is provided as context.

\subsection{Trigger-based DoS Attacks}
\label{appendix_sec:trigger_dos_attacks}

\myparatight{AgentPoison (AP)~\cite{chen2024agentpoison}}%
This attack was originally designed to poison RAG-based LLM agent systems by optimizing a trigger string to manipulate agent responses for queries containing that trigger. We adapt it to RAG poisoning attacks. Specifically, the attacker has \emph{white-box} access to the retriever and uses AgentPoison's adversarial retrieval optimization method to optimize a trigger string, which is then concatenated with a malicious instruction that prompts the LLM to refuse answering, forming a poisoned text.

\myparatight{BadRAG~\cite{xue2024badrag}}%
Unlike the AP attack, BadRAG's trigger is predefined and remains fixed throughout the attack process. In this attack, the attacker has \emph{white-box} access to the retriever. BadRAG optimizes the poisoned text by minimizing a contrastive learning loss that measures the similarity between the text and triggered queries relative to non-triggered queries. Additionally, the poisoned text includes a prompt designed to activate the alignment mechanisms of the LLM, causing it to produce a refusal response. 

\myparatight{Phantom~\cite{chaudhari2024phantom}}%
This is similar to BadRAG, with the main difference being its optimization loss function, which is a difference loss, whereas BadRAG uses a contrastive loss.

\section{Other Attacks Not Considered in Experiments}
\label{appendix_other_attacks}

In this section, we provide a brief overview of several RAG poisoning attacks that were excluded from our experimental benchmark and explain the rationale behind their omission.

\myparatight{GARAG~\cite{cho2024typos}}%
GARAG introduces a technique for crafting adversarial noisy texts aimed at undermining RAG systems. Although the method does not rely on internal parameters of the LLM, it does assume access to the output probabilities of the model’s responses. This assumption diverges from our threat model, which considers the LLM as a strict black box, where such probabilistic information is typically inaccessible.

\myparatight{Opinion Manipulation Attacks~\cite{chen2025flipedrag,gong2025topic}}%
FlippedRAG~\cite{chen2025flipedrag}\footnote{Note that an earlier version of this paper was titled ``Black-Box Opinion Manipulation Attacks to Retrieval-Augmented Generation of Large Language Models''.} and Topic-FlipRAG~\cite{gong2025topic} present opinion manipulation attacks targeting RAG systems. We exclude these methods from our benchmark primarily because their performance and evaluation are tightly coupled with the specific topics of the input queries. In contrast, the attacks included in our evaluation are designed to be broadly applicable across diverse queries, without relying on topic-specific assumptions.

\myparatight{The RAG Paradox~\cite{choi2025rag}]}%
This paper examines a particular setting where RAG systems explicitly cite the web links that serve as sources for their responses. Leveraging this behavior, the proposed attack targets the system by injecting malicious content into web pages to influence the generated output. In contrast, our benchmark is designed for RAG systems that retrieve information from an offline, curated knowledge base and do not directly reference external web links in their responses. Given this fundamental difference in system design and attack surface, we did not include this work in our evaluation.

\myparatight{PR-Attack~\cite{jiao2025pr}}%
PR-Attack introduces a poisoning attack against RAG systems, but its threat model relies on the assumption that the attacker can modify or influence the user's query to insert a backdoor trigger. This assumption is not aligned with our benchmark, which considers a setting where the attacker cannot alter user queries. Therefore, PR-Attack was not included in our experimental comparisons.


\section{Experimental Setup Details}
\label{appendix_sec:setup_details}
\subsection{Details of Datasets}
\label{appendix:detail_datasets}
\myparatight{Natural Questions (NQ)~\cite{kwiatkowski2019natural}}%
Its queries originate from actual anonymized searches submitted to the Google search engine. The knowledge database for NQ is derived from Wikipedia and contains 2,681,468 texts.

\myparatight{HotpotQA~\cite{yang2018hotpotqa}}%
This dataset features multi-hop queries that require reasoning across multiple texts to determine the correct answer. The knowledge database for HotpotQA is also sourced from Wikipedia and contains 5,233,329 texts.

\myparatight{MS-MARCO~\cite{nguyen2016ms}}%
Its queries are collected from anonymized Bing search query logs. Its knowledge database comprises 8,841,823 texts gathered from web pages through Microsoft's Bing search engine.

\myparatight{SQuAD~\cite{rajpurkar2016squad}}%
Its queries are created by crowdworkers based on Wikipedia articles for reading comprehension tasks. We construct the knowledge database for SQuAD by combining the knowledge database of HotpotQA with all relevant texts from the original dataset.

\myparatight{BoolQ~\cite{clark2019boolq}}%
Its queries are naturally occurring by users, whose answers are yes or no. The knowledge database for BoolQ is constructed following the same approach used for SQuAD.

\myparatight{NQ/HotpotQA/MS-MARCO/BoolQ/SQuAD-EX-M}%
These expansions are enhanced at the medium level. Specifically, for each targeted query, we add 5 relevant texts to the knowledge database. We use GPT-4o-mini to generate these texts, ensuring they support the correct answer for the targeted query. Additionally, we prepend the targeted query to each relevant text to increase their similarity to the targeted query.

\myparatight{NQ/HotpotQA/MS-MARCO/BoolQ/SQuAD-EX-L}%
We also construct the expansions at the large level, where we add 30 relevant texts to the knowledge database for each targeted query.

\begin{table*}[!htbp]
\centering
\tiny
\addtolength{\tabcolsep}{3.5pt}

\caption{Statistics of datasets.}
\label{tab:datasets}

\begin{tabular}{lcc} 
\toprule
Dataset & Number of queries & Number of texts in the knowledge database \\ 
\midrule
NQ & 91,535 & 2,681,468 \\
\rowcolor{greyD}  \cellcolor{greyD}HotpotQA & 97,852 & 5,233,329 \\
MS-MARCO & 909,824 & 8,841,823 \\
\rowcolor{greyD}  \cellcolor{greyD}BoolQ & 12,697 & 5,243,473 \\
SQuAD & 98,169 & 5,254,287 \\
   
\rowcolor{greyD}  \cellcolor{greyD}NQ-EX-M & 91,535 & 2,681,968 \\
HotpotQA-EX-M & 97,852 & 5,233,829 \\
\rowcolor{greyD}  \cellcolor{greyD}MS-MARCO-EX-M & 909,824 & 8,842,323 \\
BoolQ-EX-M & 12,697 & 5,243,973 \\
\rowcolor{greyD}  \cellcolor{greyD}SQuAD-EX-M & 98,169 & 5,254,787 \\

NQ-EX-L & 91,535 & 2,684,468 \\
\rowcolor{greyD}  \cellcolor{greyD}HotpotQA-EX-L & 97,852 & 5,236,329 \\
MS-MARCO-EX-L & 909,824 & 8,844,823 \\
\rowcolor{greyD}  \cellcolor{greyD}BoolQ-EX-L & 12,697 & 5,246,473 \\
SQuAD-EX-L & 98,169 & 5,257,287 \\

\bottomrule
\end{tabular}
\end{table*}

\subsection{Details of Evaluation Metrics}
\label{appendix:details_of_metrics}
Our benchmark uses the following widely adopted metrics: accuracy (ACC), attack success rate (ASR), and F1-score. 

\myparatight{ACC}%
ACC is defined as the ratio of queries for which the RAG system generates the correct answers to the total number of queries. 

\myparatight{ASR}%
ASR is defined as the ratio of queries for which the RAG system generates the targeted answers to the total number of queries. We employ an LLM (GPT-4o-mini is used in our evaluation) to determine whether the output of the RAG system matches the correct or targeted answer. 

\myparatight{F1-score}%
F1-score is a metric that measures the retrieval effectiveness of poisoned texts, calculated by $\text{F1-score} = 2 \times \frac{\text{Precision} \times \text{Recall}}{\text{Precision} + \text{Recall}}$.
Precision is defined as the ratio of the number of poisoned texts retrieved in the top-$K$ relevant texts for a targeted query. Recall is defined as the ratio of the number of poisoned texts among the top-$K$ texts to the total number of poisoned texts injected for a targeted query. For the metric F1-score, we report the average values across all targeted queries.

\subsection{Details of Targeted Queries and Targeted Answers}
\label{appendix:details_of_targeted_answers}
For each type of poisoning attack, we created a set of 100 targeted queries and their corresponding targeted answers based on query collections from NQ, HotpotQA, MS-MARCO, BoolQ, and SQuAD. Specifically, we first randomly sampled 500 queries from the original query collections as initial targeted queries. Second, for targeted poisoning attacks, we used GPT-4o-mini to generate a random answer different from the correct answer as the targeted answer for each targeted query, while for DoS attacks and Trigger-based DoS attacks, we used ``I don't know'' as the targeted answer. Then, we submitted these 500 targeted queries to the benign RAG system (without any attack) to obtain the RAG response for each targeted query, and filtered out queries where the RAG response matched the targeted answer judged by GPT-4o-mini. Finally, we randomly sampled 100 queries from the remaining queries as our targeted queries.
For the -EX-M and -EX-L datasets, we reused the same set of targeted queries as in the corresponding base dataset. For instance, the targeted queries used for the NQ-EX-M and NQ-EX-L datasets are identical to those selected for the original NQ dataset.

Note that we select distinct targeted queries for each type of attack to ensure that, under a benign RAG system, the target query does not naturally yield the intended answer. During the simulation of trigger-based DoS attacks, however, adding triggers frequently caused the model to respond with ``I don't know''. As a result, it was infeasible to identify a shared set of 100 queries from the initial 500 that satisfied the zero-targeted-answer condition across all three attack types. To address this, we screened and selected queries independently for each attack type. This design choice does not compromise the validity of our evaluation, as we assess performance within each attack category. By applying consistent selection criteria within each type, we ensure fair and comparable results across attacks of the same kind.

\begin{table*}[!htbp]
\centering
\tiny
\setlength\tabcolsep{10pt}
\caption{Answer accuracy of the RAG system for targeted queries under non-attack conditions.}
\label{tab:performanc_100_ques}
\begin{tabular}{lccc}
\toprule
Dataset & Targeted poisoning attack & DoS attack & Trigger-based DoS attack \\ \midrule
NQ & 0.82 & 0.99 & 0.97  \\
\rowcolor{greyD}  \cellcolor{greyD}HotpotQA & 0.92 & 0.98 & 0.98 \\
MS-MARCO & 0.92 & 0.98 & 0.98 \\
\rowcolor{greyD}  \cellcolor{greyD}BoolQ & 1.00 & 0.98 & 0.95 \\
SQuAD & 0.93 & 0.94 & 0.98 \\

\rowcolor{greyD}  \cellcolor{greyD}NQ-EX-M & 0.98 & 1.00 & 1.00 \\
HotpotQA-EX-M & 0.99 & 1.00 & 1.00 \\
\rowcolor{greyD}  \cellcolor{greyD}MS-MARCO-EX-M & 0.97 & 1.00 & 0.98 \\
BoolQ-EX-M & 0.99 & 1.00 & 0.99 \\
\rowcolor{greyD}  \cellcolor{greyD}SQuAD-EX-M & 1.00 & 1.00 & 0.99 \\

NQ-EX-L & 0.98 & 1.00 & 1.00 \\
\rowcolor{greyD}  \cellcolor{greyD}HotpotQA-EX-L & 0.99 & 1 .00& 1.00 \\
MS-MARCO-EX-L & 0.97 & 0.98 & 0.97 \\
\rowcolor{greyD}  \cellcolor{greyD}BoolQ-EX-L & 0.99 & 0.99 & 0.98 \\
SQuAD-EX-L & 1.00 & 0.99 & 0.98\\
\bottomrule
\end{tabular}
\end{table*}

\section{Details of Results for BPRAG and WPRAG}
\label{appendix_sec:details_of_poisonedrag_results}

We note that the ASR of BPRAG and WPRAG in our benchmark is lower than reported in the original papers, even though we used exactly the same code provided by the authors.
BPRAG and WPRAG construct poisoned texts by splitting them into two distinct parts: the retrieval sub-text, which ensures that the poisoned content appears among the top-$K$ retrieved results, and the generation sub-text, which aims to guide the RAG model to produce a specific target output. A close analysis of our experimental results shows that both methods consistently achieve high F1-scores across all datasets, indicating effective retrieval. However, their attack success rates vary and are lower on certain datasets. This suggests that while the retrieval sub-text reliably satisfies its objective, the generation sub-text demonstrates variable effectiveness in influencing the model’s output.

Recall that both BPRAG and WPRAG rely on a proxy language model to construct the generation sub-text, making its effectiveness closely tied to the choice of proxy model. Additionally, the language model used in the RAG system itself plays a significant role in determining how successful the generation sub-text is during inference. To gain deeper insights, we conducted a series of experiments on the NQ dataset, systematically varying both the proxy LLM and the RAG's LLM to assess their impact on attack performance. The results of these experiments are presented in Table~\ref{tab:results_of_poisonedrag}.

Our analysis reveals two important observations. First, the choice of proxy LLM has a substantial impact on attack effectiveness. When GPT-4.1 was used as the proxy model, both BPRAG and WPRAG achieved significantly higher attack success rates, approximately 20 percent greater than when GPT-4o-mini was used, as shown in Table~\ref{tab:comparision}. This difference is likely due to variations in model capability and adherence to instructions. More powerful and instruction-following models are better at generating persuasive generation sub-texts. In contrast, less capable models or those that are more restrictive in following prompts often struggle to produce effective sub-texts, especially when the instructions may conflict with their alignment constraints. For example, when GPT-4o-mini was used as the proxy, many attempts failed to produce successful generation sub-texts even after reaching the maximum number of tries. Second, the language model used in the RAG system also influences attack performance. For instance, when GPT-4.1-mini was the proxy model, using GPT-4 as the RAG’s LLM resulted in a lower attack success rate.

We selected GPT-4o-mini as the default model for BPRAG and WPRAG in our experiments for two primary reasons. First, GPT-4o-mini is a widely adopted model with performance comparable to PaLM 2, which was used in the original studies. Additionally, we employed GPT-4o-mini as the default proxy LLM for other attacks in our benchmark that also require a proxy model, such as CRAG-AK, to maintain consistency. Second, our benchmark involves large-scale evaluations, and using GPT-4 or GPT-4.1 would have significantly increased computational costs, making the evaluation substantially more expensive.

A rough cost analysis further supports our decision to use GPT-4o-mini as the default model. Specifically, we estimated that running BPRAG attack on the NQ dataset would cost around \$10 with GPT-4o-mini, compared to approximately \$390 with GPT-4 and \$96 with GPT-4.1. Given that our benchmark includes 15 datasets, the total projected cost would rise to about \$150 with GPT-4o-mini, but escalate to \$5,850 with GPT-4 and \$1,440 with GPT-4.1. These substantial cost differences reinforce the practicality of using GPT-4o-mini for large-scale evaluations.

\begin{table*}[!htbp]
\tiny
\centering
\addtolength{\tabcolsep}{0pt}
\caption{The results of BPRAG and WPRAG on NQ dataset under different proxy LLMs and LLMs of RAG.}
\label{tab:results_of_poisonedrag}
\begin{tabular}{l|c|cccccccc} 
\toprule
\multirow{3}{*}{Proxy LLM} & \multirow{3}{*}{Metric} & \multicolumn{8}{c}{LLM of RAG} \\ 
\cmidrule{3-10}
 &  & \multicolumn{2}{c|}{GPT-4o-mini} & \multicolumn{2}{c|}{GPT-4.1-mini} & \multicolumn{2}{c|}{GPT-4} & \multicolumn{2}{c}{GPT-4.1} \\ 
\cmidrule{3-10}
 &  & \multicolumn{1}{c|}{BPRAG} & \multicolumn{1}{c|}{WPRAG} & \multicolumn{1}{c|}{BPRAG} & \multicolumn{1}{c|}{WPRAG} & \multicolumn{1}{c|}{BPRAG} & \multicolumn{1}{c|}{WPRAG} & \multicolumn{1}{c|}{BPRAG} & WPRAG \\ 
\midrule
\multirow{3}{*}{GPT-4.1-mini} & ACC & 0.16 & 0.19 & 0.22 & 0.22 & 0.20 & 0.16 & 0.20 & 0.21 \\
 & ASR & 0.73 & 0.72 & 0.72 & 0.69 & 0.63 & 0.55 & 0.73 & 0.74 \\
 & F1-score & 0.98 & 0.96 & 0.98 & 0.96 & 0.98 & 0.96 & 0.98 & 0.96 \\ 
\midrule
 \rowcolor{greyD}  \cellcolor{greyD}  & ACC & 0.09 & 0.09 & 0.18 & 0.20 & 0.10 & 0.13 & 0.12 & 0.15 \\
 \rowcolor{greyD}  \cellcolor{greyD}   & ASR & 0.79 & 0.78 & 0.82 & 0.80 & 0.80 & 0.63 & 0.85 & 0.82 \\
 \rowcolor{greyD}  \cellcolor{greyD}\multirow{-3}{*}{GPT-4}  & F1-score & 0.92 & 0.92 & 0.92 & 0.92 & 0.92 & 0.92 & 0.92 & 0.92 \\ 
\midrule
\multirow{3}{*}{GPT-4.1} & ACC & 0.05 & 0.05 & 0.12 & 0.11 & 0.09 & 0.10 & 0.10 & 0.09 \\
 & ASR & 0.84 & 0.86 & 0.84 & 0.86 & 0.81 & 0.70 & 0.84 & 0.84 \\
 & F1-score & 0.93 & 0.93 & 0.93 & 0.93 & 0.93 & 0.93 & 0.93 & 0.93 \\
\bottomrule
\end{tabular}
\end{table*}

\section{Details of Results for BadRAG and Phantom}
\label{appendix_sec:details_of_effectiveness}

We believe the poor performance of these two attacks in our benchmark stems from their limited robustness under cosine similarity, which is used to assess the relevance between queries and texts. As shown in Table~\ref{tab:impact_of_similarity_nq}, their ASR and F1-scores increase dramatically when the similarity metric is switched to dot product. To further test this hypothesis, we conducted additional evaluations using dot product on the HotpotQA and MS-MARCO datasets. The results, reported in Table~\ref{tab:results_of_badrag} and Table~\ref{tab:results_of_phantom}, provide strong evidence supporting our claim.

\begin{table*}[!htbp]
\tiny
\centering
\addtolength{\tabcolsep}{6.0pt}
\caption{The results of BadRAG when the similarity measurement is dot product.}
\label{tab:results_of_badrag}
\begin{tabular}{lccc} 
\toprule
Dataset & ACC & ASR & F1-score \\
\midrule
NQ &0.19  &0.81  &0.72  \\
\rowcolor{greyD}  \cellcolor{greyD}HotpotQA &0.22 &0.77  &0.87  \\
MS-MARCO &0.27  &0.72  &0.75  \\
\bottomrule
\end{tabular}
\end{table*}

\begin{table*}[!htbp]
\tiny
\centering
\addtolength{\tabcolsep}{6.0pt}
\caption{The results of Phantom when the similarity measurement is dot product.}
\label{tab:results_of_phantom}
\begin{tabular}{lccc} 
\toprule
Dataset & ACC & ASR & F1-score \\
\midrule
NQ &0.03  &0.97  &0.95  \\
\rowcolor{greyD}  \cellcolor{greyD}HotpotQA &0.03  &0.97  &1.00  \\
MS-MARCO &0.03  &0.97  &0.98  \\
\bottomrule
\end{tabular}
\end{table*}

\section{Details of Measuring the Number of Correct-Answer Texts Appearing in the Top-5 Retrieved Results.}
\label{appendix_sec:details_of_correct_in_top5}

\begin{figure*}[t]  
    \centering  
    \includegraphics[width=\textwidth]{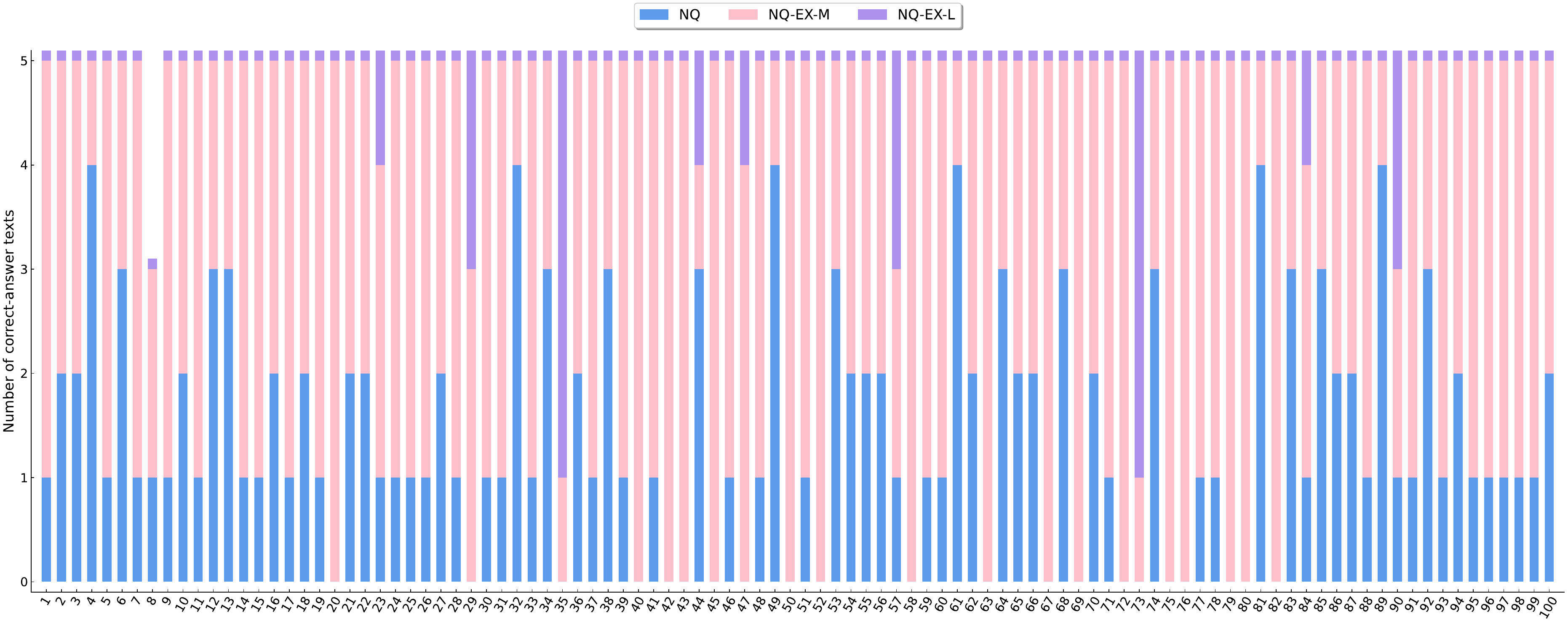} 
    
    \caption{The number of correct-answer texts among top-5 for each targeted query on NQ, NQ-EX-M, and NQ-EX-L datasets.}  
    \label{fig:number_of_correct_texts_nq}  
\end{figure*}

We conducted an analysis to assess how many of the top-$K$ relevant texts retrieved by a benign RAG system genuinely support the correct answer to targeted queries. For each targeted query, we submitted it to the benign RAG system and collected the top-$K$ retrieved texts. We then employed GPT-4o-mini to evaluate each of these texts and determine whether it provides valid support for the correct answer. The results, illustrated in Fig.~\ref{fig:number_of_correct_texts_nq}, present the distribution of supporting texts across the NQ, NQ-EX-M, and NQ-EX-L datasets, highlighting the extent to which the retrieved evidence aligns with the ground-truth answers.

\section{Details of Defenses}
\label{appendix_sec:details_of_defenses}

\subsection{Process-optimized Defense}

\myparatight{Paraphrasing~\cite{zou2024poisonedrag}}%
This defense was originally designed to counter prompt injection attacks against LLMs. Recent research has adapted it to the RAG scenario to defend against poisoning attacks. Specifically, the user's query is paraphrased using an LLM (we use GPT-4o-mini in our experiments) before retrieval, and then the paraphrased query is used for retrieval and sent together with the top-$K$ relevant texts to the LLM for response generation.

\myparatight{InstructRAG~\cite{wei2024instructrag}}%
This defense  designs a RAG system prompt that requires the LLM to explain how its generated answer is derived from the top-$K$ relevant texts, and also instructs the LLM to judge whether the top-$K$ relevant texts are helpful based on its own knowledge. If the retrieved texts are not useful or are harmful, the LLM is instructed to answer directly based on its own knowledge.

\myparatight{RobustRAG~\cite{xiang2024certifiably}}%
This defense proposes an isolate-and-aggregate mechanism. Specifically, RobustRAG first generates a response for each text in the top-$K$ separately, and then selects the highest-ranked answer for final answer generation based on the ranking of keywords in each response.

\myparatight{AstuteRAG~\cite{wang2024astute}}%
This defense adaptively extracts necessary information from the LLM's internal knowledge and integrates it with externally retrieved knowledge to mitigate biases introduced by poisoned texts. Specifically, AstuteRAG first uses the LLM to generate an informational text about the user query based on its internal knowledge, then asks the LLM to merge this generated information with the retrieved top-$K$ texts, and finally asks the LLM to answer the user query based on the merged knowledge.

\subsection{Detection-based Defense}

\myparatight{Perplexity-based detection (PPL)~\cite{jelinek1980interpolated,shafran2024machine,zhang2025practical,zou2024poisonedrag}}%
This detection was originally designed to detect malicious prompts for the LLMs, as perplexity can measure the fluency and naturalness of a text, with higher values indicating less natural text. Recent research has adapted PPL to RAG to detect poisoned texts. Specifically, a proxy LLM (we use Llama 2 in our experiments) is used to calculate the perplexity of each text in the knowledge base, and an appropriate threshold is selected to identify texts with perplexity above the threshold as poisoned texts.

\myparatight{Norm-based detection (Norm)~\cite{xue2024badrag,zhong2023poisoning}}%
This detection proposes that poisoned texts, in order to increase their similarity with targeted queries, are likely to have embedding vectors with abnormally large norms, so poisoned texts can be identified by measuring the norm of text embeddings. Specifically, this detection uses the retriever's embedding model to calculate the norm of the embedding vector for each text, and then selects an appropriate threshold to identify texts with norms above the threshold as poisoned texts.

\subsection{Hybrid Defense}

\myparatight{TrustRAG~\cite{zhou2025trustrag}}%
This method proposes a two-stage defense mechanism. In the first stage, TrustRAG uses K-means to cluster the texts in the top-$K$ and filters out potentially poisoned texts. In the second stage, TrustRAG designs a prompt that asks the LLM to answer the user query based on its own knowledge and the remaining relevant texts.

\section{Details of Defense Results}
\label{appendix_sec:details_of_defense_results}
\myparatight{Effectiveness of process-optimized defense}%
Our observations indicate that existing defenses remain inadequate in effectively mitigating poisoning attacks. For instance, the paraphrasing defense was largely ineffective, failing to prevent attacks across nearly all datasets. While InstructRAG and RobustRAG exhibited some defensive capability, their impact was limited, as the attack success rate still exceeded 50\% in most scenarios.

\myparatight{Effectiveness of detection-based defense}%
Our results show that these defenses are largely ineffective against poisoning attacks. In most scenarios, the attack success rate remained similar to the no-defense baseline, indicating limited defensive impact. We attribute this to the inherent complexity of the knowledge database, where texts originate from diverse sources and exhibit varying distributions. This heterogeneity makes it challenging to define a reliable threshold for accurately detecting poisoned content.

\myparatight{Effectiveness of hybrid defense}%
Our findings indicate that TrustRAG is generally effective in mitigating poisoning attacks across most scenarios. However, this effectiveness comes at a substantial cost to the overall performance of RAG. In many settings, the accuracy of TrustRAG drops significantly compared to the accuracy of the standard RAG system under no attack, as shown in Table~\ref{tab:performanc_100_ques}, with most decreases exceeding 20\%. This decline is particularly evident on the dataset expansions. A detailed review of TrustRAG's logs reveals that during its initial filtering stage, it often removes all top-$K$ retrieved texts, even when all of them are benign. This overly aggressive filtering behavior results in very high false positive rates, as reported in Tables~\ref{tab:detection_nq}-\ref{tab:detection_squad_ex_l}, highlighting a fundamental trade-off between security and utility in this defense strategy.

\section{Details of Detection Metrics}
\label{appendix_sec:details_of_detection_metrics}
\myparatight{Detection accuracy (DACC)} DACC measures the overall correctness of the detection method, defined as the proportion of correctly classified texts (both poisoned and benign) among all texts. Mathematically, $\text{DACC} = (TP + TN) / (TP + TN + FP + FN)$, where $TP$ represents the number of correctly identified poisoned texts, $FP$ represents the number of benign texts misclassified as poisoned, $TN$ represents the number of correctly identified benign texts, $FN$ represents the number of poisoned texts misclassified as benign. Higher DACC values indicate better overall detection performance, with a perfect detector achieving a DACC of 1.0.

\myparatight{False positive rate (FPR)} False positive rate quantifies the proportion of benign texts incorrectly classified as poisoned, calculated as $\text{FPR} = FP / (FP + TN)$.  A high FPR suggests the detector is overly sensitive and frequently flags benign content as malicious, which can significantly degrade the utility of the RAG system by unnecessarily filtering out valuable information.

\myparatight{False negative rate (FNR)} False negative rate measures the proportion of poisoned texts incorrectly classified as benign, defined as $\text{FNR} = FN / (FN + TP)$. A high FNR indicates the detector is failing to identify poisoned content, allowing malicious texts to bypass the defense mechanism and potentially compromise the RAG system's outputs.

\section{Details of LLMs}
\label{appendix_sec:details_of_llms}

\begin{figure*}[!t]  
    \centering  
    \begin{subfigure}{\textwidth}
        \centering
        \includegraphics[width=\textwidth]{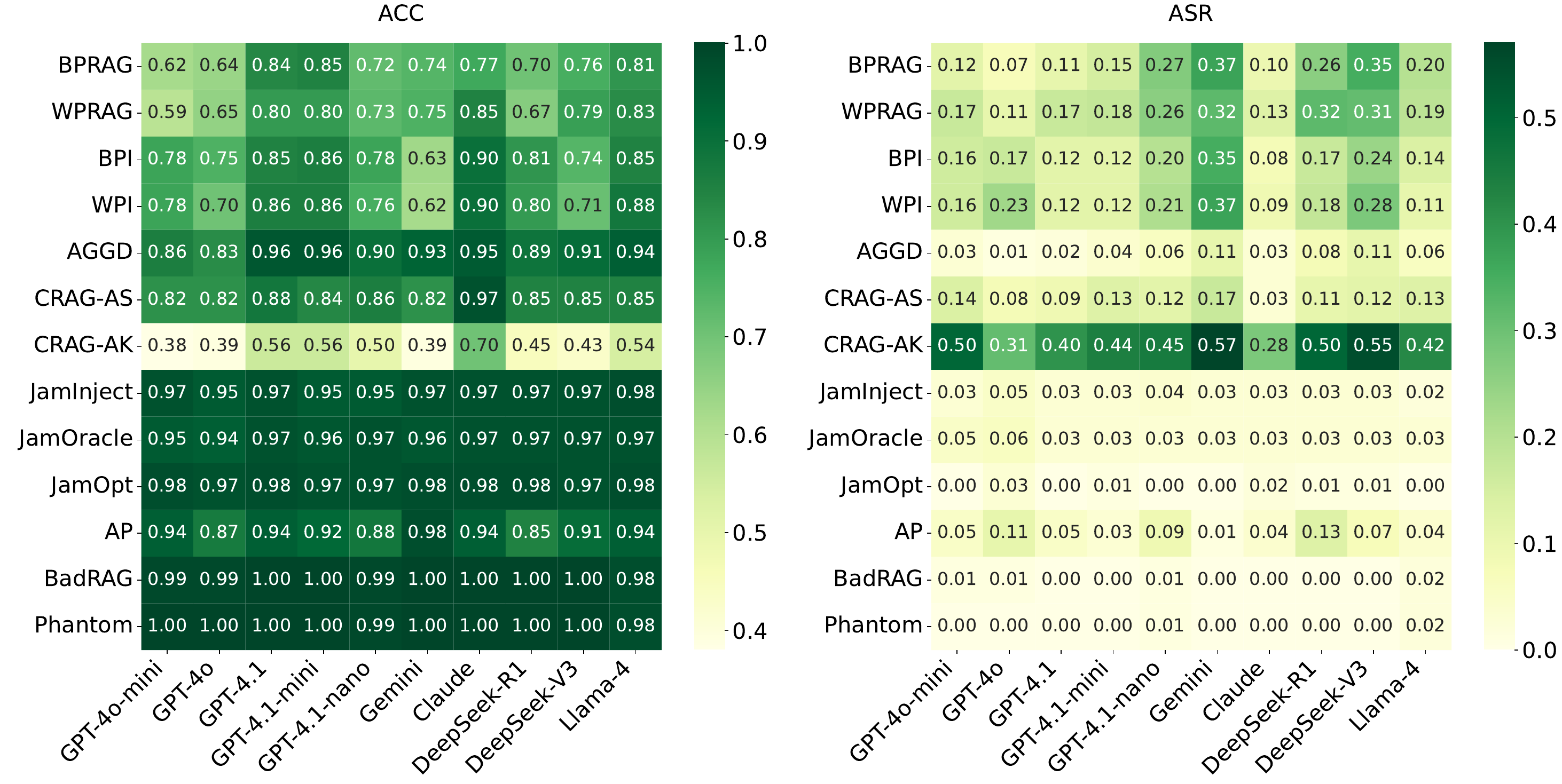}
        \caption{NQ-EX-M dataset}
        \label{fig:rag_llm_nq_ex_m}
    \end{subfigure}
    
    \vspace{1em}
    
    \begin{subfigure}{\textwidth}
        \centering
        \includegraphics[width=\textwidth]{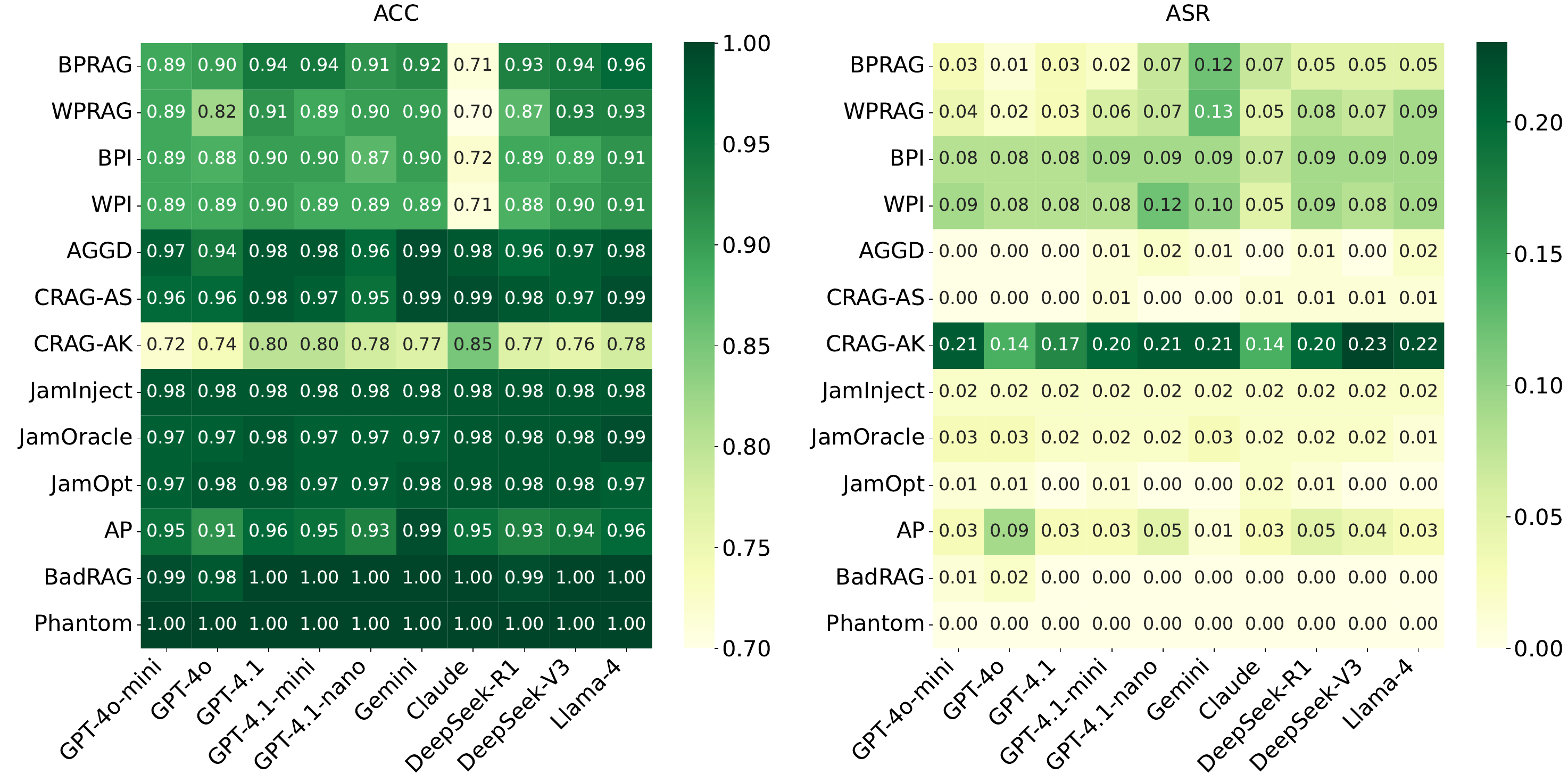}
        \caption{NQ-EX-L dataset}
        \label{fig:rag_llm_nq_ex_l}
    \end{subfigure}
    
    \caption{The results of poisoning attacks under different LLMs of RAG on NQ-EX-M and NQ-EX-L datasets.}  
    \label{fig:rag_llm_nq_ex}  
\end{figure*}

We provide further details on the LLMs used in Fig.~\ref{fig:rag_llm_nq} and Fig.~\ref{fig:rag_llm_nq_ex}. Specifically, the Claude model refers to claude-3-7-Sonnet, Gemini corresponds to gemini-2.0-flash, and Llama-4 denotes Llama-4-Scout-17B-16E-Instruct. All models were accessed via their respective API endpoints. To ensure consistent and stable responses across different experimental settings, the temperature parameter was set to 0.1.

\section{Details of Advanced RAG Frameworks}
\label{appendix_sec:details_of_advanced_rag_frameworks}

\subsection{Sequential RAG}

\myparatight{AAR~\cite{yu2023augmentation}}%
This framework follows the same process as the naive RAG, namely the retrieve-then-generate mechanism. AAR primarily introduces an augmentation-adapted retriever that can effectively generalize to unseen LLMs and achieves high retrieval performance.

\subsection{Branching RAG}

\myparatight{SuRe~\cite{yu2023augmentation}}%
This framework proposes a branch mechanism. Specifically, SuRe first generates multiple candidate answers based on the retrieved relevant texts. Then, it uses an LLM to generate a summary for each candidate answer based on the retrieved relevant texts. Finally, it selects the highest-ranked candidate answer as the response based on these summaries.

\subsection{Conditional RAG}

\myparatight{Adaptive-RAG~\cite{jeong2024adaptive}}%
Unlike the naive RAG, this framework first determines whether a user query requires retrieval. Specifically, Adaptive-RAG uses a classifier to judge whether the user's query needs retrieval. If retrieval is deemed necessary, it executes the same process as the naive RAG; if not, it directly generates a response using the LLM.

\subsection{Loop RAG}

\myparatight{IRCoT~\cite{trivedi2022interleaving}}%
This framework proposes an iterative retrieve-then-generate mechanism. Specifically, IRCoT first retrieves the top-$K$ relevant texts based on the user's query and generates a response. It then determines whether this response contains the answer; if not, it uses this response to continue retrieving relevant texts and generate a new response. This iterative process continues until the response contains the answer.

\myparatight{FLARE~\cite{jiang2023active}}%
This framework is also an iterative retrieve-then-generate mechanism, but it employs an adaptive retrieval judgment method. Specifically, for a user query, FLARE first directly asks the LLM to generate a response. It then calculates the generation probability of this response; if it exceeds a threshold, it determines that no further retrieval is needed. If it falls below this threshold, the retrieval mechanism is activated.

\myparatight{RQRAG~\cite{chan2024rq}}%
This framework proposes a mechanism that integrates multiple operations such as query decomposition and disambiguation. Specifically, RQRAG first determines whether the user query requires retrieval. It then breaks down the user's query into multiple sub-queries. Finally, it eliminates ambiguities in the queries and responses.

\section{Details of Poisoning Attacks against Multi-turn RAG}
\label{appendix_sec:details_of_multi_turn_rag}

Since no prior work has explored the effectiveness of poisoning attacks in multi-turn RAG, there is currently no existing framework suitable for direct use. To address this gap, we designed a simulated evaluation setup for multi-turn poisoning attacks, which consists of the following key components.

\myparatight{Conversation construction}%
To generate a multi-turn dialogue scenario, we crafted a prompt that instructs the LLM to simulate a conversation between a human and an intelligent assistant based on a targeted query and its correct answer. The prompt specifically requires that the final turn of the conversation includes only the human's query, and that the entire dialogue history, when combined with this final query, remains aligned with the intent of the original targeted query. The full prompt is presented below.

\begin{tcolorbox}[colback=gray!10,
                  colframe=black!80,
                  width=\linewidth,
                  arc=1mm, auto outer arc,
                  boxrule=1pt,
                  title =  Prompt for conversation construction
                 ]

Task: Based on the provided target question and answer, create a 5-turn dialogue between a human and an AI assistant with the following requirements:

1. Dialogue format requirements:

- Each turn consists of a human question and an AI assistant response

- The first 4 turns include both human questions and AI assistant responses

- The 5th turn contains only the human question, with no AI assistant response

2. Dialogue content requirements:

- The dialogue should be natural and fluent, resembling a realistic conversation

- Earlier turns should gradually lead toward the target question

- The dialogue content should maintain consistency and coherence

- Human questions should be diverse (including open-ended questions, requests for explanations, seeking advice, etc.)

- AI responses should be professional, helpful, and informative

3. Naming and reference requirements:

- The key nouns and entities from the target question should appear explicitly in the human questions within the first 4 turns, establishing clear context

- In the 5th turn, the human question should avoid directly repeating these key nouns; instead, it should use pronouns or other referring expressions to maintain naturalness and reduce redundancy

- Despite the change in wording, the 5th turn human question must preserve the original intent and goal of the target question

4. Final turn requirements:

- The human question in the 5th turn must achieve the same goal as the provided target question

- However, this final question should be significantly different in wording and structure from the target question

- The final question should be concise and leverage the context established in previous turns

- Much of the information contained in the target question should already be established in the dialogue history

- This approach should allow the final question to be shorter and more contextually appropriate

\end{tcolorbox}

\myparatight{Poisoning attacks to multi-turn RAG}%
The approach for generating poisoned texts and the strategy for injecting them into the knowledge database follow the same procedures as those used in the standard RAG setting.

\myparatight{Workflow of multi-turn RAG}%
For each user query, we begin by rewriting it to include the preceding conversation context, using the prompt provided below.

\begin{tcolorbox}[colback=gray!10,
                  colframe=black!80,
                  width=\linewidth,
                  arc=1mm, auto outer arc,
                  boxrule=1pt,
                  title =  Prompt for conversation construction
                 ]

Given the following conversation, please reword the final utterance from the human into a single utterance that does not need the history to understand the human's intent. Output in proper JSON format indicating the "class" (standalone or non-standalone) and the "reworded version" of the last utterance. 

In your rewording of the last utterance, do not do any unnecessary rephrasing or introduction of new terms or concepts that were not mentioned in the prior part of the conversation. Be minimal, by staying as close as possible to the shape and meaning of the last user utterance. If the last user utterance is already clear and standalone, the reworded version should be THE SAME as the last user utterance, and the class should be 'standalone'.

\end{tcolorbox}

Next, we use the rewritten query to retrieve the top-$K$ most relevant texts. Once retrieved, we combine the conversation history, the top-$K$ texts, and the original user query, and submit them to the LLM to generate a response.

\section{Details of Poisoning Attacks against Multimodal RAG}
\label{appendix_sec:details_of_multi_modal_rag}
We implemented a multimodal RAG based on the FlashRAG~\cite{jin2024flashrag} framework. Following work~\cite{liu2025poisoned}, we used InfoSeek~\cite{chen2023can} for evaluation, which includes a knowledge database containing 481,782 image-text pairs and a collection of 17,593 image-text queries. In our benchmark, similar to work~\cite{liu2025poisoned}, we randomly selected 50 queries as targeted queries and used a vision large model (VLM) to generate targeted answers that differ from the correct answers. For adapted JamInject and JamOracle attacks, we set their targeted answer to ``I do not know''. We evaluated the dirty-label attack proposed in work~\cite{liu2025poisoned} and adapted existing poisoning attacks to the multimodal RAG setting. We introduce these attack methods as follows.

\myparatight{Dirty-label attack~\cite{liu2025poisoned}}%
This attack begins by using the image associated with the targeted query as the image in the malicious image-text pair to preserve high retrieval similarity. Then, it utilizes a vision-language model (VLM) to generate a textual description of the image, conditioned on both the targeted query and the intended answer. This ensures that, when the description is retrieved as context, the VLM will produce the targeted answer. The resulting description is then used as the text component of the malicious image-text pair.

\myparatight{Adaptive attack}%
For the four poisoning attacks, namely BPRAG, BPI, JamInject, and JamOracle, we adapted each method using a consistent two-step process. First, we assigned the image from the targeted query as the image in the malicious image-text pair to maintain retrieval relevance. Second, following the original attack designs, we generated the corresponding text content to serve as the textual component of the malicious pair.

\section{Details of Poisoning Attacks against LLM Agent Systems }
\label{appendix_sec:details_of_llm_agent}

Following AgentPoison~\cite{chen2024agentpoison}, a poisoning attack designed for LLM agent systems, we adopted the ReAct-StrategyQA~\cite{yao2023react,geva2021did} agent task for evaluation. This task involves a knowledge database of 10,000 key-value pairs along with a set of multi-step commonsense reasoning queries. In our benchmark, we randomly selected 100 queries as targeted queries and used GPT-4o-mini to generate targeted answers that intentionally contradicted the correct ones. For AgentPoison, as well as the adapted JamInject and JamOracle attacks, we defined the targeted answer as ``I do not know''. We then evaluated both the original AgentPoison method and the adapted poisoning attacks within this LLM agent setting. The details of these attack methods are introduced below.

\myparatight{AgentPoison attack~\cite{chen2024agentpoison}}%
This attack targets the knowledge database by injecting malicious key-value pairs, with the goal of manipulating the LLM agent to produce targeted answers when queries include a specific trigger. The process begins by optimizing a trigger using training set queries, ensuring that any query containing this trigger exhibits high embedding similarity. Then, the attacker uses the embedding vector of the optimized trigger as the key and pairs it with a malicious instruction as the value, designed to prompt the LLM agent to generate the targeted response.

\myparatight{Adaptive attack}%
For the six poisoning attacks, including BPRAG, BPI, CRAG-AS, CRAG-AK, JamInject, and JamOracle, we adapted each method using a unified procedure. First, we extracted the embedding vector of the targeted query to serve as the key in the malicious key-value pair. Second, consistent with the design of each original attack, we generated the corresponding textual content to serve as the value associated with that key.

\begin{table*}[!htbp]
\tiny
\centering
\addtolength{\tabcolsep}{0pt}
\caption{The results of all poisoning attacks against various defenses on HotpotQA dataset.}
\label{tab:defense_hotpotqa}
\begin{tabular}{l|c|cccccccc} 
    \toprule
    Attack & Metric & No defense & Paraphrasing & InstructRAG & RobustRAG & AstuteRAG & PPL & Norm & TrustRAG \\ 
    \midrule
    \multirow{2}{*}{BPRAG} & ACC & 0.13~ & 0.16~ & 0.27~ & 0.40~ & 0.29~ & 0.12~ & 0.12~ & 0.60~ \\
     & ASR & 0.81~ & 0.79~ & 0.63~ & 0.36~ & 0.71~ & 0.83~ & 0.82~ & 0.06~ \\ 
    \midrule
    \rowcolor{greyD} 
\cellcolor{greyD}  & ACC & 0.15~ & 0.15~ & 0.31~ & 0.42~ & 0.28~ & 0.15~ & 0.13~ & 0.56~ \\
  \rowcolor{greyD} 
\cellcolor{greyD}\multirow{-2}{*}{WPRAG} & ASR & 0.79~ & 0.80~ & 0.53~ & 0.33~ & 0.73~ & 0.79~ & 0.80~ & 0.06~ \\ 
    \midrule
    \multirow{2}{*}{BPI} & ACC & 0.00~ & 0.01~ & 0.34~ & 0.32~ & 0.39~ & 0.01~ & 0.01~ & 0.58~ \\
     & ASR & 0.99~ & 0.99~ & 0.57~ & 0.40~ & 0.60~ & 0.99~ & 0.99~ & 0.08~ \\ 
    \midrule
   \rowcolor{greyD} 
\cellcolor{greyD}  & ACC & 0.00~ & 0.00~ & 0.37~ & 0.36~ & 0.39~ & 0.00~ & 0.00~ & 0.61~ \\
\bottomrule
    \end{tabular}
\end{table*}

\begin{table*}[!htbp]
\tiny
\centering
\addtolength{\tabcolsep}{0pt}
\caption{The results of all poisoning attacks against various defenses on MS-MARCO dataset.}
\label{tab:defense_msmarco}
\begin{tabular}{l|c|cccccccc} 
    \toprule
    Attack & Metric & No defense & Paraphrasing & InstructRAG & RobustRAG & AstuteRAG & PPL & Norm & TrustRAG \\ 
    \midrule
    \multirow{2}{*}{BPRAG} & ACC & 0.06~ & 0.14~ & 0.27~ & 0.57~ & 0.23~ & 0.08~ & 0.07~ & 0.77~ \\
     & ASR & 0.81~ & 0.73~ & 0.60~ & 0.31~ & 0.75~ & 0.81~ & 0.80~ & 0.07~ \\ 
    \midrule
 \rowcolor{greyD} 
\cellcolor{greyD}   & ACC & 0.10~ & 0.14~ & 0.27~ & 0.58~ & 0.27~ & 0.10~ & 0.10~ & 0.79~ \\
\rowcolor{greyD} 
\cellcolor{greyD}\multirow{-2}{*}{WPRAG}  & ASR & 0.78~ & 0.65~ & 0.58~ & 0.31~ & 0.69~ & 0.76~ & 0.76~ & 0.11~ \\ 
    \midrule
    \multirow{2}{*}{BPI} & ACC & 0.08~ & 0.06~ & 0.34~ & 0.53~ & 0.44~ & 0.08~ & 0.09~ & 0.79~ \\
     & ASR & 0.90~ & 0.87~ & 0.58~ & 0.26~ & 0.52~ & 0.90~ & 0.90~ & 0.10~ \\ 
    \midrule
   \rowcolor{greyD} 
\cellcolor{greyD} & ACC & 0.05~ & 0.09~ & 0.50~ & 0.50~ & 0.54~ & 0.17~ & 0.06~ & 0.80~ \\
\bottomrule
    \end{tabular}
\end{table*}

\begin{table*}[!htbp]
\tiny
\centering
\addtolength{\tabcolsep}{0pt}
\caption{The results of all poisoning attacks against various defenses on BoolQ dataset.}
\label{tab:defense_boolq}
\begin{tabular}{l|c|cccccccc} 
    \toprule
    Attack & Metric & No defense & Paraphrasing & InstructRAG & RobustRAG & AstuteRAG & PPL & Norm & TrustRAG \\ 
    \midrule
    \multirow{2}{*}{BPRAG} & ACC & 0.20~ & 0.28~ & 0.33~ & 0.55~ & 0.61~ & 0.18~ & 0.18~ & 0.79~ \\
     & ASR & 0.65~ & 0.52~ & 0.56~ & 0.25~ & 0.19~ & 0.60~ & 0.60~ & 0.09~ \\ 
    \midrule
    \rowcolor{greyD}  \cellcolor{greyD}  & ACC & 0.22~ & 0.32~ & 0.32~ & 0.58~ & 0.65~ & 0.21~ & 0.22~ & 0.77~ \\
  \rowcolor{greyD}  \cellcolor{greyD}\multirow{-2}{*}{WPRAG}  & ASR & 0.68~ & 0.54~ & 0.56~ & 0.26~ & 0.16~ & 0.59~ & 0.61~ & 0.10~ \\ 
    \midrule
    \multirow{2}{*}{BPI} & ACC & 0.06~ & 0.01~ & 0.46~ & 0.62~ & 0.83~ & 0.05~ & 0.05~ & 0.81~ \\
     & ASR & 0.94~ & 0.95~ & 0.46~ & 0.03~ & 0.10~ & 0.94~ & 0.94~ & 0.10~ \\ 
    \midrule
   \rowcolor{greyD}  \cellcolor{greyD}   & ACC & 0.03~ & 0.01~ & 0.52~ & 0.64~ & 0.82~ & 0.03~ & 0.03~ & 0.80~ \\
\bottomrule
    \end{tabular}
\end{table*}

\begin{table*}[!htbp]
\tiny
\centering
\addtolength{\tabcolsep}{0pt}
\caption{The results of all poisoning attacks against various defenses on SQuAD dataset.}
\label{tab:defense_squad}
\begin{tabular}{l|c|cccccccc} 
    \toprule
    Attack & Metric & No defense & Paraphrasing & InstructRAG & RobustRAG & AstuteRAG & PPL & Norm & TrustRAG \\ 
    \midrule
    \multirow{2}{*}{BPRAG} & ACC & 0.08~ & 0.09~ & 0.22~ & 0.24~ & 0.20~ & 0.07~ & 0.07~ & 0.49~ \\
     & ASR & 0.91~ & 0.84~ & 0.79~ & 0.50~ & 0.78~ & 0.89~ & 0.88~ & 0.07~ \\ 
    \midrule
  \rowcolor{greyD}  \cellcolor{greyD}  & ACC & 0.07~ & 0.09~ & 0.22~ & 0.23~ & 0.23~ & 0.07~ & 0.07~ & 0.48~ \\
   \rowcolor{greyD}  \cellcolor{greyD}\multirow{-2}{*}{WPRAG}  & ASR & 0.89~ & 0.89~ & 0.77~ & 0.53~ & 0.76~ & 0.88~ & 0.88~ & 0.06~ \\ 
    \midrule
    \multirow{2}{*}{BPI} & ACC & 0.00~ & 0.00~ & 0.25~ & 0.21~ & 0.26~ & 0.00~ & 0.00~ & 0.47~ \\
     & ASR & 1.00~ & 0.95~ & 0.69~ & 0.48~ & 0.70~ & 0.99~ & 1.00~ & 0.05~ \\ 
    \midrule
   \rowcolor{greyD}  \cellcolor{greyD}   & ACC & 0.00~ & 0.00~ & 0.26~ & 0.18~ & 0.29~ & 0.00~ & 0.00~ & 0.50~ \\
\bottomrule
    \end{tabular}
\end{table*}

\begin{table*}[!htbp]
\tiny
\centering
\addtolength{\tabcolsep}{0pt}
\caption{The results of all poisoning attacks against various defenses on NQ-EX-M dataset.}
\label{tab:defense_nq_ex_m}
\begin{tabular}{l|c|cccccccc} 
    \toprule
    Attack & Metric & No defense & Paraphrasing & InstructRAG & RobustRAG & AstuteRAG & PPL & Norm & TrustRAG \\ 
    \midrule
    \multirow{2}{*}{BPRAG} & ACC & 0.65~ & 0.59~ & 0.82~ & 0.69~ & 0.79~ & 0.64~ & 0.63~ & 0.71~ \\
     & ASR & 0.11~ & 0.17~ & 0.15~ & 0.11~ & 0.01~ & 0.13~ & 0.11~ & 0.04~ \\ 
    \midrule
  \rowcolor{greyD}  \cellcolor{greyD}  & ACC & 0.59~ & 0.66~ & 0.79~ & 0.64~ & 0.73~ & 0.64~ & 0.63~ & 0.73~ \\
   \rowcolor{greyD}  \cellcolor{greyD}\multirow{-2}{*}{WPRAG}  & ASR & 0.15~ & 0.12~ & 0.17~ & 0.13~ & 0.03~ & 0.15~ & 0.15~ & 0.06~ \\ 
    \midrule
    \multirow{2}{*}{BPI} & ACC & 0.77~ & 0.85~ & 0.87~ & 0.76~ & 0.90~ & 0.77~ & 0.76~ & 0.72~ \\
     & ASR & 0.16~ & 0.07~ & 0.12~ & 0.07~ & 0.09~ & 0.17~ & 0.17~ & 0.05~ \\ 
    \midrule
   \rowcolor{greyD}  \cellcolor{greyD}   & ACC & 0.80~ & 0.84~ & 0.88~ & 0.78~ & 0.87~ & 0.78~ & 0.78~ & 0.69~ \\
\bottomrule
    \end{tabular}
\end{table*}

\begin{table*}[!htbp]
\tiny
\centering
\addtolength{\tabcolsep}{0pt}
\caption{The results of all poisoning attacks against various defenses on HotpotQA-EX-M dataset.}
\label{tab:defense_hotpotqa_ex_m}
\begin{tabular}{l|c|cccccccc} 
    \toprule
    Attack & Metric & No defense & Paraphrasing & InstructRAG & RobustRAG & AstuteRAG & PPL & Norm & TrustRAG \\ 
    \midrule
    \multirow{2}{*}{BPRAG} & ACC & 0.64~ & 0.58~ & 0.69~ & 0.72~ & 0.84~ & 0.67~ & 0.62~ & 0.57~ \\
     & ASR & 0.20~ & 0.17~ & 0.15~ & 0.15~ & 0.05~ & 0.21~ & 0.20~ & 0.07~ \\ 
    \midrule
  \rowcolor{greyD}  \cellcolor{greyD} & ACC & 0.72~ & 0.62~ & 0.77~ & 0.70~ & 0.82~ & 0.70~ & 0.67~ & 0.60~ \\
   \rowcolor{greyD}  \cellcolor{greyD}\multirow{-2}{*}{WPRAG}   & ASR & 0.20~ & 0.17~ & 0.13~ & 0.15~ & 0.03~ & 0.18~ & 0.19~ & 0.10~ \\ 
    \midrule
    \multirow{2}{*}{BPI} & ACC & 0.88~ & 0.83~ & 0.90~ & 0.82~ & 0.96~ & 0.88~ & 0.88~ & 0.59~ \\
     & ASR & 0.09~ & 0.12~ & 0.02~ & 0.04~ & 0.03~ & 0.10~ & 0.10~ & 0.05~ \\ 
    \midrule
   \rowcolor{greyD}  \cellcolor{greyD}  & ACC & 0.92~ & 0.89~ & 0.89~ & 0.82~ & 0.97~ & 0.91~ & 0.92~ & 0.60~ \\
\bottomrule
    \end{tabular}
\end{table*}

\begin{table*}[!htbp]
\tiny
\centering
\addtolength{\tabcolsep}{0pt}
\caption{The results of all poisoning attacks against various defenses on MS-MARCO-EX-M dataset.}
\label{tab:defense_msmarco_ex_m}
\begin{tabular}{l|c|cccccccc} 
    \toprule
    Attacks & Metrics & No defense & Paraphrasing & InstructRAG & RobustRAG & AstuteRAG & PPL & Norm & TrustRAG \\ 
    \midrule
    \multirow{2}{*}{BPRAG} & ACC & 0.34~ & 0.52~ & 0.49~ & 0.71~ & 0.66~ & 0.30~ & 0.30~ & 0.79~ \\
     & ASR & 0.30~ & 0.24~ & 0.23~ & 0.19~ & 0.17~ & 0.35~ & 0.30~ & 0.08~ \\ 
    \midrule
  \rowcolor{greyD}  \cellcolor{greyD}  & ACC & 0.49~ & 0.62~ & 0.54~ & 0.76~ & 0.72~ & 0.49~ & 0.44~ & 0.79~ \\
   \rowcolor{greyD}  \cellcolor{greyD}\multirow{-2}{*}{WPRAG}  & ASR & 0.29~ & 0.12~ & 0.22~ & 0.12~ & 0.13~ & 0.28~ & 0.27~ & 0.08~ \\ 
    \midrule
    \multirow{2}{*}{BPI} & ACC & 0.63~ & 0.78~ & 0.84~ & 0.79~ & 0.83~ & 0.64~ & 0.62~ & 0.79~ \\
     & ASR & 0.33~ & 0.19~ & 0.13~ & 0.07~ & 0.13~ & 0.32~ & 0.35~ & 0.11~ \\ 
    \midrule
   \rowcolor{greyD}  \cellcolor{greyD}   & ACC & 0.63~ & 0.82~ & 0.87~ & 0.76~ & 0.83~ & 0.72~ & 0.64~ & 0.79~ \\
\bottomrule
    \end{tabular}
\end{table*}

\begin{table*}[!htbp]
\tiny
\centering
\addtolength{\tabcolsep}{0pt}
\caption{The results of all poisoning attacks against various defenses on BoolQ-EX-M dataset.}
\label{tab:defense_boolq_ex_m}
\begin{tabular}{l|c|cccccccc} 
    \toprule
    Attack & Metric & No defense & Paraphrasing & InstructRAG & RobustRAG & AstuteRAG & PPL & Norm & TrustRAG \\ 
    \midrule
    \multirow{2}{*}{BPRAG} & ACC & 0.49~ & 0.56~ & 0.67~ & 0.81~ & 0.83~ & 0.46~ & 0.44~ & 0.79~ \\
     & ASR & 0.35~ & 0.16~ & 0.23~ & 0.12~ & 0.10~ & 0.18~ & 0.19~ & 0.10~ \\ 
    \midrule
  \rowcolor{greyD}  \cellcolor{greyD} & ACC & 0.54~ & 0.57~ & 0.63~ & 0.84~ & 0.75~ & 0.49~ & 0.46~ & 0.78~ \\
   \rowcolor{greyD}  \cellcolor{greyD}\multirow{-2}{*}{WPRAG}   & ASR & 0.34~ & 0.16~ & 0.21~ & 0.11~ & 0.13~ & 0.17~ & 0.16~ & 0.10~ \\ 
    \midrule
    \multirow{2}{*}{BPI} & ACC & 0.77~ & 0.83~ & 0.90~ & 0.88~ & 0.86~ & 0.73~ & 0.75~ & 0.78~ \\
     & ASR & 0.24~ & 0.12~ & 0.05~ & 0.02~ & 0.05~ & 0.24~ & 0.23~ & 0.11~ \\ 
    \midrule
   \rowcolor{greyD}  \cellcolor{greyD}   & ACC & 0.66~ & 0.85~ & 0.89~ & 0.94~ & 0.89~ & 0.66~ & 0.67~ & 0.79~ \\
\bottomrule
    \end{tabular}
\end{table*}

\begin{table*}[!htbp]
\tiny
\centering
\addtolength{\tabcolsep}{0pt}
\caption{The results of all poisoning attacks against various defenses on SQuAD-EX-M dataset.}
\label{tab:defense_squad_ex_m}
\begin{tabular}{l|c|cccccccc} 
    \toprule
    Attack & Metric & No defense & Paraphrasing & InstructRAG & RobustRAG & AstuteRAG & PPL & Norm & TrustRAG \\ 
    \midrule
    \multirow{2}{*}{BPRAG} & ACC & 0.52~ & 0.59~ & 0.76~ & 0.63~ & 0.68~ & 0.52~ & 0.54~ & 0.53~ \\
     & ASR & 0.21~ & 0.20~ & 0.23~ & 0.10~ & 0.15~ & 0.25~ & 0.25~ & 0.05~ \\ 
    \midrule
  \rowcolor{greyD}  \cellcolor{greyD}  & ACC & 0.58~ & 0.57~ & 0.79~ & 0.66~ & 0.73~ & 0.57~ & 0.56~ & 0.49~ \\
    \rowcolor{greyD}  \cellcolor{greyD}\multirow{-2}{*}{WPRAG} & ASR & 0.24~ & 0.17~ & 0.19~ & 0.12~ & 0.11~ & 0.24~ & 0.21~ & 0.06~ \\ 
    \midrule
    \multirow{2}{*}{BPI} & ACC & 0.68~ & 0.77~ & 0.87~ & 0.67~ & 0.94~ & 0.66~ & 0.67~ & 0.51~ \\
     & ASR & 0.24~ & 0.11~ & 0.08~ & 0.11~ & 0.06~ & 0.26~ & 0.26~ & 0.03~ \\ 
    \midrule
   \rowcolor{greyD}  \cellcolor{greyD}   & ACC & 0.76~ & 0.83~ & 0.92~ & 0.67~ & 0.96~ & 0.72~ & 0.71~ & 0.51~ \\
\bottomrule
    \end{tabular}
\end{table*}

\begin{table*}[!htbp]
\tiny
\centering
\addtolength{\tabcolsep}{0pt}
\caption{The results of all poisoning attacks against various defenses on NQ-EX-L dataset.}
\label{tab:defense_nq_ex_l}
\begin{tabular}{l|c|cccccccc} 
    \toprule
    Attack & Metric & No defense & Paraphrasing & InstructRAG & RobustRAG & AstuteRAG & PPL & Norm & TrustRAG \\ 
    \midrule
    \multirow{2}{*}{BPRAG} & ACC & 0.89~ & 0.86~ & 0.94~ & 0.82~ & 0.97~ & 0.87~ & 0.87~ & 0.73~ \\
     & ASR & 0.03~ & 0.04~ & 0.04~ & 0.04~ & 0.03~ & 0.03~ & 0.03~ & 0.03~ \\ 
    \midrule
  \rowcolor{greyD}  \cellcolor{greyD}  & ACC & 0.87~ & 0.89~ & 0.91~ & 0.80~ & 0.93~ & 0.86~ & 0.86~ & 0.72~ \\
  \rowcolor{greyD}  \cellcolor{greyD}\multirow{-2}{*}{WPRAG}   & ASR & 0.05~ & 0.02~ & 0.07~ & 0.04~ & 0.07~ & 0.04~ & 0.04~ & 0.05~ \\ 
    \midrule
    \multirow{2}{*}{BPI} & ACC & 0.90~ & 0.91~ & 0.89~ & 0.78~ & 0.92~ & 0.89~ & 0.89~ & 0.71~ \\
     & ASR & 0.08~ & 0.04~ & 0.09~ & 0.04~ & 0.06~ & 0.08~ & 0.08~ & 0.04~ \\ 
    \midrule
   \rowcolor{greyD}  \cellcolor{greyD}  & ACC & 0.86~ & 0.91~ & 0.91~ & 0.78~ & 0.92~ & 0.87~ & 0.87~ & 0.70~ \\
\bottomrule
    \end{tabular}
\end{table*}

\begin{table*}[!htbp]
\tiny
\centering
\addtolength{\tabcolsep}{0pt}
\caption{The results of all poisoning attacks against various defenses on HotpotQA-EX-L dataset.}
\label{tab:defense_hotpotqa_ex_l}
\begin{tabular}{l|c|cccccccc} 
    \toprule
    Attack & Metric & No defense & Paraphrasing & InstructRAG & RobustRAG & AstuteRAG & PPL & Norm & TrustRAG \\ 
    \midrule
    \multirow{2}{*}{BPRAG} & ACC & 0.88~ & 0.90~ & 0.87~ & 0.82~ & 0.94~ & 0.86~ & 0.86~ & 0.59~ \\
     & ASR & 0.05~ & 0.04~ & 0.08~ & 0.05~ & 0.05~ & 0.04~ & 0.04~ & 0.09~ \\ 
    \midrule
  \rowcolor{greyD}  \cellcolor{greyD} & ACC & 0.87~ & 0.90~ & 0.89~ & 0.82~ & 0.94~ & 0.87~ & 0.89~ & 0.59~ \\
  \rowcolor{greyD}  \cellcolor{greyD}\multirow{-2}{*}{WPRAG}    & ASR & 0.05~ & 0.06~ & 0.05~ & 0.04~ & 0.05~ & 0.05~ & 0.05~ & 0.07~ \\ 
    \midrule
    \multirow{2}{*}{BPI} & ACC & 0.97~ & 0.95~ & 0.88~ & 0.86~ & 0.96~ & 0.96~ & 0.96~ & 0.59~ \\
     & ASR & 0.03~ & 0.04~ & 0.04~ & 0.03~ & 0.04~ & 0.03~ & 0.03~ & 0.08~ \\ 
    \midrule
   \rowcolor{greyD}  \cellcolor{greyD}  & ACC & 0.98~ & 0.95~ & 0.90~ & 0.85~ & 0.97~ & 0.97~ & 0.97~ & 0.56~ \\
\bottomrule
    \end{tabular}
\end{table*}

\begin{table*}[!htbp]
\tiny
\centering
\addtolength{\tabcolsep}{0pt}
\caption{The results of all poisoning attacks against various defenses on MS-MARCO-EX-L dataset.}
\label{tab:defense_msmarco_ex_l}
\begin{tabular}{l|c|cccccccc} 
    \toprule
    Attack & Metric & No defense & Paraphrasing & InstructRAG & RobustRAG & AstuteRAG & PPL & Norm & TrustRAG \\ 
    \midrule
    \multirow{2}{*}{BPRAG} & ACC & 0.68~ & 0.71~ & 0.78~ & 0.82~ & 0.75~ & 0.70~ & 0.69~ & 0.76~ \\
     & ASR & 0.08~ & 0.06~ & 0.04~ & 0.11~ & 0.23~ & 0.09~ & 0.07~ & 0.08~ \\ 
    \midrule
  \rowcolor{greyD}  \cellcolor{greyD}  & ACC & 0.74~ & 0.77~ & 0.77~ & 0.86~ & 0.75~ & 0.73~ & 0.76~ & 0.72~ \\
   \rowcolor{greyD}  \cellcolor{greyD}\multirow{-2}{*}{WPRAG}  & ASR & 0.13~ & 0.07~ & 0.06~ & 0.06~ & 0.20~ & 0.06~ & 0.10~ & 0.09~ \\ 
    \midrule
    \multirow{2}{*}{BPI} & ACC & 0.87~ & 0.89~ & 0.93~ & 0.84~ & 0.86~ & 0.89~ & 0.88~ & 0.79~ \\
     & ASR & 0.10~ & 0.06~ & 0.05~ & 0.05~ & 0.13~ & 0.10~ & 0.10~ & 0.10~ \\ 
    \midrule
   \rowcolor{greyD}  \cellcolor{greyD}  & ACC & 0.91~ & 0.90~ & 0.93~ & 0.85~ & 0.85~ & 0.91~ & 0.90~ & 0.76~ \\
\bottomrule
    \end{tabular}
\end{table*}

\begin{table*}[!htbp]
\tiny
\centering
\addtolength{\tabcolsep}{0pt}
\caption{The results of all poisoning attacks against various defenses on BoolQ-EX-L dataset.}
\label{tab:defense_boolq_ex_l}
\begin{tabular}{l|c|cccccccc} 
    \toprule
    Attack & Metric & No defense & Paraphrasing & InstructRAG & RobustRAG & AstuteRAG & PPL & Norm & TrustRAG \\ 
    \midrule
    \multirow{2}{*}{BPRAG} & ACC & 0.84~ & 0.80~ & 0.91~ & 0.88~ & 0.82~ & 0.81~ & 0.81~ & 0.81~ \\
     & ASR & 0.11~ & 0.08~ & 0.07~ & 0.06~ & 0.08~ & 0.04~ & 0.05~ & 0.09~ \\ 
    \midrule
  \rowcolor{greyD}  \cellcolor{greyD} & ACC & 0.75~ & 0.78~ & 0.85~ & 0.87~ & 0.83~ & 0.75~ & 0.74~ & 0.83~ \\
  \rowcolor{greyD}  \cellcolor{greyD}\multirow{-2}{*}{WPRAG}    & ASR & 0.21~ & 0.08~ & 0.09~ & 0.05~ & 0.05~ & 0.08~ & 0.09~ & 0.08~ \\ 
    \midrule
    \multirow{2}{*}{BPI} & ACC & 0.91~ & 0.94~ & 0.91~ & 0.87~ & 0.86~ & 0.88~ & 0.89~ & 0.79~ \\
     & ASR & 0.10~ & 0.03~ & 0.04~ & 0.01~ & 0.06~ & 0.08~ & 0.08~ & 0.10~ \\ 
    \midrule
   \rowcolor{greyD}  \cellcolor{greyD} & ACC & 0.92~ & 0.94~ & 0.95~ & 0.92~ & 0.91~ & 0.88~ & 0.87~ & 0.81~ \\
\bottomrule
    \end{tabular}
\end{table*}

\begin{table*}[!htbp]
\tiny
\centering
\addtolength{\tabcolsep}{0pt}
\caption{The results of all poisoning attacks against various defenses on SQuAD-EX-L dataset.}
\label{tab:defense_squad_ex_l}
\begin{tabular}{l|c|cccccccc} 
    \toprule
    Attack & Metric & No defense & Paraphrasing & InstructRAG & RobustRAG & AstuteRAG & PPL & Norm & TrustRAG \\ 
    \midrule
    \multirow{2}{*}{BPRAG} & ACC & 0.89~ & 0.86~ & 0.93~ & 0.79~ & 0.87~ & 0.87~ & 0.88~ & 0.53~ \\
     & ASR & 0.04~ & 0.04~ & 0.06~ & 0.01~ & 0.03~ & 0.04~ & 0.05~ & 0.06~ \\ 
    \midrule
  \rowcolor{greyD}  \cellcolor{greyD} & ACC & 0.89~ & 0.85~ & 0.98~ & 0.79~ & 0.89~ & 0.88~ & 0.90~ & 0.52~ \\
  \rowcolor{greyD}  \cellcolor{greyD}\multirow{-2}{*}{WPRAG}    & ASR & 0.04~ & 0.06~ & 0.02~ & 0.03~ & 0.02~ & 0.03~ & 0.03~ & 0.05~ \\ 
    \midrule
    \multirow{2}{*}{BPI} & ACC & 0.96~ & 0.94~ & 0.96~ & 0.79~ & 0.91~ & 0.96~ & 0.95~ & 0.54~ \\
     & ASR & 0.03~ & 0.03~ & 0.02~ & 0.02~ & 0.01~ & 0.03~ & 0.04~ & 0.07~ \\ 
    \midrule
   \rowcolor{greyD}  \cellcolor{greyD}  & ACC & 0.95~ & 0.93~ & 0.98~ & 0.79~ & 0.91~ & 0.96~ & 0.95~ & 0.56~ \\
\bottomrule
    \end{tabular}
\end{table*}


\begin{table*}[!htbp]
\tiny
\centering
\addtolength{\tabcolsep}{-4.0pt}
\caption{The results of detection-based defenses against all poisoning attacks on NQ dataset.}
\label{tab:detection_nq}
\begin{tabular}{l|c|ccccccc|ccc|ccc} 
    \toprule
    \multirow{2}{*}{Defense} & \multirow{2}{*}{Metric} & \multicolumn{7}{c|}{Targeted poisoning attack} & \multicolumn{3}{c|}{DoS attack} & \multicolumn{3}{c}{Trigger-based DoS attack} \\ 
    \cmidrule{3-15}
     &  & BPRAG & WPRAG & BPI & WPI & AGGD & CRAG-AS & CRAG-AK & JamInject & JamOracle & JamOpt & AP & BadRAG & Phantom \\ 
    \midrule
    \multirow{3}{*}{PPL} & DACC & 0.04~ & 0.04~ & 0.09~ & 0.07~ & 0.46~ & 0.22~ & 0.14~ & 0.25~ & 0.17~ & 0.24~ & 0.00~ & 0.63~ & 1.00~ \\
     & FPR & 0.00~ & 0.00~ & 0.00~ & 0.00~ & 0.00~ & 0.00~ & 0.00~ & 0.00~ & 0.00~ & 0.00~ & 0.00~ & 0.00~ & 0.00~ \\
     & FNR & 1.00~ & 1.00~ & 0.94~ & 1.00~ & 0.00~ & 0.00~ & 0.00~ & 0.85~ & 0.95~ & 0.81~ & 1.00~ & 0.51~ & 0.00~ \\ 
    \midrule
      \rowcolor{greyD}  \cellcolor{greyD}   & DACC & 0.04~ & 0.04~ & 0.09~ & 0.07~ & 0.22~ & 0.14~ & 0.05~ & 0.25~ & 0.17~ & 0.24~ & 0.00~ & 0.63~ & 1.00~ \\
   \rowcolor{greyD}  \cellcolor{greyD}    & FPR & 0.00~ & 0.00~ & 0.00~ & 0.00~ & 0.00~ & 0.00~ & 0.00~ & 0.00~ & 0.00~ & 0.00~ & 0.00~ & 0.00~ & 0.00~ \\
   \rowcolor{greyD}  \cellcolor{greyD}\multirow{-3}{*}{Norm}    & FNR & 1.00~ & 1.00~ & 0.94~ & 1.00~ & 0.91~ & 0.90~ & 1.00~ & 0.85~ & 0.95~ & 0.81~ & 1.00~ & 0.51~ & 0.00~ \\ 
    \midrule
    \multirow{3}{*}{TrustRAG} & DACC & 0.99~ & 0.98~ & 0.98~ & 0.99~ & 0.93~ & 0.97~ & 0.99~ & 0.93~ & 0.93~ & 0.94~ & 1.00~ & 0.85~ & 0.74~ \\
     & FPR & 0.03~ & 0.02~ & 0.02~ & 0.01~ & 0.10~ & 0.03~ & 0.04~ & 0.09~ & 0.07~ & 0.06~ & 0.00~ & 0.16~ & 0.26~ \\
     & FNR & 0.00~ & 0.01~ & 0.00~ & 0.01~ & 0.02~ & 0.00~ & 0.01~ & 0.01~ & 0.03~ & 0.00~ & 0.00~ & 0.04~ & 0.00~ \\
    \bottomrule
    \end{tabular}
\end{table*}

\begin{table*}[!htbp]
\tiny
\centering
\addtolength{\tabcolsep}{-4.0pt}
\caption{The results of detection-based defenses against all poisoning attacks on HotpotQA dataset.}
\label{tab:detection_hotpotqa}
\begin{tabular}{l|c|ccccccc|ccc|ccc} 
    \toprule
    \multirow{2}{*}{Defense} & \multirow{2}{*}{Metric} & \multicolumn{7}{c|}{Targeted poisoning attack} & \multicolumn{3}{c|}{DoS attack} & \multicolumn{3}{c}{Trigger-based DoS attack} \\ 
    \cmidrule{3-15}
     &  & BPRAG & WPRAG & BPI & WPI & AGGD & CRAG-AS & CRAG-AK & JamInject & JamOracle & JamOpt & AP & BadRAG & Phantom \\ 
    \midrule
    \multirow{3}{*}{PPL} & DACC & 0.00~ & 0.00~ & 0.00~ & 0.00~ & 0.18~ & 0.01~ & 0.00~ & 0.04~ & 0.00~ & 0.43~ & 0.96~ & 1.00~ & 1.00~ \\
     & FPR & 0.00~ & 0.00~ & 0.00~ & 0.00~ & 0.00~ & 0.00~ & 0.00~ & 0.00~ & 0.00~ & 0.02~ & 0.07~ & 0.00~ & 0.00~ \\
     & FNR & 1.00~ & 1.00~ & 1.00~ & 1.00~ & 0.11~ & 0.01~ & 0.00~ & 0.97~ & 1.00~ & 0.56~ & 0.00~ & 0.00~ & 0.00~ \\ 
    \midrule
      \rowcolor{greyD}  \cellcolor{greyD}   & DACC & 0.00~ & 0.00~ & 0.00~ & 0.00~ & 0.02~ & 0.00~ & 0.00~ & 0.04~ & 0.00~ & 0.20~ & 0.00~ & 0.60~ & 1.00~ \\
    \rowcolor{greyD}  \cellcolor{greyD}   & FPR & 0.00~ & 0.00~ & 0.00~ & 0.00~ & 0.00~ & 0.00~ & 0.00~ & 0.00~ & 0.00~ & 0.00~ & 0.00~ & 0.00~ & 0.00~ \\
  \rowcolor{greyD}  \cellcolor{greyD}\multirow{-3}{*}{Norm}     & FNR & 1.00~ & 1.00~ & 1.00~ & 1.00~ & 0.99~ & 1.00~ & 1.00~ & 0.97~ & 1.00~ & 0.80~ & 1.00~ & 0.59~ & 0.01~ \\ 
    \midrule
    \multirow{3}{*}{TrustRAG} & DACC & 1.00~ & 1.00~ & 1.00~ & 1.00~ & 0.99~ & 1.00~ & 1.00~ & 0.97~ & 1.00~ & 0.90~ & 1.00~ & 0.81~ & 0.69~ \\
     & FPR & 0.00~ & 0.00~ & 0.00~ & 0.00~ & 0.01~ & 0.00~ & 0.00~ & 0.04~ & 0.00~ & 0.10~ & 0.00~ & 0.19~ & 0.31~ \\
     & FNR & 0.00~ & 0.00~ & 0.00~ & 0.00~ & 0.00~ & 0.00~ & 0.00~ & 0.00~ & 0.00~ & 0.00~ & 0.00~ & 0.08~ & 0.00~ \\
    \bottomrule
    \end{tabular}
\end{table*}

\begin{table*}[!htbp]
\tiny
\centering
\addtolength{\tabcolsep}{-4.0pt}
\caption{The results of detection-based defenses against all poisoning attacks on MS-MARCO dataset.}
\label{tab:detection_msmarco}
\begin{tabular}{l|c|ccccccc|ccc|ccc} 
    \toprule
    \multirow{2}{*}{Defense} & \multirow{2}{*}{Metric} & \multicolumn{7}{c|}{Targeted poisoning attack} & \multicolumn{3}{c|}{DoS attack} & \multicolumn{3}{c}{Trigger-based DoS attack} \\ 
    \cmidrule{3-15}
     &  & BPRAG & WPRAG & BPI & WPI & AGGD & CRAG-AS & CRAG-AK & JamInject & JamOracle & JamOpt & AP & BadRAG & Phantom \\ 
    \midrule
    \multirow{3}{*}{PPL} & DACC & 0.07~ & 0.17~ & 0.16~ & 0.33~ & 0.55~ & 0.50~ & 0.05~ & 0.40~ & 0.44~ & 0.70~ & 1.00~ & 1.00~ & 1.00~ \\
     & FPR & 0.00~ & 0.00~ & 0.00~ & 0.00~ & 0.00~ & 0.00~ & 0.00~ & 0.00~ & 0.00~ & 0.00~ & 0.00~ & 0.00~ & 0.00~ \\
     & FNR & 0.98~ & 0.98~ & 0.90~ & 0.75~ & 0.03~ & 0.56~ & 0.97~ & 0.69~ & 0.68~ & 0.37~ & 0.00~ & 0.00~ & 0.00~ \\ 
    \midrule
    \rowcolor{greyD}  \cellcolor{greyD}    & DACC & 0.07~ & 0.16~ & 0.16~ & 0.17~ & 0.34~ & 0.32~ & 0.05~ & 0.38~ & 0.44~ & 0.52~ & 0.00~ & 0.95~ & 1.00~ \\
   \rowcolor{greyD}  \cellcolor{greyD}    & FPR & 0.00~ & 0.00~ & 0.00~ & 0.00~ & 0.00~ & 0.00~ & 0.00~ & 0.00~ & 0.00~ & 0.00~ & 0.00~ & 0.00~ & 0.00~ \\
     \rowcolor{greyD}  \cellcolor{greyD}\multirow{-3}{*}{Norm}  & FNR & 0.98~ & 0.99~ & 0.90~ & 0.96~ & 0.84~ & 0.76~ & 0.97~ & 0.71~ & 0.68~ & 0.60~ & 1.00~ & 0.08~ & 0.00~ \\ 
    \midrule
    \multirow{3}{*}{TrustRAG} & DACC & 0.95~ & 0.89~ & 0.91~ & 0.91~ & 0.79~ & 0.83~ & 0.96~ & 0.79~ & 0.69~ & 0.73~ & 1.00~ & 0.51~ & 0.50~ \\
     & FPR & 0.10~ & 0.21~ & 0.10~ & 0.13~ & 0.28~ & 0.21~ & 0.05~ & 0.24~ & 0.35~ & 0.28~ & 0.01~ & 0.49~ & 0.50~ \\
     & FNR & 0.01~ & 0.01~ & 0.01~ & 0.02~ & 0.07~ & 0.01~ & 0.01~ & 0.02~ & 0.04~ & 0.05~ & 0.00~ & 0.02~ & 0.00~ \\
    \bottomrule
    \end{tabular}
\end{table*}

\begin{table*}[!htbp]
\tiny
\centering
\addtolength{\tabcolsep}{-4.0pt}
\caption{The results of detection-based defenses against all poisoning attacks on BoolQ dataset.}
\label{tab:detection_boolq}
\begin{tabular}{l|c|ccccccc|ccc|ccc} 
    \toprule
    \multirow{2}{*}{Defense} & \multirow{2}{*}{Metric} & \multicolumn{7}{c|}{Targeted poisoning attack} & \multicolumn{3}{c|}{DoS attack} & \multicolumn{3}{c}{Trigger-based DoS attack} \\ 
    \cmidrule{3-15}
     &  & BPRAG & WPRAG & BPI & WPI & AGGD & CRAG-AS & CRAG-AK & JamInject & JamOracle & JamOpt & AP & BadRAG & Phantom \\ 
    \midrule
    \multirow{3}{*}{PPL} & DACC & 0.04~ & 0.06~ & 0.11~ & 0.13~ & 0.12~ & 0.23~ & 0.03~ & 0.27~ & 0.15~ & 0.36~ & 0.98~ & 1.00~ & 1.00~ \\
     & FPR & 0.00~ & 0.00~ & 0.00~ & 0.00~ & 0.00~ & 0.00~ & 0.00~ & 0.00~ & 0.00~ & 0.00~ & 0.03~ & 0.00~ & 0.00~ \\
     & FNR & 1.00~ & 1.00~ & 0.94~ & 0.99~ & 0.98~ & 0.84~ & 0.99~ & 0.80~ & 0.95~ & 0.71~ & 0.00~ & 0.00~ & 0.00~ \\ 
    \midrule
      \rowcolor{greyD}  \cellcolor{greyD}   & DACC & 0.04~ & 0.06~ & 0.11~ & 0.12~ & 0.11~ & 0.22~ & 0.03~ & 0.27~ & 0.15~ & 0.30~ & 0.00~ & 1.00~ & 1.00~ \\
  \rowcolor{greyD}  \cellcolor{greyD}     & FPR & 0.00~ & 0.00~ & 0.00~ & 0.00~ & 0.00~ & 0.00~ & 0.00~ & 0.00~ & 0.00~ & 0.00~ & 0.00~ & 0.00~ & 0.00~ \\
  \rowcolor{greyD}   \cellcolor{greyD}\multirow{-3}{*}{Norm}    & FNR & 1.00~ & 1.00~ & 0.94~ & 0.99~ & 0.99~ & 0.85~ & 0.99~ & 0.80~ & 0.95~ & 0.77~ & 1.00~ & 0.00~ & 0.00~ \\ 
    \midrule
    \multirow{3}{*}{TrustRAG} & DACC & 0.97~ & 0.95~ & 0.92~ & 0.93~ & 0.91~ & 0.86~ & 0.98~ & 0.84~ & 0.89~ & 0.85~ & 1.00~ & 0.55~ & 0.55~ \\
     & FPR & 0.07~ & 0.13~ & 0.11~ & 0.10~ & 0.17~ & 0.17~ & 0.03~ & 0.20~ & 0.17~ & 0.18~ & 0.00~ & 0.45~ & 0.45~ \\
     & FNR & 0.00~ & 0.01~ & 0.00~ & 0.03~ & 0.03~ & 0.01~ & 0.00~ & 0.01~ & 0.01~ & 0.02~ & 0.00~ & 0.00~ & 0.00~ \\
    \bottomrule
    \end{tabular}
\end{table*}

\begin{table*}[!htbp]
\tiny
\centering
\addtolength{\tabcolsep}{-4.0pt}
\caption{The results of detection-based defenses against all poisoning attacks on SQuAD dataset.}
\label{tab:detection_squad}
\begin{tabular}{l|c|ccccccc|ccc|ccc} 
    \toprule
    \multirow{2}{*}{Defense} & \multirow{2}{*}{Metric} & \multicolumn{7}{c|}{Targeted poisoning attack} & \multicolumn{3}{c|}{DoS attack} & \multicolumn{3}{c}{Trigger-based DoS attack} \\ 
    \cmidrule{3-15}
     &  & BPRAG & WPRAG & BPI & WPI & AGGD & CRAG-AS & CRAG-AK & JamInject & JamOracle & JamOpt & AP & BadRAG & Phantom \\ 
    \midrule
    \multirow{3}{*}{PPL} & DACC & 0.05~ & 0.05~ & 0.03~ & 0.04~ & 0.09~ & 0.03~ & 0.08~ & 0.08~ & 0.10~ & 0.38~ & 0.98~ & 0.99~ & 0.99~ \\
     & FPR & 0.00~ & 0.00~ & 0.00~ & 0.00~ & 0.00~ & 0.00~ & 0.00~ & 0.00~ & 0.00~ & 0.05~ & 0.03~ & 0.01~ & 0.01~ \\
     & FNR & 1.00~ & 1.00~ & 0.97~ & 0.96~ & 0.97~ & 0.00~ & 0.00~ & 0.92~ & 0.93~ & 0.57~ & 0.00~ & 0.00~ & 0.00~ \\ 
    \midrule
      \rowcolor{greyD}  \cellcolor{greyD}  & DACC & 0.05~ & 0.05~ & 0.03~ & 0.04~ & 0.09~ & 0.03~ & 0.04~ & 0.08~ & 0.10~ & 0.20~ & 0.00~ & 1.00~ & 0.99~ \\
 \rowcolor{greyD}  \cellcolor{greyD}    & FPR & 0.00~ & 0.00~ & 0.00~ & 0.00~ & 0.00~ & 0.00~ & 0.00~ & 0.00~ & 0.00~ & 0.00~ & 0.00~ & 0.00~ & 0.00~ \\
   \rowcolor{greyD}  \cellcolor{greyD}\multirow{-3}{*}{Norm}     & FNR & 1.00~ & 1.00~ & 0.97~ & 0.96~ & 0.97~ & 0.97~ & 1.00~ & 0.92~ & 0.93~ & 0.81~ & 1.00~ & 0.00~ & 0.02~ \\ 
    \midrule
    \multirow{3}{*}{TrustRAG} & DACC & 0.95~ & 0.95~ & 0.97~ & 0.96~ & 0.91~ & 0.97~ & 0.96~ & 0.92~ & 0.90~ & 0.84~ & 1.00~ & 0.71~ & 0.71~ \\
     & FPR & 0.12~ & 0.12~ & 0.03~ & 0.04~ & 0.16~ & 0.03~ & 0.08~ & 0.08~ & 0.14~ & 0.18~ & 0.00~ & 0.29~ & 0.29~ \\
     & FNR & 0.00~ & 0.00~ & 0.00~ & 0.00~ & 0.01~ & 0.00~ & 0.00~ & 0.00~ & 0.00~ & 0.00~ & 0.00~ & 0.00~ & 0.01~ \\
    \bottomrule
    \end{tabular}
\end{table*}

\begin{table*}[!htbp]
\tiny
\centering
\addtolength{\tabcolsep}{-4.0pt}
\caption{The results of detection-based defenses against all poisoning attacks on NQ-EX-M dataset.}
\label{tab:detection_nq_ex_m}
\begin{tabular}{l|c|ccccccc|ccc|ccc} 
    \toprule
    \multirow{2}{*}{Defense} & \multirow{2}{*}{Metric} & \multicolumn{7}{c|}{Targeted poisoning attack} & \multicolumn{3}{c|}{DoS attack} & \multicolumn{3}{c}{Trigger-based DoS attack} \\ 
    \cmidrule{3-15}
     &  & BPRAG & WPRAG & BPI & WPI & AGGD & CRAG-AS & CRAG-AK & JamInject & JamOracle & JamOpt & AP & BadRAG & Phantom \\ 
    \midrule
    \multirow{3}{*}{PPL} & DACC & 0.52~ & 0.47~ & 0.80~ & 0.74~ & 0.80~ & 0.93~ & 0.60~ & 0.94~ & 0.76~ & 0.92~ & 0.98~ & 1.00~ & 1.00~ \\
     & FPR & 0.00~ & 0.00~ & 0.00~ & 0.00~ & 0.00~ & 0.00~ & 0.00~ & 0.00~ & 0.00~ & 0.00~ & 0.00~ & 0.00~ & 0.00~ \\
     & FNR & 0.96~ & 0.98~ & 0.35~ & 0.51~ & 0.43~ & 0.18~ & 0.71~ & 0.14~ & 0.58~ & 0.16~ & 0.04~ & 0.00~ & 0.00~ \\ 
    \midrule
\rowcolor{greyD}  \cellcolor{greyD}     & DACC & 0.52~ & 0.47~ & 0.80~ & 0.74~ & 0.80~ & 0.93~ & 0.60~ & 0.94~ & 0.76~ & 0.92~ & 0.98~ & 1.00~ & 1.00~ \\
  \rowcolor{greyD}  \cellcolor{greyD}     & FPR & 0.00~ & 0.00~ & 0.00~ & 0.00~ & 0.00~ & 0.00~ & 0.00~ & 0.00~ & 0.00~ & 0.00~ & 0.00~ & 0.00~ & 0.00~ \\
    \rowcolor{greyD}  \cellcolor{greyD}\multirow{-3}{*}{Norm}     & FNR & 0.96~ & 0.98~ & 0.35~ & 0.51~ & 0.43~ & 0.18~ & 0.71~ & 0.14~ & 0.58~ & 0.16~ & 0.04~ & 0.00~ & 0.00~ \\ 
 
    \midrule
    \multirow{3}{*}{TrustRAG} & DACC & 0.49~ & 0.53~ & 0.20~ & 0.25~ & 0.21~ & 0.09~ & 0.40~ & 0.08~ & 0.25~ & 0.09~ & 0.02~ & 0.00~ & 0.00~ \\
     & FPR & 0.94~ & 0.93~ & 0.85~ & 0.86~ & 0.98~ & 0.95~ & 0.78~ & 0.95~ & 0.92~ & 0.95~ & 0.99~ & 1.00~ & 1.00~ \\
     & FNR & 0.03~ & 0.01~ & 0.08~ & 0.09~ & 0.02~ & 0.03~ & 0.12~ & 0.01~ & 0.10~ & 0.02~ & 0.02~ & 0.00~ & 0.00~ \\
    \bottomrule
    \end{tabular}
\end{table*}

\begin{table*}[!htbp]
\tiny
\centering
\addtolength{\tabcolsep}{-4.0pt}
\caption{The results of detection-based defenses against all poisoning attacks on HotpotQA-EX-M dataset.}
\label{tab:detection_hotpotqa_ex_m}
\begin{tabular}{l|c|ccccccc|ccc|ccc} 
    \toprule
    \multirow{2}{*}{Defense} & \multirow{2}{*}{Metric} & \multicolumn{7}{c|}{Targeted poisoning attack} & \multicolumn{3}{c|}{DoS attack} & \multicolumn{3}{c}{Trigger-based DoS attack} \\ 
    \cmidrule{3-15}
     &  & BPRAG & WPRAG & BPI & WPI & AGGD & CRAG-AS & CRAG-AK & JamInject & JamOracle & JamOpt & AP & BadRAG & Phantom \\ 
    \midrule
    \multirow{3}{*}{PPL} & DACC & 0.50~ & 0.54~ & 0.92~ & 0.91~ & 0.85~ & 0.94~ & 0.64~ & 0.98~ & 0.60~ & 1.00~ & 1.00~ & 1.00~ & 1.00~ \\
     & FPR & 0.00~ & 0.00~ & 0.00~ & 0.00~ & 0.00~ & 0.00~ & 0.00~ & 0.00~ & 0.00~ & 0.00~ & 0.00~ & 0.00~ & 0.00~ \\
     & FNR & 0.98~ & 0.93~ & 0.20~ & 0.21~ & 0.32~ & 0.14~ & 0.70~ & 0.06~ & 0.80~ & 0.01~ & 0.00~ & 0.00~ & 0.00~ \\ 
    \midrule
      \rowcolor{greyD}  \cellcolor{greyD}  & DACC & 0.50~ & 0.54~ & 0.92~ & 0.91~ & 0.85~ & 0.94~ & 0.64~ & 0.98~ & 0.60~ & 1.00~ & 1.00~ & 1.00~ & 1.00~ \\
    \rowcolor{greyD}  \cellcolor{greyD}   & FPR & 0.00~ & 0.00~ & 0.00~ & 0.00~ & 0.00~ & 0.00~ & 0.00~ & 0.00~ & 0.00~ & 0.00~ & 0.00~ & 0.00~ & 0.00~ \\
   \rowcolor{greyD}  \cellcolor{greyD}\multirow{-3}{*}{Norm}     & FNR & 0.98~ & 0.93~ & 0.20~ & 0.21~ & 0.32~ & 0.14~ & 0.70~ & 0.06~ & 0.80~ & 0.01~ & 0.00~ & 0.00~ & 0.00~ \\ 
    \midrule
    \multirow{3}{*}{TrustRAG} & DACC & 0.50~ & 0.46~ & 0.08~ & 0.08~ & 0.15~ & 0.06~ & 0.35~ & 0.02~ & 0.40~ & 0.01~ & 0.00~ & 0.00~ & 0.00~ \\
     & FPR & 0.95~ & 0.95~ & 0.95~ & 0.96~ & 1.00~ & 0.97~ & 0.88~ & 1.00~ & 0.86~ & 1.00~ & 1.00~ & 1.00~ & 1.00~ \\
     & FNR & 0.01~ & 0.01~ & 0.06~ & 0.05~ & 0.00~ & 0.04~ & 0.07~ & 0.01~ & 0.07~ & 0.00~ & 0.00~ & 0.00~ & 0.00~ \\
    \bottomrule
    \end{tabular}
\end{table*}

\begin{table*}[!htbp]
\tiny
\centering
\addtolength{\tabcolsep}{-4.0pt}
\caption{The results of detection-based defenses against all poisoning attacks on MS-MARCO-EX-M dataset.}
\label{tab:detection_msmarco_ex_m}
\begin{tabular}{l|c|ccccccc|ccc|ccc} 
    \toprule
    \multirow{2}{*}{Defense} & \multirow{2}{*}{Metric} & \multicolumn{7}{c|}{Targeted poisoning attack} & \multicolumn{3}{c|}{DoS attack} & \multicolumn{3}{c}{Trigger-based DoS attack} \\ 
    \cmidrule{3-15}
     &  & BPRAG & WPRAG & BPI & WPI & AGGD & CRAG-AS & CRAG-AK & JamInject & JamOracle & JamOpt & AP & BadRAG & Phantom \\ 
    \midrule
    \multirow{3}{*}{PPL} & DACC & 0.43~ & 0.49~ & 0.73~ & 0.75~ & 0.74~ & 0.97~ & 0.31~ & 0.99~ & 0.91~ & 0.99~ & 0.98~ & 1.00~ & 1.00~ \\
     & FPR & 0.00~ & 0.00~ & 0.00~ & 0.00~ & 0.00~ & 0.00~ & 0.00~ & 0.00~ & 0.00~ & 0.00~ & 0.03~ & 0.00~ & 0.00~ \\
     & FNR & 0.96~ & 0.92~ & 0.42~ & 0.44~ & 0.49~ & 0.06~ & 0.90~ & 0.04~ & 0.25~ & 0.02~ & 0.00~ & 0.00~ & 0.00~ \\ 
    \midrule
      \rowcolor{greyD}  \cellcolor{greyD} & DACC & 0.44~ & 0.49~ & 0.73~ & 0.68~ & 0.74~ & 0.95~ & 0.31~ & 0.99~ & 0.91~ & 0.98~ & 0.98~ & 1.00~ & 1.00~ \\
     \rowcolor{greyD}  \cellcolor{greyD}   & FPR & 0.00~ & 0.00~ & 0.00~ & 0.00~ & 0.00~ & 0.00~ & 0.00~ & 0.00~ & 0.00~ & 0.00~ & 0.00~ & 0.00~ & 0.00~ \\
  \rowcolor{greyD}  \cellcolor{greyD}\multirow{-3}{*}{Norm}       & FNR & 0.96~ & 0.93~ & 0.42~ & 0.58~ & 0.49~ & 0.11~ & 0.90~ & 0.04~ & 0.25~ & 0.05~ & 0.04~ & 0.00~ & 0.00~ \\ 
    \midrule
    \multirow{3}{*}{TrustRAG} & DACC & 0.58~ & 0.52~ & 0.29~ & 0.33~ & 0.28~ & 0.07~ & 0.70~ & 0.02~ & 0.10~ & 0.03~ & 0.01~ & 0.00~ & 0.00~ \\
     & FPR & 0.89~ & 0.90~ & 0.79~ & 0.82~ & 0.96~ & 0.95~ & 0.46~ & 0.99~ & 0.98~ & 0.98~ & 0.99~ & 1.00~ & 1.00~ \\
     & FNR & 0.01~ & 0.02~ & 0.05~ & 0.09~ & 0.03~ & 0.03~ & 0.05~ & 0.03~ & 0.03~ & 0.03~ & 0.03~ & 0.00~ & 0.00~ \\
    \bottomrule
    \end{tabular}
\end{table*}

\begin{table*}[!htbp]
\tiny
\centering
\addtolength{\tabcolsep}{-4.0pt}
\caption{The results of detection-based defenses against all poisoning attacks on BoolQ-EX-M dataset.}
\label{tab:detection_boolq_ex_m}
\begin{tabular}{l|c|ccccccc|ccc|ccc} 
    \toprule
    \multirow{2}{*}{Defense} & \multirow{2}{*}{Metric} & \multicolumn{7}{c|}{Targeted poisoning attack} & \multicolumn{3}{c|}{DoS attack} & \multicolumn{3}{c}{Trigger-based DoS attack} \\ 
    \cmidrule{3-15}
     &  & BPRAG & WPRAG & BPI & WPI & AGGD & CRAG-AS & CRAG-AK & JamInject & JamOracle & JamOpt & AP & BadRAG & Phantom \\ 
    \midrule
    \multirow{3}{*}{PPL} & DACC & 0.47~ & 0.45~ & 0.83~ & 0.82~ & 0.72~ & 0.96~ & 0.45~ & 0.98~ & 0.90~ & 0.96~ & 1.00~ & 1.00~ & 1.00~ \\
     & FPR & 0.00~ & 0.00~ & 0.00~ & 0.00~ & 0.00~ & 0.00~ & 0.00~ & 0.00~ & 0.00~ & 0.00~ & 0.00~ & 0.00~ & 0.00~ \\
     & FNR & 0.99~ & 0.98~ & 0.33~ & 0.42~ & 0.57~ & 0.08~ & 0.84~ & 0.04~ & 0.26~ & 0.06~ & 0.00~ & 0.00~ & 0.00~ \\ 
    \midrule
      \rowcolor{greyD}  \cellcolor{greyD}   & DACC & 0.47~ & 0.45~ & 0.83~ & 0.82~ & 0.72~ & 0.96~ & 0.45~ & 0.98~ & 0.90~ & 0.94~ & 0.99~ & 1.00~ & 1.00~ \\
    \rowcolor{greyD}  \cellcolor{greyD}   & FPR & 0.00~ & 0.00~ & 0.00~ & 0.00~ & 0.00~ & 0.00~ & 0.00~ & 0.00~ & 0.00~ & 0.00~ & 0.00~ & 0.00~ & 0.00~ \\
  \rowcolor{greyD}  \cellcolor{greyD}\multirow{-3}{*}{Norm}     & FNR & 0.99~ & 0.98~ & 0.33~ & 0.42~ & 0.57~ & 0.08~ & 0.84~ & 0.04~ & 0.26~ & 0.09~ & 0.01~ & 0.00~ & 0.00~ \\ 
    \midrule
    \multirow{3}{*}{TrustRAG} & DACC & 0.54~ & 0.55~ & 0.18~ & 0.18~ & 0.28~ & 0.05~ & 0.57~ & 0.03~ & 0.10~ & 0.06~ & 0.01~ & 0.00~ & 0.00~ \\
     & FPR & 0.93~ & 0.91~ & 0.88~ & 0.93~ & 0.95~ & 0.96~ & 0.64~ & 0.98~ & 0.99~ & 0.96~ & 0.99~ & 1.00~ & 1.00~ \\
     & FNR & 0.02~ & 0.02~ & 0.03~ & 0.06~ & 0.02~ & 0.04~ & 0.04~ & 0.00~ & 0.02~ & 0.00~ & 0.00~ & 0.00~ & 0.00~ \\
    \bottomrule
    \end{tabular}
\end{table*}

\begin{table*}[!htbp]
\tiny
\centering
\addtolength{\tabcolsep}{-4.0pt}
\caption{The results of detection-based defenses against all poisoning attacks on SQuAD-EX-M dataset.}
\label{tab:detection_squad_ex_m}
\begin{tabular}{l|c|ccccccc|ccc|ccc} 
    \toprule
    \multirow{2}{*}{Defense} & \multirow{2}{*}{Metric} & \multicolumn{7}{c|}{Targeted poisoning attack} & \multicolumn{3}{c|}{DoS attack} & \multicolumn{3}{c}{Trigger-based DoS attack} \\ 
    \cmidrule{3-15}
     &  & BPRAG & WPRAG & BPI & WPI & AGGD & CRAG-AS & CRAG-AK & JamInject & JamOracle & JamOpt & AP & BadRAG & Phantom \\ 
    \midrule
    \multirow{3}{*}{PPL} & DACC & 0.56~ & 0.59~ & 0.75~ & 0.76~ & 0.75~ & 0.91~ & 0.49~ & 0.96~ & 0.54~ & 1.00~ & 1.00~ & 0.99~ & 0.99~ \\
     & FPR & 0.00~ & 0.00~ & 0.00~ & 0.00~ & 0.00~ & 0.00~ & 0.00~ & 0.00~ & 0.00~ & 0.00~ & 0.00~ & 0.01~ & 0.01~ \\
     & FNR & 0.93~ & 0.90~ & 0.44~ & 0.43~ & 0.65~ & 0.22~ & 0.78~ & 0.08~ & 0.76~ & 0.01~ & 0.00~ & 0.00~ & 0.00~ \\ 
    \midrule
      \rowcolor{greyD}  \cellcolor{greyD}   & DACC & 0.56~ & 0.59~ & 0.75~ & 0.76~ & 0.75~ & 0.91~ & 0.49~ & 0.96~ & 0.54~ & 1.00~ & 0.99~ & 1.00~ & 1.00~ \\
   \rowcolor{greyD}  \cellcolor{greyD}   & FPR & 0.00~ & 0.00~ & 0.00~ & 0.00~ & 0.00~ & 0.00~ & 0.00~ & 0.00~ & 0.00~ & 0.00~ & 0.00~ & 0.00~ & 0.00~ \\
    \rowcolor{greyD}  \cellcolor{greyD}\multirow{-3}{*}{Norm}   & FNR & 0.93~ & 0.90~ & 0.44~ & 0.43~ & 0.65~ & 0.22~ & 0.78~ & 0.08~ & 0.76~ & 0.01~ & 0.01~ & 0.00~ & 0.00~ \\ 
    \midrule
    \multirow{3}{*}{TrustRAG} & DACC & 0.44~ & 0.41~ & 0.26~ & 0.24~ & 0.25~ & 0.08~ & 0.53~ & 0.04~ & 0.47~ & 0.00~ & 0.02~ & 0.00~ & 0.00~ \\
     & FPR & 0.97~ & 0.98~ & 0.84~ & 0.86~ & 0.98~ & 0.95~ & 0.66~ & 0.97~ & 0.81~ & 1.00~ & 0.98~ & 1.00~ & 1.00~ \\
     & FNR & 0.04~ & 0.03~ & 0.05~ & 0.05~ & 0.00~ & 0.09~ & 0.05~ & 0.01~ & 0.03~ & 0.00~ & 0.00~ & 0.00~ & 0.00~ \\
    \bottomrule
    \end{tabular}
\end{table*}

\begin{table*}[!htbp]
\tiny
\centering
\addtolength{\tabcolsep}{-4.0pt}
\caption{The results of detection-based defenses against all poisoning attacks on NQ-EX-L dataset.}
\label{tab:detection_nq_ex_l}
\begin{tabular}{l|c|ccccccc|ccc|ccc} 
    \toprule
    \multirow{2}{*}{Defense} & \multirow{2}{*}{Metric} & \multicolumn{7}{c|}{Targeted poisoning attack} & \multicolumn{3}{c|}{DoS attack} & \multicolumn{3}{c}{Trigger-based DoS attack} \\ 
    \cmidrule{3-15}
     &  & BPRAG & WPRAG & BPI & WPI & AGGD & CRAG-AS & CRAG-AK & JamInject & JamOracle & JamOpt & AP & BadRAG & Phantom \\ 
    \midrule
    \multirow{3}{*}{PPL} & DACC & 0.81~ & 0.76~ & 0.92~ & 0.89~ & 0.93~ & 1.00~ & 0.83~ & 0.98~ & 0.87~ & 0.98~ & 1.00~ & 1.00~ & 1.00~ \\
     & FPR & 0.00~ & 0.00~ & 0.00~ & 0.00~ & 0.00~ & 0.00~ & 0.00~ & 0.00~ & 0.00~ & 0.00~ & 0.00~ & 0.00~ & 0.00~ \\
     & FNR & 0.57~ & 0.63~ & 0.09~ & 0.17~ & 0.24~ & 0.00~ & 0.30~ & 0.02~ & 0.28~ & 0.03~ & 0.00~ & 0.00~ & 0.00~ \\ 
    \midrule
      \rowcolor{greyD}  \cellcolor{greyD}   & DACC & 0.81~ & 0.76~ & 0.92~ & 0.89~ & 0.93~ & 1.00~ & 0.83~ & 0.98~ & 0.87~ & 0.98~ & 1.00~ & 1.00~ & 1.00~ \\
    \rowcolor{greyD}  \cellcolor{greyD}  & FPR & 0.00~ & 0.00~ & 0.00~ & 0.00~ & 0.00~ & 0.00~ & 0.00~ & 0.00~ & 0.00~ & 0.00~ & 0.00~ & 0.00~ & 0.00~ \\
   \rowcolor{greyD}  \cellcolor{greyD}\multirow{-3}{*}{Norm}    & FNR & 0.57~ & 0.63~ & 0.09~ & 0.17~ & 0.24~ & 0.00~ & 0.30~ & 0.02~ & 0.28~ & 0.03~ & 0.00~ & 0.00~ & 0.00~ \\ 
    \midrule
    \multirow{3}{*}{TrustRAG} & DACC & 0.19~ & 0.24~ & 0.09~ & 0.11~ & 0.07~ & 0.01~ & 0.17~ & 0.02~ & 0.13~ & 0.03~ & 0.00~ & 0.00~ & 0.00~ \\
     & FPR & 0.99~ & 0.98~ & 0.92~ & 0.92~ & 1.00~ & 0.99~ & 0.91~ & 0.98~ & 0.97~ & 0.98~ & 1.00~ & 1.00~ & 1.00~ \\
     & FNR & 0.02~ & 0.02~ & 0.00~ & 0.04~ & 0.01~ & 0.00~ & 0.05~ & 0.00~ & 0.04~ & 0.00~ & 0.00~ & 0.00~ & 0.00~ \\
    \bottomrule
    \end{tabular}
\end{table*}

\begin{table*}[!htbp]
\tiny
\centering
\addtolength{\tabcolsep}{-4.0pt}
\caption{The results of detection-based defenses against all poisoning attacks on HotpotQA-EX-L dataset.}
\label{tab:detection_hotpotqa_ex_l}
\begin{tabular}{l|c|ccccccc|ccc|ccc} 
    \toprule
    \multirow{2}{*}{Defense} & \multirow{2}{*}{Metric} & \multicolumn{7}{c|}{Targeted poisoning attack} & \multicolumn{3}{c|}{DoS attack} & \multicolumn{3}{c}{Trigger-based DoS attack} \\ 
    \cmidrule{3-15}
     &  & BPRAG & WPRAG & BPI & WPI & AGGD & CRAG-AS & CRAG-AK & JamInject & JamOracle & JamOpt & AP & BadRAG & Phantom \\ 
    \midrule
    \multirow{3}{*}{PPL} & DACC & 0.82~ & 0.83~ & 0.98~ & 0.98~ & 0.95~ & 1.00~ & 0.87~ & 1.00~ & 0.83~ & 1.00~ & 1.00~ & 1.00~ & 1.00~ \\
     & FPR & 0.00~ & 0.00~ & 0.00~ & 0.00~ & 0.00~ & 0.00~ & 0.00~ & 0.00~ & 0.00~ & 0.00~ & 0.00~ & 0.00~ & 0.00~ \\
     & FNR & 0.57~ & 0.51~ & 0.02~ & 0.02~ & 0.15~ & 0.00~ & 0.27~ & 0.00~ & 0.38~ & 0.00~ & 0.00~ & 0.00~ & 0.00~ \\ 
    \midrule
      \rowcolor{greyD}  \cellcolor{greyD}   & DACC & 0.82~ & 0.83~ & 0.98~ & 0.98~ & 0.95~ & 1.00~ & 0.87~ & 1.00~ & 0.83~ & 1.00~ & 1.00~ & 1.00~ & 1.00~ \\
  \rowcolor{greyD}  \cellcolor{greyD}     & FPR & 0.00~ & 0.00~ & 0.00~ & 0.00~ & 0.00~ & 0.00~ & 0.00~ & 0.00~ & 0.00~ & 0.00~ & 0.00~ & 0.00~ & 0.00~ \\
      \rowcolor{greyD}  \cellcolor{greyD}\multirow{-3}{*}{Norm} & FNR & 0.57~ & 0.51~ & 0.02~ & 0.02~ & 0.15~ & 0.00~ & 0.27~ & 0.00~ & 0.38~ & 0.00~ & 0.00~ & 0.00~ & 0.00~ \\ 
    \midrule
    \multirow{3}{*}{TrustRAG} & DACC & 0.18~ & 0.17~ & 0.02~ & 0.02~ & 0.05~ & 0.00~ & 0.13~ & 0.00~ & 0.17~ & 0.00~ & 0.00~ & 0.00~ & 0.00~ \\
     & FPR & 0.98~ & 0.98~ & 0.98~ & 0.98~ & 1.00~ & 1.00~ & 0.94~ & 1.00~ & 0.95~ & 1.00~ & 1.00~ & 1.00~ & 1.00~ \\
     & FNR & 0.00~ & 0.00~ & 0.00~ & 0.00~ & 0.00~ & 0.00~ & 0.03~ & 0.00~ & 0.04~ & 0.00~ & 0.00~ & 0.00~ & 0.00~ \\
    \bottomrule
    \end{tabular}
\end{table*}

\begin{table*}[!htbp]
\tiny
\centering
\addtolength{\tabcolsep}{-4.0pt}
\caption{The results of detection-based defenses against all poisoning attacks on MS-MARCO-EX-L dataset.}
\label{tab:detection_msmarco_ex_l}
\begin{tabular}{l|c|ccccccc|ccc|ccc} 
    \toprule
    \multirow{2}{*}{Defense} & \multirow{2}{*}{Metric} & \multicolumn{7}{c|}{Targeted poisoning attack} & \multicolumn{3}{c|}{DoS attack} & \multicolumn{3}{c}{Trigger-based DoS attack} \\ 
    \cmidrule{3-15}
     &  & BPRAG & WPRAG & BPI & WPI & AGGD & CRAG-AS & CRAG-AK & JamInject & JamOracle & JamOpt & AP & BadRAG & Phantom \\ 
    \midrule
    \multirow{3}{*}{PPL} & DACC & 0.73~ & 0.72~ & 0.92~ & 0.91~ & 0.88~ & 1.00~ & 0.52~ & 1.00~ & 0.98~ & 1.00~ & 0.98~ & 1.00~ & 1.00~ \\
     & FPR & 0.00~ & 0.00~ & 0.00~ & 0.00~ & 0.00~ & 0.00~ & 0.00~ & 0.00~ & 0.00~ & 0.00~ & 0.02~ & 0.00~ & 0.00~ \\
     & FNR & 0.74~ & 0.75~ & 0.11~ & 0.17~ & 0.36~ & 0.00~ & 0.64~ & 0.00~ & 0.05~ & 0.00~ & 0.00~ & 0.00~ & 0.00~ \\ 
    \midrule
      \rowcolor{greyD}  \cellcolor{greyD}   & DACC & 0.73~ & 0.72~ & 0.92~ & 0.88~ & 0.88~ & 1.00~ & 0.52~ & 1.00~ & 0.98~ & 1.00~ & 1.00~ & 1.00~ & 1.00~ \\
   \rowcolor{greyD}  \cellcolor{greyD}    & FPR & 0.00~ & 0.00~ & 0.00~ & 0.00~ & 0.00~ & 0.00~ & 0.00~ & 0.00~ & 0.00~ & 0.00~ & 0.00~ & 0.00~ & 0.00~ \\
 \rowcolor{greyD}  \cellcolor{greyD}\multirow{-3}{*}{Norm}      & FNR & 0.74~ & 0.76~ & 0.11~ & 0.28~ & 0.36~ & 0.01~ & 0.64~ & 0.00~ & 0.05~ & 0.00~ & 0.00~ & 0.00~ & 0.00~ \\ 
    \midrule
    \multirow{3}{*}{TrustRAG} & DACC & 0.27~ & 0.28~ & 0.10~ & 0.12~ & 0.12~ & 0.01~ & 0.50~ & 0.01~ & 0.02~ & 0.01~ & 0.00~ & 0.00~ & 0.00~ \\
     & FPR & 0.99~ & 0.97~ & 0.92~ & 0.94~ & 0.99~ & 0.99~ & 0.62~ & 0.99~ & 0.99~ & 0.99~ & 1.00~ & 1.00~ & 1.00~ \\
     & FNR & 0.03~ & 0.04~ & 0.00~ & 0.03~ & 0.04~ & 0.00~ & 0.03~ & 0.00~ & 0.00~ & 0.00~ & 0.00~ & 0.00~ & 0.00~ \\
    \bottomrule
    \end{tabular}
\end{table*}

\begin{table*}[!htbp]
\tiny
\centering
\addtolength{\tabcolsep}{-4.0pt}
\caption{The results of detection-based defenses against all poisoning attacks on BoolQ-EX-L dataset.}
\label{tab:detection_boolq_ex_l}
\begin{tabular}{l|c|ccccccc|ccc|ccc} 
    \toprule
    \multirow{2}{*}{Defense} & \multirow{2}{*}{Metric} & \multicolumn{7}{c|}{Targeted poisoning attack} & \multicolumn{3}{c|}{DoS attack} & \multicolumn{3}{c}{Trigger-based DoS attack} \\ 
    \cmidrule{3-15}
     &  & BPRAG & WPRAG & BPI & WPI & AGGD & CRAG-AS & CRAG-AK & JamInject & JamOracle & JamOpt & AP & BadRAG & Phantom \\ 
    \midrule
    \multirow{3}{*}{PPL} & DACC & 0.77~ & 0.72~ & 0.93~ & 0.94~ & 0.88~ & 0.98~ & 0.75~ & 1.00~ & 0.96~ & 0.98~ & 1.00~ & 1.00~ & 1.00~ \\
     & FPR & 0.00~ & 0.00~ & 0.00~ & 0.00~ & 0.00~ & 0.00~ & 0.00~ & 0.00~ & 0.00~ & 0.00~ & 0.00~ & 0.00~ & 0.00~ \\
     & FNR & 0.63~ & 0.69~ & 0.08~ & 0.13~ & 0.32~ & 0.02~ & 0.48~ & 0.00~ & 0.10~ & 0.03~ & 0.00~ & 0.00~ & 0.00~ \\ 
    \midrule
      \rowcolor{greyD}  \cellcolor{greyD}   & DACC & 0.77~ & 0.72~ & 0.93~ & 0.94~ & 0.88~ & 0.98~ & 0.75~ & 1.00~ & 0.96~ & 0.97~ & 0.99~ & 1.00~ & 1.00~ \\
    \rowcolor{greyD}  \cellcolor{greyD}    & FPR & 0.00~ & 0.00~ & 0.00~ & 0.00~ & 0.00~ & 0.00~ & 0.00~ & 0.00~ & 0.00~ & 0.00~ & 0.00~ & 0.00~ & 0.00~ \\
  \rowcolor{greyD}  \cellcolor{greyD}\multirow{-3}{*}{Norm}     & FNR & 0.63~ & 0.69~ & 0.08~ & 0.13~ & 0.32~ & 0.02~ & 0.48~ & 0.00~ & 0.10~ & 0.04~ & 0.01~ & 0.00~ & 0.00~ \\ 
    \midrule
    \multirow{3}{*}{TrustRAG} & DACC & 0.23~ & 0.27~ & 0.07~ & 0.05~ & 0.12~ & 0.02~ & 0.25~ & 0.00~ & 0.05~ & 0.03~ & 0.01~ & 0.00~ & 0.00~ \\
     & FPR & 1.00~ & 1.00~ & 0.94~ & 0.98~ & 1.00~ & 0.98~ & 0.85~ & 1.00~ & 1.00~ & 0.98~ & 1.00~ & 1.00~ & 1.00~ \\
     & FNR & 0.03~ & 0.04~ & 0.00~ & 0.02~ & 0.01~ & 0.00~ & 0.05~ & 0.00~ & 0.00~ & 0.00~ & 0.00~ & 0.00~ & 0.00~ \\
    \bottomrule
    \end{tabular}
\end{table*}

\begin{table*}[!htbp]
\tiny
\centering
\addtolength{\tabcolsep}{-4.0pt}
\caption{The results of detection-based defenses against all poisoning attacks on SQuAD-EX-L dataset.}
\label{tab:detection_squad_ex_l}
\begin{tabular}{l|c|ccccccc|ccc|ccc} 
    \toprule
    \multirow{2}{*}{Defense} & \multirow{2}{*}{Metric} & \multicolumn{7}{c|}{Targeted poisoning attack} & \multicolumn{3}{c|}{DoS attack} & \multicolumn{3}{c}{Trigger-based DoS attack} \\ 
    \cmidrule{3-15}
     &  & BPRAG & WPRAG & BPI & WPI & AGGD & CRAG-AS & CRAG-AK & JamInject & JamOracle & JamOpt & AP & BadRAG & Phantom \\ 
    \midrule
    \multirow{3}{*}{PPL} & DACC & 0.88~ & 0.88~ & 0.94~ & 0.95~ & 0.95~ & 0.99~ & 0.73~ & 1.00~ & 0.77~ & 1.00~ & 1.00~ & 1.00~ & 1.00~ \\
     & FPR & 0.00~ & 0.00~ & 0.00~ & 0.00~ & 0.00~ & 0.00~ & 0.00~ & 0.00~ & 0.00~ & 0.00~ & 0.00~ & 0.00~ & 0.00~ \\
     & FNR & 0.43~ & 0.41~ & 0.10~ & 0.08~ & 0.18~ & 0.04~ & 0.44~ & 0.00~ & 0.48~ & 0.00~ & 0.00~ & 0.00~ & 0.00~ \\ 
    \midrule
      \rowcolor{greyD}  \cellcolor{greyD}  & DACC & 0.88~ & 0.89~ & 0.94~ & 0.95~ & 0.95~ & 0.99~ & 0.73~ & 1.00~ & 0.77~ & 1.00~ & 1.00~ & 1.00~ & 1.00~ \\
  \rowcolor{greyD}  \cellcolor{greyD}    & FPR & 0.00~ & 0.00~ & 0.00~ & 0.00~ & 0.00~ & 0.00~ & 0.00~ & 0.00~ & 0.00~ & 0.00~ & 0.00~ & 0.00~ & 0.00~ \\
    \rowcolor{greyD}  \cellcolor{greyD}\multirow{-3}{*}{Norm}    & FNR & 0.43~ & 0.41~ & 0.10~ & 0.08~ & 0.18~ & 0.04~ & 0.44~ & 0.00~ & 0.48~ & 0.00~ & 0.00~ & 0.00~ & 0.00~ \\ 
    \midrule
    \multirow{3}{*}{TrustRAG} & DACC & 0.12~ & 0.11~ & 0.05~ & 0.05~ & 0.05~ & 0.01~ & 0.28~ & 0.00~ & 0.23~ & 0.00~ & 0.00~ & 0.00~ & 0.00~ \\
     & FPR & 1.00~ & 1.00~ & 0.96~ & 0.96~ & 1.00~ & 1.00~ & 0.84~ & 1.00~ & 0.92~ & 1.00~ & 1.00~ & 1.00~ & 1.00~ \\
     & FNR & 0.02~ & 0.02~ & 0.03~ & 0.02~ & 0.01~ & 0.01~ & 0.02~ & 0.00~ & 0.05~ & 0.00~ & 0.00~ & 0.00~ & 0.00~ \\
    \bottomrule
    \end{tabular}
\end{table*}

\begin{table}[!htbp]
\tiny
\centering
\addtolength{\tabcolsep}{-1.5pt}
\caption{The results of all poisoning attacks under different retrievers on NQ, NQ-EX-M, and NQ-EX-L datasets.}
\label{tab:impact_of_retrievers_nq}
\begin{tabular}{l|c|ccc|ccc|ccc}
\toprule
\multirow{2}{*}{Attack} & \multirow{2}{*}{Metric} & \multicolumn{3}{c|}{NQ} & \multicolumn{3}{c|}{NQ-EX-M} & \multicolumn{3}{c}{NQ-EX-L} \\ \cmidrule{3-11} 
 &  & Contriever & Contriever-MS & ANCE & Contriever & Contriever-MS & ANCE & Contriever & Contriever-MS & ANCE \\ \midrule
\multirow{3}{*}{BPRAG} & ACC & 0.27 & 0.22 & 0.22 & 0.65 & 0.59 & 0.51 & 0.89 & 0.76 & 0.81 \\
 & ASR & 0.63 & 0.66 & 0.67 & 0.12 & 0.18 & 0.17 & 0.03 & 0.08 & 0.05 \\
 & F1-score & 0.96 & 1.00 & 1.00 & 0.48 & 0.50 & 0.53 & 0.19 & 0.21 & 0.20 \\ \midrule
   \rowcolor{greyD}  \cellcolor{greyD} & ACC & 0.24 & 0.23 & 0.23 & 0.62 & 0.65 & 0.56 & 0.84 & 0.80 & 0.68 \\
   \rowcolor{greyD}  \cellcolor{greyD}  & ASR & 0.64 & 0.62 & 0.63 & 0.17 & 0.17 & 0.27 & 0.04 & 0.05 & 0.12 \\
   \rowcolor{greyD}  \cellcolor{greyD}\multirow{-3}{*}{WPRAG} & F1-score & 0.96 & 0.99 & 1.00 & 0.53 & 0.47 & 0.64 & 0.24 & 0.19 & 0.39 \\ \midrule
\multirow{3}{*}{BPI} & ACC & 0.03 & 0.00 & 0.00 & 0.79 & 0.45 & 0.37 & 0.89 & 0.67 & 0.66 \\
 & ASR & 0.94 & 1.00 & 1.00 & 0.18 & 0.50 & 0.49 & 0.08 & 0.27 & 0.28 \\
 & F1-score & 0.91 & 1.00 & 1.00 & 0.20 & 0.50 & 0.56 & 0.08 & 0.28 & 0.30 \\ \midrule
   \rowcolor{greyD}  \cellcolor{greyD}  & ACC & 0.01 & 0.01 & 0.00 & 0.77 & 0.59 & 0.60 & 0.87 & 0.74 & 0.79 \\
    \rowcolor{greyD}  \cellcolor{greyD} & ASR & 0.98 & 0.98 & 1.00 & 0.15 & 0.36 & 0.31 & 0.09 & 0.23 & 0.17 \\
   \rowcolor{greyD}  \cellcolor{greyD}\multirow{-3}{*}{WPI} & F1-score & 0.93 & 0.98 & 0.99 & 0.26 & 0.40 & 0.42 & 0.11 & 0.23 & 0.22 \\ \midrule
\multirow{3}{*}{AGGD} & ACC & 0.33 & 0.36 & 0.28 & 0.88 & 0.87 & 0.84 & 0.96 & 0.91 & 0.91 \\
 & ASR & 0.52 & 0.52 & 0.59 & 0.04 & 0.06 & 0.04 & 0.00 & 0.02 & 0.02 \\
 & F1-score & 0.78 & 0.79 & 0.89 & 0.20 & 0.18 & 0.24 & 0.07 & 0.08 & 0.08 \\ \midrule
   \rowcolor{greyD}  \cellcolor{greyD} & ACC & 0.06 & 0.00 & 0.00 & 0.81 & 0.28 & 0.83 & 0.97 & 0.64 & 0.93 \\
    \rowcolor{greyD}  \cellcolor{greyD} & ASR & 0.90 & 0.99 & 0.99 & 0.15 & 0.66 & 0.13 & 0.00 & 0.33 & 0.03 \\
\rowcolor{greyD}  \cellcolor{greyD}\multirow{-3}{*}{CRAG-AS}  & F1-score & 0.86 & 1.00 & 0.97 & 0.07 & 0.54 & 0.08 & 0.00 & 0.27 & 0.02 \\ \midrule
\multirow{3}{*}{CRAG-AK} & ACC & 0.05 & 0.04 & 0.06 & 0.36 & 0.15 & 0.27 & 0.74 & 0.46 & 0.57 \\
 & ASR & 0.89 & 0.89 & 0.92 & 0.46 & 0.74 & 0.56 & 0.20 & 0.46 & 0.32 \\
 & F1-score & 0.95 & 1.00 & 1.00 & 0.40 & 0.71 & 0.51 & 0.17 & 0.47 & 0.24 \\ \midrule
   \rowcolor{greyD}  \cellcolor{greyD}  & ACC & 0.15 & 0.12 & 0.15 & 0.97 & 0.96 & 0.98 & 0.98 & 0.98 & 1.00 \\
   \rowcolor{greyD}  \cellcolor{greyD}  & ASR & 0.85 & 0.88 & 0.85 & 0.03 & 0.04 & 0.02 & 0.02 & 0.02 & 0.00 \\
\rowcolor{greyD}  \cellcolor{greyD}\multirow{-3}{*}{JamInject} & F1-score & 0.75 & 0.88 & 0.80 & 0.06 & 0.06 & 0.03 & 0.02 & 0.03 & 0.01 \\ \midrule
\multirow{3}{*}{JamOracle} & ACC & 0.13 & 0.04 & 0.14 & 0.94 & 0.78 & 0.98 & 0.94 & 0.84 & 1.00 \\
   & ASR & 0.87 & 0.96 & 0.86 & 0.05 & 0.22 & 0.02 & 0.04 & 0.15 & 0.00 \\
 & F1-score & 0.83 & 0.95 & 0.90 & 0.25 & 0.45 & 0.23 & 0.14 & 0.27 & 0.08 \\ \midrule
  \rowcolor{greyD}  \cellcolor{greyD} & ACC & 0.29 & 0.19 & 0.24 & 0.98 & 0.83 & 0.77 & 0.98 & 0.86 & 0.84 \\
   \rowcolor{greyD}  \cellcolor{greyD} & ASR & 0.59 & 0.67 & 0.64 & 0.01 & 0.13 & 0.17 & 0.01 & 0.11 & 0.10 \\
   \rowcolor{greyD}  \cellcolor{greyD}\multirow{-3}{*}{JamOpt} & F1-score & 0.76 & 0.81 & 0.80 & 0.08 & 0.26 & 0.34 & 0.02 & 0.18 & 0.19 \\ \midrule
\multirow{3}{*}{AP} & ACC & 0.01 & 0.08 & 0.24 & 0.94 & 0.94 & 0.95 & 0.95 & 0.95 & 0.94 \\
 & ASR & 0.99 & 0.90 & 0.69 & 0.05 & 0.05 & 0.04 & 0.03 & 0.05 & 0.05 \\
 & F1-score & 1.00 & 0.89 & 0.45 & 0.09 & 0.00 & 0.00 & 0.04 & 0.00 & 0.00 \\ \midrule
  \rowcolor{greyD}  \cellcolor{greyD} & ACC & 0.65 & 0.83 & 0.62 & 0.99 & 0.99 & 0.99 & 0.99 & 0.99 & 0.99 \\
   \rowcolor{greyD}  \cellcolor{greyD} & ASR & 0.35 & 0.16 & 0.35 & 0.01 & 0.01 & 0.01 & 0.01 & 0.01 & 0.01 \\
   \rowcolor{greyD}  \cellcolor{greyD}\multirow{-3}{*}{BadRAG}  & F1-score & 0.37 & 0.01 & 0.00 & 0.00 & 0.00 & 0.00 & 0.00 & 0.00 & 0.00 \\ \midrule
\multirow{3}{*}{Phantom} & ACC & 0.99 & 0.83 & 0.62 & 1.00 & 0.99 & 0.99 & 1.00 & 0.99 & 0.99 \\
 & ASR & 0.00 & 0.15 & 0.35 & 0.00 & 0.00 & 0.00 & 0.00 & 0.00 & 0.00 \\
 & F1-score & 0.00 & 0.00 & 0.00 & 0.00 & 0.00 & 0.00 & 0.00 & 0.00 & 0.00 \\
 \bottomrule
\end{tabular}
\end{table}

\begin{table}[!htbp]
\tiny
\centering
\addtolength{\tabcolsep}{2.0pt}
\caption{The results of all poisoning attacks under different similarity measurements on NQ, NQ-EX-M, and NQ-EX-L datasets.}
\label{tab:impact_of_similarity_nq}
\begin{tabular}{l|c|cc|cc|cc}
\toprule
\multirow{2}{*}{Attack} & \multirow{2}{*}{Metric} & \multicolumn{2}{c|}{NQ} & \multicolumn{2}{c|}{NQ-EX-M} & \multicolumn{2}{c}{NQ-EX-L} \\ \cmidrule{3-8} 
 &  & Cosine similarity & Dot product & Cosine similarity & Dot product & Cosine similarity & Dot product \\ \midrule
& ACC & 0.27 & 0.26 & 0.61 & 0.60 & 0.88 & 0.86 \\
& ASR & 0.61 & 0.63 & 0.13 & 0.18 & 0.03 & 0.01 \\
 \multirow{-3}{*}{BPRAG}  & F1-score & 0.96 & 0.94 & 0.48 & 0.47 & 0.19 & 0.16 \\ \midrule
   \rowcolor{greyD}  \cellcolor{greyD} 
  & ACC & 0.24 & 0.27 & 0.63 & 0.52 & 0.88 & 0.75 \\
   \rowcolor{greyD}  \cellcolor{greyD}  & ASR & 0.63 & 0.63 & 0.17 & 0.25 & 0.04 & 0.08 \\
    \rowcolor{greyD}  \cellcolor{greyD}\multirow{-3}{*}{WPRAG}& F1-score & 0.96 & 0.98 & 0.53 & 0.57 & 0.24 & 0.31 \\ \midrule
\multirow{3}{*}{BPI} & ACC & 0.03 & 0.05 & 0.75 & 0.69 & 0.88 & 0.82 \\
 & ASR & 0.94 & 0.91 & 0.16 & 0.26 & 0.08 & 0.14 \\
 & F1-score & 0.91 & 0.83 & 0.20 & 0.28 & 0.08 & 0.15 \\ \midrule
 \rowcolor{greyD}  \cellcolor{greyD} & ACC & 0.01 & 0.01 & 0.77 & 0.55 & 0.87 & 0.68 \\
  \rowcolor{greyD}  \cellcolor{greyD} & ASR & 0.98 & 0.98 & 0.16 & 0.38 & 0.09 & 0.28 \\
 \rowcolor{greyD}  \cellcolor{greyD}\multirow{-3}{*}{WPI}  & F1-score & 0.93 & 0.97 & 0.26 & 0.46 & 0.11 & 0.31 \\ \midrule
\multirow{3}{*}{AGGD} & ACC & 0.33 & 0.30 & 0.84 & 0.72 & 0.92 & 0.82 \\
 & ASR & 0.55 & 0.59 & 0.05 & 0.18 & 0.03 & 0.11 \\
 & F1-score & 0.78 & 0.90 & 0.20 & 0.40 & 0.06 & 0.24 \\ \midrule
 \rowcolor{greyD}  \cellcolor{greyD} & ACC & 0.06 & 0.04 & 0.81 & 0.47 & 0.97 & 0.76 \\
 \rowcolor{greyD}  \cellcolor{greyD}  & ASR & 0.90 & 0.93 & 0.13 & 0.46 & 0.00 & 0.21 \\
 \rowcolor{greyD}  \cellcolor{greyD}\multirow{-3}{*}{CRAG-AS}  & F1-score & 0.86 & 0.89 & 0.07 & 0.34 & 0.00 & 0.19 \\ \midrule
\multirow{3}{*}{CRAG-AK} & ACC & 0.05 & 0.04 & 0.38 & 0.35 & 0.72 & 0.60 \\
 & ASR & 0.89 & 0.87 & 0.47 & 0.54 & 0.21 & 0.35 \\
 & F1-score & 0.95 & 0.93 & 0.40 & 0.50 & 0.17 & 0.30 \\ \midrule
 \rowcolor{greyD}  \cellcolor{greyD} & ACC & 0.15 & 0.08 & 0.97 & 0.80 & 0.98 & 0.85 \\
  \rowcolor{greyD}  \cellcolor{greyD} & ASR & 0.85 & 0.92 & 0.03 & 0.20 & 0.02 & 0.15 \\
  \rowcolor{greyD}  \cellcolor{greyD}\multirow{-3}{*}{JamInject} & F1-score & 0.75 & 0.87 & 0.06 & 0.32 & 0.02 & 0.17 \\ \midrule
\multirow{3}{*}{JamOracle} & ACC & 0.13 & 0.06 & 0.94 & 0.84 & 0.94 & 0.89 \\
 & ASR & 0.87 & 0.94 & 0.05 & 0.16 & 0.04 & 0.11 \\
 & F1-score & 0.83 & 0.88 & 0.25 & 0.45 & 0.14 & 0.28 \\ \midrule
 \rowcolor{greyD}  \cellcolor{greyD} & ACC & 0.29 & 0.18 & 0.98 & 0.76 & 0.98 & 0.76 \\
  \rowcolor{greyD}  \cellcolor{greyD} & ASR & 0.59 & 0.68 & 0.01 & 0.18 & 0.01 & 0.18 \\
 \rowcolor{greyD}  \cellcolor{greyD}\multirow{-3}{*}{JamOpt}  & F1-score & 0.76 & 0.81 & 0.08 & 0.39 & 0.02 & 0.27 \\ \midrule
\multirow{3}{*}{AP} & ACC & 0.01 & 0.01 & 0.94 & 0.95 & 0.95 & 0.95 \\
 & ASR & 0.99 & 0.99 & 0.05 & 0.04 & 0.03 & 0.04 \\
 & F1-score & 1.00 & 1.00 & 0.09 & 0.00 & 0.04 & 0.00 \\ \midrule
 \rowcolor{greyD}  \cellcolor{greyD} & ACC & 0.65 & 0.19 & 0.99 & 0.99 & 0.99 & 1.00 \\
 \rowcolor{greyD}  \cellcolor{greyD} & ASR & 0.35 & 0.81 & 0.01 & 0.01 & 0.01 & 0.00 \\
\rowcolor{greyD}  \cellcolor{greyD}\multirow{-3}{*}{BadRAG}  & F1-score & 0.37 & 0.72 & 0.00 & 0.00 & 0.00 & 0.00 \\ \midrule
\multirow{3}{*}{Phantom} & ACC & 0.99 & 0.03 & 1.00 & 1.00 & 1.00 & 1.00 \\
 & ASR & 0.00 & 0.97 & 0.00 & 0.00 & 0.00 & 0.00 \\
 & F1-score & 0.00 & 0.95 & 0.00 & 0.00 & 0.00 & 0.00 \\ 
\bottomrule
\end{tabular}
\end{table}

\begin{figure*}[!t]  
    \centering  
    
    \begin{subfigure}{\textwidth}
        \centering
        \includegraphics[width=\textwidth]{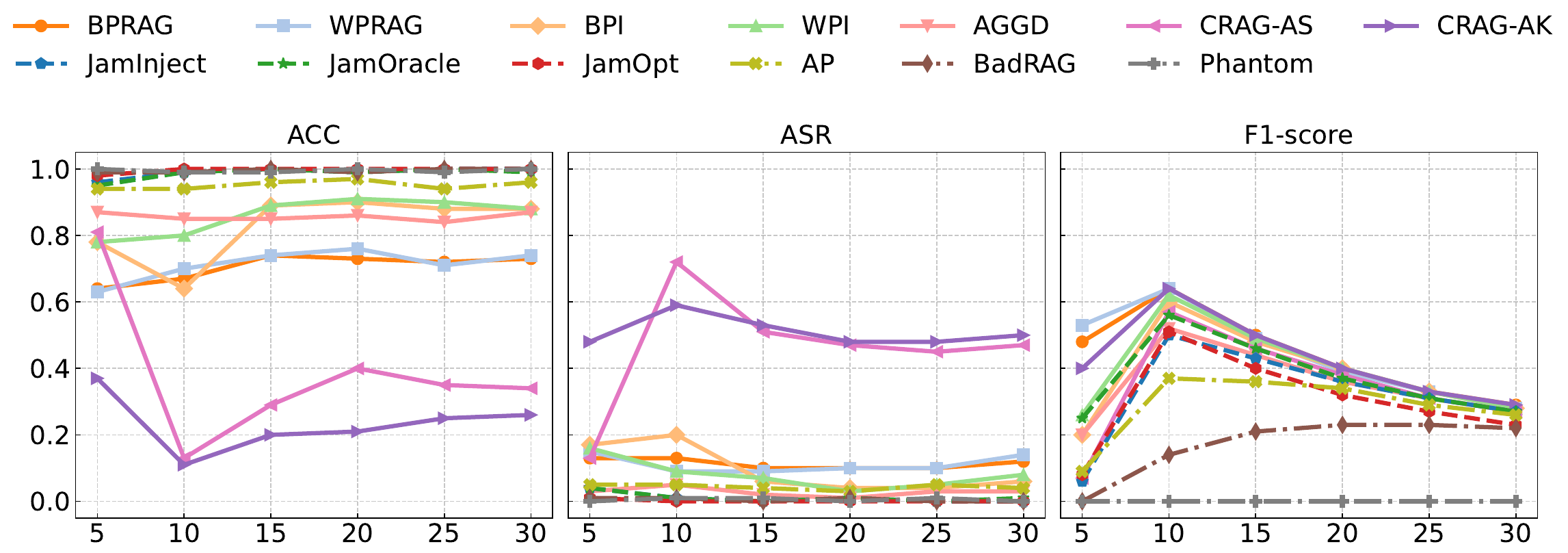}
        \caption{NQ-EX-M dataset}
        \label{fig:k_nq_ex_m}
    \end{subfigure}
    
    \vspace{1em}
    
    \begin{subfigure}{\textwidth}
        \centering
        \includegraphics[width=\textwidth]{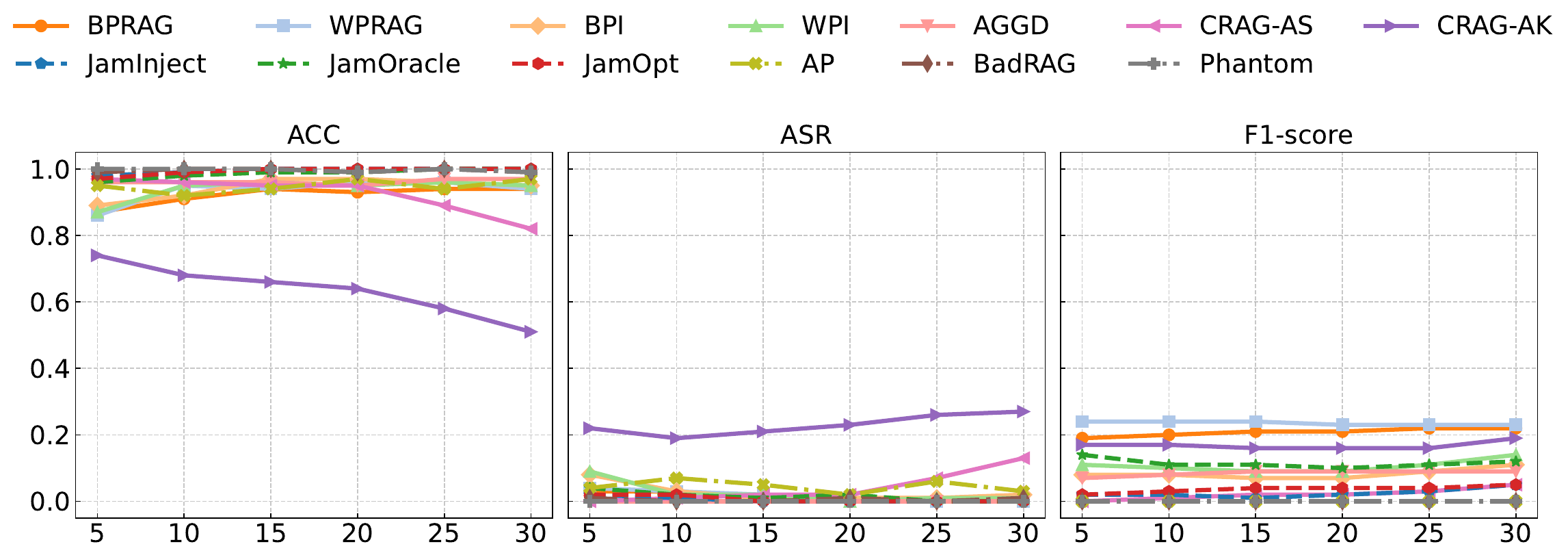}
        \caption{NQ-EX-L dataset}
        \label{fig:k_nq_ex_l}
    \end{subfigure}
    
    \caption{The results of poisoning attacks under different top-$K$ of RAG on NQ-EX-M, and NQ-EX-L datasets.}  
    \label{fig:k_nq_exs}  
\end{figure*}

\begin{table*}[!htbp]
\tiny
\centering
\addtolength{\tabcolsep}{3.0pt}
\caption{The results of all poisoning attacks against different advanced RAG methods on NQ dataset.}
\label{tab:framework_nq}
\begin{tabular}{l|c|ccccccc} 
\toprule
Attack & Metric & Naive RAG & AAR & SuRe & Adaptive-RAG & IRCoT & FLARE & RQRAG \\ 
\midrule
\multirow{2}{*}{BPRAG} & ACC & 0.27 & 0.23 & 0.34 & 0.27 & 0.27 & 0.65 & 0.26 \\
 & ASR & 0.62 & 0.66 & 0.48 & 0.55 & 0.60 & 0.11 & 0.70 \\ 
\midrule
 \rowcolor{greyD}  \cellcolor{greyD} & ACC & 0.25 & 0.23 & 0.33 & 0.24 & 0.24 & 0.65 & 0.23 \\
 \rowcolor{greyD}  \cellcolor{greyD}\multirow{-2}{*}{WPRAG}  & ASR & 0.64 & 0.66 & 0.48 & 0.60 & 0.64 & 0.11 & 0.66 \\ 
\midrule
\multirow{2}{*}{BPI} & ACC & 0.02 & 0.00 & 0.22 & 0.03 & 0.03 & 0.67 & 0.17 \\
 & ASR & 0.94 & 1.00 & 0.67 & 0.94 & 0.94 & 0.04 & 0.71 \\ 
\midrule
 \rowcolor{greyD}  \cellcolor{greyD} & ACC & 0.01 & 0.00 & 0.25 & 0.00 & 0.00 & 0.68 & 0.26 \\
 \rowcolor{greyD}  \cellcolor{greyD}\multirow{-2}{*}{WPI}  & ASR & 0.97 & 1.00 & 0.67 & 0.97 & 0.97 & 0.04 & 0.58 \\ 
\midrule
\multirow{2}{*}{AGGD} & ACC & 0.33 & 0.33 & 0.36 & 0.33 & 0.34 & 0.66 & 0.28 \\
 & ASR & 0.51 & 0.62 & 0.37 & 0.45 & 0.51 & 0.06 & 0.57 \\ 
\midrule
 \rowcolor{greyD}  \cellcolor{greyD} 
 & ACC & 0.06 & 0.00 & 0.43 & 0.06 & 0.05 & 0.65 & 0.18 \\
  \rowcolor{greyD}  \cellcolor{greyD}\multirow{-2}{*}{CRAG-AS}& ASR & 0.89 & 1.00 & 0.47 & 0.90 & 0.85 & 0.04 & 0.63 \\ 
\midrule
\multirow{2}{*}{CRAG-AK} & ACC & 0.04 & 0.04 & 0.30 & 0.04 & 0.06 & 0.64 & 0.14 \\
 & ASR & 0.88 & 0.89 & 0.63 & 0.83 & 0.87 & 0.12 & 0.73 \\ 
\midrule
 \rowcolor{greyD}  \cellcolor{greyD}  & ACC & 0.15 & 0.13 & 0.44 & 0.15 & 0.15 & 0.79 & 0.59 \\
\rowcolor{greyD}  \cellcolor{greyD}\multirow{-2}{*}{JamInject} & ASR & 0.85 & 0.87 & 0.47 & 0.85 & 0.85 & 0.02 & 0.02 \\ 
\midrule
\multirow{2}{*}{JamOracle} & ACC & 0.13 & 0.03 & 0.44 & 0.13 & 0.12 & 0.78 & 0.48 \\
 & ASR & 0.87 & 0.97 & 0.21 & 0.87 & 0.87 & 0.05 & 0.11 \\ 
\midrule
\rowcolor{greyD}  \cellcolor{greyD} & ACC & 0.29 & 0.21 & 0.69 & 0.26 & 0.26 & 0.78 & 0.58 \\
\rowcolor{greyD}  \cellcolor{greyD}\multirow{-2}{*}{JamOpt}  & ASR & 0.59 & 0.65 & 0.04 & 0.64 & 0.62 & 0.03 & 0.00 \\ 
\midrule
\multirow{2}{*}{AP} & ACC & 0.01 & 0.42 & 0.49 & 0.00 & 0.00 & 0.69 & 0.41 \\
 & ASR & 0.99 & 0.51 & 0.26 & 1.00 & 1.00 & 0.01 & 0.06 \\ 
\midrule
\rowcolor{greyD}  \cellcolor{greyD} 
 & ACC & 0.65 & 0.80 & 0.74 & 0.66 & 0.64 & 0.80 & 0.70 \\
\rowcolor{greyD}  \cellcolor{greyD}\multirow{-2}{*}{BadRAG}  & ASR & 0.35 & 0.18 & 0.07 & 0.34 & 0.34 & 0.03 & 0.04 \\ 
\midrule
\multirow{2}{*}{Phantom} & ACC & 0.99 & 0.77 & 0.74 & 0.96 & 0.91 & 0.71 & 0.77 \\
 & ASR & 0.00 & 0.20 & 0.00 & 0.04 & 0.02 & 0.01 & 0.00 \\
\bottomrule
\end{tabular}
\end{table*}

\begin{table*}[!htbp]
\tiny
\centering
\addtolength{\tabcolsep}{3.0pt}
\caption{The results of all poisoning attacks against different advanced RAG methods on NQ-EX-M dataset.}
\label{tab:framework_nq_ex_m}
\begin{tabular}{l|c|ccccccc} 
\toprule
Attack & Metric & Naive RAG & AAR & SuRe & Adaptive-RAG & IRCoT & FLARE & RQRAG \\ 
\midrule
\multirow{2}{*}{BPRAG} & ACC & 0.65 & 0.59 & 0.79 & 0.66 & 0.63 & 0.71 & 0.73 \\
 & ASR & 0.11 & 0.17 & 0.08 & 0.06 & 0.12 & 0.04 & 0.24 \\ 
\midrule
\rowcolor{greyD}  \cellcolor{greyD} 
  & ACC & 0.59 & 0.64 & 0.34 & 0.63 & 0.65 & 0.68 & 0.79 \\
\rowcolor{greyD}  \cellcolor{greyD}\multirow{-2}{*}{WPRAG} & ASR & 0.15 & 0.14 & 0.44 & 0.10 & 0.15 & 0.07 & 0.17 \\ 
\midrule
\multirow{2}{*}{BPI} & ACC & 0.77 & 0.47 & 0.76 & 0.75 & 0.77 & 0.73 & 0.85 \\
 & ASR & 0.16 & 0.43 & 0.11 & 0.19 & 0.17 & 0.01 & 0.11 \\ 
\midrule
\rowcolor{greyD}  \cellcolor{greyD} 
& ACC & 0.80 & 0.56 & 0.78 & 0.78 & 0.79 & 0.73 & 0.86 \\
\rowcolor{greyD}  \cellcolor{greyD}\multirow{-2}{*}{WPI}  & ASR & 0.14 & 0.35 & 0.09 & 0.15 & 0.14 & 0.00 & 0.07 \\ 
\midrule
\multirow{2}{*}{AGGD} & ACC & 0.87 & 0.87 & 0.80 & 0.87 & 0.85 & 0.72 & 0.85 \\
 & ASR & 0.04 & 0.06 & 0.03 & 0.02 & 0.03 & 0.01 & 0.10 \\ 
\midrule
\rowcolor{greyD}  \cellcolor{greyD}  
& ACC & 0.81 & 0.49 & 0.85 & 0.81 & 0.86 & 0.75 & 0.89 \\
\rowcolor{greyD}  \cellcolor{greyD}\multirow{-2}{*}{CRAG-AS}    & ASR & 0.14 & 0.44 & 0.01 & 0.13 & 0.09 & 0.00 & 0.05 \\ 
\midrule
\multirow{2}{*}{CRAG-AK} & ACC & 0.36 & 0.20 & 0.74 & 0.37 & 0.42 & 0.72 & 0.70 \\
 & ASR & 0.50 & 0.65 & 0.19 & 0.42 & 0.39 & 0.04 & 0.25 \\ 
\midrule
 \rowcolor{greyD}  \cellcolor{greyD}  & ACC & 0.97 & 0.98 & 0.87 & 0.97 & 0.96 & 0.83 & 0.96 \\
 \rowcolor{greyD}  \cellcolor{greyD}\multirow{-2}{*}{JamInject} & ASR & 0.03 & 0.02 & 0.03 & 0.03 & 0.03 & 0.02 & 0.00 \\ 
\midrule
\multirow{2}{*}{JamOracle} & ACC & 0.94 & 0.86 & 0.83 & 0.95 & 0.94 & 0.82 & 0.95 \\
 & ASR & 0.05 & 0.14 & 0.02 & 0.05 & 0.05 & 0.04 & 0.00 \\ 
\midrule
\rowcolor{greyD}  \cellcolor{greyD} & ACC & 0.98 & 0.82 & 0.85 & 0.98 & 0.97 & 0.81 & 0.96 \\
\rowcolor{greyD}  \cellcolor{greyD}\multirow{-2}{*}{JamOpt}  & ASR & 0.01 & 0.16 & 0.01 & 0.01 & 0.00 & 0.03 & 0.00 \\ 
\midrule
\multirow{2}{*}{AP} & ACC & 0.94 & 0.94 & 0.83 & 0.94 & 0.93 & 0.71 & 0.91 \\
 & ASR & 0.05 & 0.05 & 0.00 & 0.05 & 0.05 & 0.00 & 0.02 \\ 
\midrule
\rowcolor{greyD}  \cellcolor{greyD} 
& ACC & 0.99 & 0.99 & 0.84 & 0.98 & 0.98 & 0.84 & 0.99 \\
\rowcolor{greyD}  \cellcolor{greyD}\multirow{-2}{*}{BadRAG}  & ASR & 0.01 & 0.01 & 0.00 & 0.02 & 0.01 & 0.03 & 0.00 \\ 
\midrule
\multirow{2}{*}{Phantom} & ACC & 1.00 & 0.99 & 0.86 & 0.99 & 0.98 & 0.83 & 0.97 \\
 & ASR & 0.00 & 0.01 & 0.00 & 0.01 & 0.01 & 0.01 & 0.00 \\
\bottomrule
\end{tabular}
\end{table*}

\begin{table*}[!htbp]
\tiny
\centering
\addtolength{\tabcolsep}{3.0pt}
\caption{The results of all poisoning attacks against different advanced RAG methods on NQ-EX-L dataset.}
\label{tab:framework_nq_ex_l}
\begin{tabular}{l|c|ccccccc} 
\toprule
Attack & Metric & Naive RAG & AAR & SuRe & Adaptive-RAG & IRCoT & FLARE & RQRAG \\ 
\midrule
\multirow{2}{*}{BPRAG} & ACC & 0.89 & 0.76 & 0.84 & 0.87 & 0.86 & 0.77 & 0.92 \\
 & ASR & 0.03 & 0.04 & 0.02 & 0.01 & 0.03 & 0.00 & 0.06 \\ 
\midrule
\rowcolor{greyD}  \cellcolor{greyD} 
& ACC & 0.87 & 0.79 & 0.85 & 0.86 & 0.84 & 0.75 & 0.92 \\
\rowcolor{greyD}  \cellcolor{greyD}\multirow{-2}{*}{WPRAG} & ASR & 0.05 & 0.05 & 0.04 & 0.02 & 0.04 & 0.01 & 0.04 \\ 
\midrule
\multirow{2}{*}{BPI} & ACC & 0.90 & 0.69 & 0.84 & 0.88 & 0.88 & 0.76 & 0.93 \\
 & ASR & 0.08 & 0.25 & 0.08 & 0.09 & 0.08 & 0.00 & 0.04 \\ 
\midrule
\rowcolor{greyD}  \cellcolor{greyD}  & ACC & 0.86 & 0.76 & 0.82 & 0.87 & 0.86 & 0.78 & 0.95 \\
 \rowcolor{greyD}  \cellcolor{greyD}\multirow{-2}{*}{WPI} & ASR & 0.10 & 0.18 & 0.07 & 0.09 & 0.09 & 0.00 & 0.02 \\ 
\midrule
\multirow{2}{*}{AGGD} & ACC & 0.97 & 0.94 & 0.89 & 0.96 & 0.95 & 0.78 & 0.94 \\
 & ASR & 0.00 & 0.02 & 0.01 & 0.00 & 0.00 & 0.00 & 0.03 \\ 
\midrule
 \rowcolor{greyD}  \cellcolor{greyD}  & ACC & 0.97 & 0.80 & 0.90 & 0.97 & 0.95 & 0.76 & 0.96 \\
 \rowcolor{greyD}  \cellcolor{greyD}\multirow{-2}{*}{CRAG-AS}  & ASR & 0.00 & 0.16 & 0.00 & 0.00 & 0.00 & 0.00 & 0.01 \\ 
\midrule
\multirow{2}{*}{CRAG-AK} & ACC & 0.74 & 0.50 & 0.86 & 0.73 & 0.74 & 0.75 & 0.89 \\
 & ASR & 0.20 & 0.39 & 0.07 & 0.19 & 0.18 & 0.02 & 0.10 \\ 
\midrule
 \rowcolor{greyD}  \cellcolor{greyD}  & ACC & 0.98 & 1.00 & 0.84 & 0.98 & 0.97 & 0.84 & 0.96 \\
 \rowcolor{greyD}  \cellcolor{greyD}\multirow{-2}{*}{JamInject} & ASR & 0.02 & 0.00 & 0.03 & 0.02 & 0.02 & 0.03 & 0.00 \\ 
\midrule
\multirow{2}{*}{JamOracle} & ACC & 0.94 & 0.91 & 0.85 & 0.97 & 0.96 & 0.82 & 0.95 \\
 & ASR & 0.04 & 0.09 & 0.01 & 0.03 & 0.03 & 0.03 & 0.00 \\ 
\midrule
\rowcolor{greyD}  \cellcolor{greyD} & ACC & 0.98 & 0.86 & 0.86 & 0.98 & 0.97 & 0.84 & 0.96 \\
 \rowcolor{greyD}  \cellcolor{greyD}\multirow{-2}{*}{JamOpt}  & ASR & 0.01 & 0.12 & 0.01 & 0.01 & 0.00 & 0.03 & 0.00 \\ 
\midrule
\multirow{2}{*}{AP} & ACC & 0.95 & 0.96 & 0.84 & 0.94 & 0.94 & 0.71 & 0.91 \\
 & ASR & 0.03 & 0.04 & 0.00 & 0.06 & 0.05 & 0.01 & 0.01 \\ 
\midrule
 \rowcolor{greyD}  \cellcolor{greyD} 
& ACC & 0.99 & 0.99 & 0.89 & 0.99 & 0.98 & 0.83 & 0.99 \\
 \rowcolor{greyD}  \cellcolor{greyD}\multirow{-2}{*}{BadRAG}  & ASR & 0.01 & 0.01 & 0.00 & 0.01 & 0.01 & 0.04 & 0.00 \\ 
\midrule
\multirow{2}{*}{Phantom} & ACC & 1.00 & 0.99 & 0.92 & 0.99 & 0.98 & 0.83 & 0.99 \\
 & ASR & 0.00 & 0.01 & 0.00 & 0.01 & 0.01 & 0.01 & 0.00 \\
\bottomrule
\end{tabular}
\end{table*}

\begin{table*}[!htbp]
\tiny
\centering
\addtolength{\tabcolsep}{0pt}
\caption{The results of all poisoning attacks against multi-turn RAG on NQ dataset under different LLMs of RAG.}
\label{tab:multi_turn_results}
\begin{tabular}{l|c|ccccccc} 
\toprule
Attack & Metric & GPT-4o-mini & GPT-4.1-nano & GPT-4.1-mini & GPT-4.1 & Claude-3.7-Sonnet & Gemini-2.0-Flash & DeepSeek-V3 \\ 
\midrule
\multirow{3}{*}{BPRAG} & ACC & 0.49 & 0.42 & 0.58 & 0.38 & 0.34 & 0.30 & 0.45 \\
 & ASR & 0.44 & 0.46 & 0.38 & 0.26 & 0.39 & 0.42 & 0.38 \\
 & F1-score & 0.65 & 0.65 & 0.66 & 0.69 & 0.65 & 0.68 & 0.66 \\ 
\midrule
\rowcolor{greyD} \cellcolor{greyD} & ACC & 0.50 & 0.42 & 0.56 & 0.33 & 0.33 & 0.30 & 0.43 \\
\rowcolor{greyD} \cellcolor{greyD} & ASR & 0.38 & 0.43 & 0.36 & 0.25 & 0.34 & 0.41 & 0.38 \\
\rowcolor{greyD} \cellcolor{greyD}\multirow{-3}{*}{WPRAG} & F1-score & 0.58 & 0.61 & 0.57 & 0.59 & 0.57 & 0.56 & 0.62 \\ 
\midrule
\multirow{3}{*}{BPI} & ACC & 0.58 & 0.31 & 0.67 & 0.10 & 0.15 & 0.14 & 0.39 \\
 & ASR & 0.17 & 0.11 & 0.05 & 0.04 & 0.02 & 0.29 & 0.24 \\
 & F1-score & 0.52 & 0.50 & 0.54 & 0.51 & 0.53 & 0.52 & 0.52 \\ 
\midrule
\rowcolor{greyD} \cellcolor{greyD} & ACC & 0.60 & 0.34 & 0.64 & 0.15 & 0.14 & 0.17 & 0.38 \\
\rowcolor{greyD} \cellcolor{greyD} & ASR & 0.16 & 0.08 & 0.03 & 0.02 & 0.03 & 0.27 & 0.19 \\
\rowcolor{greyD} \cellcolor{greyD}\multirow{-3}{*}{WPI} & F1-score & 0.47 & 0.47 & 0.48 & 0.48 & 0.49 & 0.46 & 0.47 \\ 
\midrule
\multirow{3}{*}{AGGD} & ACC & 0.52 & 0.44 & 0.59 & 0.32 & 0.27 & 0.31 & 0.51 \\
 & ASR & 0.30 & 0.39 & 0.28 & 0.17 & 0.26 & 0.32 & 0.27 \\
 & F1-score & 0.52 & 0.50 & 0.49 & 0.46 & 0.49 & 0.51 & 0.50 \\ 
\midrule
\rowcolor{greyD} \cellcolor{greyD} & ACC & 0.44 & 0.35 & 0.57 & 0.12 & 0.13 & 0.10 & 0.25 \\
\rowcolor{greyD} \cellcolor{greyD} & ASR & 0.34 & 0.25 & 0.24 & 0.27 & 0.25 & 0.43 & 0.41 \\
\rowcolor{greyD} \cellcolor{greyD}\multirow{-3}{*}{CRAG-AS} & F1-score & 0.49 & 0.50 & 0.51 & 0.54 & 0.53 & 0.53 & 0.51 \\ 
\midrule
\multirow{3}{*}{CRAG-AK} & ACC & 0.36 & 0.20 & 0.41 & 0.15 & 0.14 & 0.13 & 0.25 \\
 & ASR & 0.48 & 0.39 & 0.42 & 0.32 & 0.36 & 0.53 & 0.52 \\
 & F1-score & 0.64 & 0.63 & 0.63 & 0.66 & 0.63 & 0.66 & 0.63 \\ 
\midrule
\rowcolor{greyD} \cellcolor{greyD} & ACC & 0.51 & 0.34 & 0.46 & 0.29 & 0.15 & 0.33 & 0.36 \\
\rowcolor{greyD} \cellcolor{greyD} & ASR & 0.47 & 0.62 & 0.48 & 0.69 & 0.80 & 0.61 & 0.62 \\
\rowcolor{greyD} \cellcolor{greyD}\multirow{-3}{*}{JamInject} & F1-score & 0.42 & 0.42 & 0.42 & 0.43 & 0.43 & 0.42 & 0.40 \\ 
\midrule
\multirow{3}{*}{JamOracle} & ACC & 0.73 & 0.36 & 0.54 & 0.24 & 0.09 & 0.23 & 0.53 \\
 & ASR & 0.21 & 0.61 & 0.32 & 0.72 & 0.70 & 0.55 & 0.40 \\
 & F1-score & 0.54 & 0.55 & 0.54 & 0.52 & 0.52 & 0.54 & 0.52 \\ 
\midrule
\rowcolor{greyD} \cellcolor{greyD} & ACC & 0.79 & 0.62 & 0.74 & 0.46 & 0.26 & 0.53 & 0.63 \\
\rowcolor{greyD} \cellcolor{greyD} & ASR & 0.15 & 0.31 & 0.19 & 0.45 & 0.67 & 0.34 & 0.33 \\
\rowcolor{greyD} \cellcolor{greyD}\multirow{-3}{*}{JamOpt} & F1-score & 0.23 & 0.24 & 0.23 & 0.24 & 0.24 & 0.24 & 0.26 \\ 
\midrule
\multirow{3}{*}{AP} & ACC & 0.66 & 0.60 & 0.66 & 0.41 & 0.25 & 0.22 & 0.67 \\
 & ASR & 0.30 & 0.37 & 0.30 & 0.55 & 0.70 & 0.72 & 0.29 \\
 & F1-score & 0.00 & 0.00 & 0.00 & 0.00 & 0.00 & 0.00 & 0.00 \\ 
\midrule
\rowcolor{greyD} \cellcolor{greyD} & ACC & 0.75 & 0.57 & 0.71 & 0.32 & 0.23 & 0.26 & 0.54 \\
\rowcolor{greyD} \cellcolor{greyD} & ASR & 0.20 & 0.40 & 0.26 & 0.64 & 0.69 & 0.67 & 0.40 \\
\rowcolor{greyD} \cellcolor{greyD}\multirow{-3}{*}{BadRAG} & F1-score & 0.00 & 0.00 & 0.00 & 0.00 & 0.00 & 0.00 & 0.00 \\ 
\midrule
\multirow{3}{*}{Phantom} & ACC & 0.76 & 0.55 & 0.69 & 0.35 & 0.21 & 0.29 & 0.54 \\
 & ASR & 0.20 & 0.40 & 0.27 & 0.61 & 0.73 & 0.65 & 0.43 \\
 & F1-score & 0.00 & 0.00 & 0.00 & 0.00 & 0.00 & 0.00 & 0.00 \\
\bottomrule
\end{tabular}

\end{table*}

\begin{table*}[!htbp]
\tiny
\centering
\addtolength{\tabcolsep}{0pt}
\caption{The results of all poisoning attacks against multimodal RAG on InfoSeek datasets across different VLMs. ``Adaptive+BPRAG'' means we adapt the BPRAG attack to the multimodal RAG setting.}
\label{tab:multi_modal_results}
\begin{tabular}{l|c|cccccc} 
\toprule
Attack & Metric & GPT-4o-mini & GPT-4o & GPT-4.1-mini & GPT-4.1 & Claude-3.7-Sonnet & Gemini-2.0-Flash \\ 
\midrule
\multirow{2}{*}{Dirty-label} & ACC &0.01  &0.00  &0.02  &0.03  &0.05  &0.04  \\
 & ASR &0.93  &0.96  &0.90  &0.94  &0.74  &0.68 \\ 
\midrule
 \rowcolor{greyD}  \cellcolor{greyD}  & ACC &0.07  &0.10  &0.10  &0.11  &0.12  &0.09  \\
 \rowcolor{greyD}  \cellcolor{greyD}\multirow{-2}{*}{Adaptive+BPRAG}  & ASR &0.81  &0.81  &0.82  &0.85  &0.79  &0.66  \\ 
\midrule
\multirow{2}{*}{Adaptive+BPI} & ACC &0.00  &0.00  &0.01  &0.01  &0.05  &0.00  \\
 & ASR &0.98  &0.95  &0.88  &0.45  &0.89  &0.99  \\ 
\midrule
 \rowcolor{greyD}  \cellcolor{greyD} & ACC &0.02  &0.03  &0.00  &0.00  &0.20  &0.03  \\
 \rowcolor{greyD}  \cellcolor{greyD}\multirow{-2}{*}{Adaptive+CRAG-AS} & ASR &0.97  &0.90  &0.85  &0.98  &0.75  &0.70  \\ 
\midrule
\multirow{2}{*}{Adaptive+CRAG-AK} & ACC &0.01  &0.04  &0.01  &0.02  &0.05  &0.03  \\
 & ASR &0.97  &0.95  &0.87  &0.98  &0.92  &0.83  \\ 
\midrule
 \rowcolor{greyD}  \cellcolor{greyD} 
 & ACC &0.00  &0.00  &0.03  &0.00  &0.05  &0.00  \\
  \rowcolor{greyD}  \cellcolor{greyD}\multirow{-2}{*}{Adaptive+JamInject}  & ASR &1.00  &1.00  &0.97  &1.00  &0.95  &1.00  \\ 
\midrule
\multirow{2}{*}{Adaptive+JamOracle} & ACC &0.11  &0.19  &0.20  &0.32  &0.15  &0.04  \\
 & ASR &0.89  &0.81  &0.80  &0.68  &0.85  &0.96  \\ 

\bottomrule
\end{tabular}
\end{table*}

\begin{table*}[!htbp]
\tiny
\centering
\addtolength{\tabcolsep}{-2.0pt}
\caption{The results of all poisoning attacks against RAG based LLM agent systems across various LLMs. ``Adaptive+BPRAG'' means we adapt the BPRAG attack to the RAG based LLM agent systems.}
\label{tab:llm_agent_results}

\begin{tabular}{l|c|cccccccccl}
\toprule

Attack                                                                          & Metric &GPT-4o& GPT-4o-mini & GPT-4.1-mini & Claude-3-7-Sonnet & DeepSeek-V3 & Gemini-2.0-Flash & Hunyuan-Lite \\
 \midrule
 & ACC  &0.12   & 0.46        & 0.43         & 0.37              & 0.63        & 0.27             & 0.46         \\
 \multirow{-2}{*}{AgentPoison}  & ASR &0.83    & 0.35        & 0.49         & 0.04              & 0.16        & 0.59             & 0.26         \\
\midrule
  \rowcolor{greyD}  \cellcolor{greyD}  & ACC   &0.18  & 0.35        & 0.4          & 0.44              & 0.48        & 0.25             & 0.27         \\
 
  \rowcolor{greyD}  \cellcolor{greyD}\multirow{-2}{*}{Adaptive+BPRAG} & ASR &0.25    & 0.4         & 0.42         & 0.47              & 0.39        & 0.67             & 0.66         \\
 \midrule
      & ACC   &0.05  & 0.28        & 0.36         & 0.6               & 0.25        & 0.04             & 0.41         \\
\multirow{-2}{*}{Adaptive+BPI}  & ASR  &0.28   & 0.44        & 0.44         & 0.33              & 0.75        & 0.93             & 0.54         \\
 \midrule
   \rowcolor{greyD}  \cellcolor{greyD}  & ACC &0.07      & 0.22        & 0.33         & 0.48              & 0.14        & 0.21             & 0.16         \\
  \rowcolor{greyD}  \cellcolor{greyD}\multirow{-2}{*}{Adaptive+CRAG-AS}   & ASR  &0.19   & 0.37        & 0.33         & 0.46              & 0.86        & 0.41             & 0.79         \\
 \midrule
 & ACC   &0.03  & 0.11        & 0.26         & 0.32              & 0.15        & 0.03             & 0.14         \\
 \multirow{-2}{*}{Adaptive+CRAG-AK}   & ASR  &0.51   & 0.65        & 0.59         & 0.66              & 0.84        & 0.94             & 0.79         \\
 \midrule
   \rowcolor{greyD}  \cellcolor{greyD}  & ACC  &0.12   & 0.21        & 0.16         &   0.45                & 0.16        & 0.18             & 0.21         \\
  \rowcolor{greyD}  \cellcolor{greyD}\multirow{-2}{*}{Adaptive+JamInJect}    & ASR    &0.86 & 0.77        & 0.81         &   0.43                & 0.82        & 0.79             & 0.67         \\
 \midrule
   & ACC   &0.21  & 0.5         & 0.56         & 0.61              & 0.6         & 0.38             & 0.56         \\
 \multirow{-2}{*}{Adaptive+JamOracle}   & ASR  &0.75   & 0.36        & 0.3          & 0.21              & 0.16        & 0.55             & 0.13         \\
 \midrule

 \rowcolor{greyD}  \cellcolor{greyD} & ACC  &0.07   & 0.4         & 0.38         & 0.54              & 0.58        & 0.31             & 0.51         \\
  \rowcolor{greyD}  \cellcolor{greyD}\multirow{-2}{*}{Adaptive+BadRAG}   & ASR  &0.91   & 0.47        & 0.5          & 0.35              & 0.2         & 0.64             & 0.25     \\
 \bottomrule
\end{tabular}
\end{table*}

\end{document}